\documentclass[12pt]{article}
\usepackage{pgfplots}
\usepackage{graphicx,psfrag,epsf}
\usepackage{enumerate}
\usepackage[authoryear]{natbib}
\usepackage{bibunits}
\usepackage{url} 
\usepackage{tikz}
\usepackage{collcell}
\usepackage{enumitem}
\usepackage{fancyhdr}
\usepackage{lipsum} 

\usepackage{caption}
\usepackage{diagbox}
\usepackage{rotating}
\usepackage{booktabs}

\usepackage{xr}
\makeatletter

\newcommand{\biasa}[1]{%
        \ifdim #1 pt > 1 pt
            \pgfmathsetmacro{\PercentColor}{max(min(100.0*ln(#1)/ln(70.1),100.0),0.00)} %
        \colorbox{yellow!\PercentColor!green}{\makebox[3em][l]{#1}\vphantom{Ag}}\hspace{-1.5em}
        \else
            \ifdim #1 pt < -1 pt
            \pgfmathsetmacro{\PercentColor}{max(min(100.0*ln(-#1)/ln(70.1),100.0),0.00)} %
        \colorbox{yellow!\PercentColor!green}{\makebox[3em][l]{#1}\vphantom{Ag}}\hspace{-1.5em}
        \else
        \pgfmathsetmacro{\PercentColor}{0} %
        \colorbox{yellow!\PercentColor!green}{\makebox[3em][l]{#1}\vphantom{Ag}}\hspace{-1.5em}
        \fi
        \fi
}
\newcolumntype{A}{>{\collectcell\biasa}c<{\endcollectcell}}

\newcommand{\RMSEa}[1]{%
        \ifdim #1 pt > 10 pt
            \pgfmathsetmacro{\PercentColor}{max(min(100.0*ln(#1-9)/ln(70),100.0),0.00)} %
        \colorbox{yellow!\PercentColor!green}{\makebox[4em][l]{#1}\vphantom{Ag}}
        \else
            \pgfmathsetmacro{\PercentColor}{0} %
        \colorbox{yellow!\PercentColor!green}{\makebox[4em][l]{#1}\vphantom{Ag}}
        \fi

}
\newcolumntype{B}{>{\collectcell\RMSEa}c<{\endcollectcell}}

\newcommand{\CIlengtha}[1]{%
        \ifdim #1 pt > 20 pt
            \pgfmathsetmacro{\PercentColor}{max(min(100.0*ln(#1-19)/ln(40),100.0),0.00)} %
        \hspace{-2.5em}\colorbox{yellow!\PercentColor!green}{\makebox[4em][l]{#1}\vphantom{Ag}}
        \else
            \pgfmathsetmacro{\PercentColor}{0} %
        \hspace{-2.5em}\colorbox{yellow!\PercentColor!green}{\makebox[4em][l]{#1}\vphantom{Ag}}
        \fi}
\newcolumntype{C}{>{\collectcell\CIlengtha}c<{\endcollectcell}}

\newcommand{\CI}[1]{%
    \begingroup
    \def\Value{#1}%
    \pgfmathsetmacro{\TempVal}{\Value}%
    \ifdim \TempVal pt > 96pt
        \pgfmathsetmacro{\PercentColor}{max(min((\TempVal - 96)*25, 100.0), 0.00)}%
    \else
        \ifdim \TempVal pt < 94pt
            \pgfmathsetmacro{\PercentColor}{max(min((94 - \TempVal)*25, 100.0), 0.00)}%
        \else
            \pgfmathsetmacro{\PercentColor}{0}%
        \fi
    \fi
    \hspace{-1.5em}\colorbox{yellow!\PercentColor!green}{\makebox[3em][l]{\Value\vphantom{Ag}}}%
    \endgroup
}

\newcolumntype{D}{>{\collectcell\CI}c<{\endcollectcell}}

\newcommand{\CIlengthb}[1]{%
        \ifdim #1 pt > 20 pt
            \pgfmathsetmacro{\PercentColor}{max(min(100.0*ln(#1-19)/ln(50),100.0),0.00)} %
        \hspace{-2.5em}\colorbox{yellow!\PercentColor!green}{\makebox[4em][l]{#1}\vphantom{Ag}}
        \else
            \pgfmathsetmacro{\PercentColor}{0} %
        \hspace{-2.5em}\colorbox{yellow!\PercentColor!green}{\makebox[4em][l]{#1}\vphantom{Ag}}
        \fi}
\newcolumntype{E}{>{\collectcell\CIlengthb}c<{\endcollectcell}}

\newcommand{\CIlengthc}[1]{%
        \ifdim #1 pt > 30 pt
            \pgfmathsetmacro{\PercentColor}{max(min(100.0*ln(#1-29)/ln(40),100.0),0.00)} %
        \hspace{-2.5em}\colorbox{yellow!\PercentColor!green}{\makebox[4em][l]{#1}\vphantom{Ag}}
        \else
            \pgfmathsetmacro{\PercentColor}{0} %
        \hspace{-2.5em}\colorbox{yellow!\PercentColor!green}{\makebox[4em][l]{#1}\vphantom{Ag}}
        \fi}
\newcolumntype{F}{>{\collectcell\CIlengthc}c<{\endcollectcell}}

\newcommand*{\addFileDependency}[1]{
\typeout{(#1)}
\@addtofilelist{#1}
\IfFileExists{#1}{}{\typeout{No file #1.}}
}\makeatother

\newcommand{\blind}{1}

\usepackage[margin=1in]{geometry}

\usepackage{graphicx}
\usepackage{amsmath}
\usepackage{amsthm}
\usepackage{bm}
\usepackage{algorithm}
\usepackage{algpseudocode}

\usepackage{}
\DeclareFontFamily{U}{mathx}{\hyphenchar\font45}
\DeclareFontShape{U}{mathx}{m}{n}{<-> mathx10}{}
\DeclareSymbolFont{mathx}{U}{mathx}{m}{n}
\DeclareMathAccent{\widebar}{0}{mathx}{"73}

\usepackage{amssymb}
\usepackage{amsfonts}
\usepackage{mathtools}
\usepackage[english]{babel}
\usepackage{footnote}
\makesavenoteenv{tabular}
\usepackage{graphicx,psfrag,epsf}
\usepackage{enumerate}
\usepackage{url} 
\usepackage{comment}
\usepackage{colortbl}
\usepackage{booktabs}   
\usepackage{bm}
\usepackage{enumitem}

\usepackage[T1]{fontenc}
\usepackage[latin9]{inputenc}
\usepackage{xcolor}
\usepackage{amsmath}
\usepackage{amssymb}

\makeatletter

\@ifundefined{date}{}{\date{}}

\usepackage{colortbl}

\usepackage{pdfcomment}
\usepackage{tcolorbox}
\usepackage{xcolor}
\usepackage{colortbl}
\usepackage{soul}

\definecolor{rahmen}{RGB}{0,73,114}
\definecolor{grund}{RGB}{238,241,251}  
\definecolor{schrift}{RGB}{0,73,114}
\definecolor{apurple}{RGB}{217,197,224}
\definecolor{ared}{RGB}{251,215,230}
\definecolor{ayellow}{RGB}{255,251,203}
\definecolor{azure}{RGB}{240,255,255}
\definecolor{lightblue}{RGB}{240,248,255}
\definecolor{steelblue}{RGB}{176,196,222}
\definecolor{slateggray}{RGB}{119,136,153}
\definecolor{darkgreen}{RGB}{60,179,113}
\definecolor{orange}{RGB}{255,165,0}
\definecolor{salmon}{RGB}{255,105,180}

\usepackage{latexsym}
\usepackage{multirow}\usepackage{bm}
\usepackage{url}

\usepackage{lineno}
\usepackage{subcaption}
\usepackage{caption} 
\usepackage{multirow}
\usepackage{threeparttable}
\usepackage{wasysym}
\usepackage{framed}
\usepackage{pgfgantt,rotating}
\usepackage{multirow}
\usepackage{blindtext}
\usepackage{times}
\usepackage[above]{placeins}

\usepackage[outdir=./]{epstopdf}
\usepackage{verbatim} 
\usetikzlibrary{arrows,shapes.arrows,shapes.geometric,shapes.multipart,backgrounds,decorations.pathmorphing,positioning,fit,automata}
\tikzset{
	>=stealth',
	true/.style={
		rectangle,
		draw=black, very thick,
		text width=6.5em,
		minimum height=3em,
		text centered,
		fill=gray, opacity = 0.5},
	punkt/.style={
		rectangle,
		rounded corners,
		draw=black, very thick,
		text width=6.5em,
		minimum height=3em,
		text centered},
	est/.style={
		circle,
		draw=black, very thick,
		text centered},
			estblue/.style={
		circle,
		draw=blue, very thick,
		text centered},
	shade/.style={
		circle,
		draw=white, very thick, fill=gray!50,
		text centered},
	weight/.style={
		circle,
		draw=black, very thick,
		text width=6.5em,
		minimum height=3em,
		text centered},
	pil/.style={
		->,
		thick,
		shorten <=2pt,
		shorten >=2pt,},
	pilred/.style={
		->, red,
		thick,
		shorten <=2pt,
		shorten >=2pt,},
	pilblue/.style={
		->,blue,
		thick,
		shorten <=2pt,
		shorten >=2pt,},
	pilgray/.style={
		dashed,gray,
		thick,
  linewidth=0.3mm,
		shorten <=2pt,
		shorten >=2pt,},
	double/.style={
		<->,
		thick,
		shorten <=2pt,
		shorten >=2pt,},
	dash/.style={
		dashed,
		thick,
		shorten <=2pt,
		shorten >=2pt,},
	dashdouble/.style={
		<->,
		dashed,
		thick,
		shorten <=2pt,
		shorten >=2pt,}
}
\DeclareMathOperator*{\argmin}{arg\,min}
\usepackage{setspace}

\newtheorem{theorem}{Theorem}
\newtheorem{lemma}{Lemma} 
\newtheorem{proposition}{Proposition} 
\newtheorem{remark}{Remark}

\newtheorem{assumption}{Assumption}

\newcommand{\ind}{\perp \!\!\! \perp }

\newcommand{\tred}{\textcolor{red}}

\usepackage{sectsty}
\sectionfont{\vspace{-0.8em}}  
\subsectionfont{\vspace{-0.5em}}

\def\T{{ \mathrm{\scriptscriptstyle T} }}

\usepackage{float}
\usepackage{graphicx}
\usepackage{enumerate}
\usepackage{url} 

\usepackage{xr}
\makeatletter

\def\T{{ \mathrm{\scriptscriptstyle T} }}
\pgfplotsset{compat=1.18}
\begin{document}

\def\spacingset#1{\renewcommand{\baselinestretch}%
{#1}\small\normalsize} \spacingset{1}

\if1\blind
{
  \vspace{-25pt}
  \title{\bf \Large{A Robust Framework for Two-Sample Mendelian Randomization under Population Heterogeneity}}
   \date{}
   \author{}
  \maketitle
  
  \begin{center}
  \vspace{-50pt}
  \author{\large{Dingke Tang$^{1}$, Xuming He$^{2}$, and Shu Yang$^{3}$ \footnote{Address for correspondence: Shu Yang, Department of Statistics, North Carolina State University, Raleigh, North Carolina 27607, U.S.A. Email: syang24@ncsu.edu
}\\ 
 \vspace{15pt}
 $^{1}$Department of Mathematics and Statistics, University of Ottawa,\hspace{5pt} \\
 $^{2}$Department of Statistics and Data Science, Washington University in St. Louis,\hspace{5pt}
\hspace{5pt}$^{3}$Department of Statistics, North Carolina State University}}
  \vspace{11pt}
  \end{center}
  \date{}
} \fi

\if0\blind
{
  \vspace{-35pt}
  \title{\bf \Large{A Robust Framework for Two-Sample Mendelian Randomization under Population Heterogeneity}}
   \date{}
   \author{}
  \maketitle
    \vspace{10pt}

} \fi

\vspace{-0.8cm}
\begin{abstract}
\doublespacing
Mendelian randomization is a powerful tool for causal inference in observational studies. The two-sample summary-data design, which estimates genetic associations with exposures and outcomes in separate cohorts, is the most widely used Mendelian randomization approach in large-scale genomic studies. 
However, this approach relies on a strong assumption of population \textit{homogeneity} across the two samples. In practice, available samples often differ in ancestry, demographics, socioeconomic factors, covariate adjustment, and measurement protocols. Violations of the homogeneity assumption can bias causal effect estimates and undermine the credibility of Mendelian randomization findings. We introduce a robust, model-free Mendelian randomization framework that directly addresses population \textit{heterogeneity} in the two-sample summary-data setting. Our method avoids parametric assumptions about population differences and is designed to address real-world challenges, including measurement error, weak instruments, and pleiotropy.  We show that the proposed estimator is consistent and asymptotically normal under heterogeneous designs, and may offer efficiency gains over the classic estimator even in homogeneous settings.
Through numerical simulations and a real data analysis for estimating the causal effect of body mass index on high-density lipoprotein cholesterol across ancestrally diverse populations, we demonstrate the practical utility, stability, and robustness of our approach. 
\end{abstract}

\noindent%
\spacingset{1.7} 
\newpage

\begin{bibunit}
\defaultbibliographystyle{asa}
\defaultbibliography{ref}

\section{Introduction}
The increasing availability of large-scale biobanks and genomic data has opened new frontiers for causal inference in observational research. Mendelian randomization (MR) \citep{lawlor2008mendelian} leverages genetic variants, such as single-nucleotide polymorphisms (SNPs), as instrumental variables (IVs). It emulates randomized experiments by exploiting the random assortment of alleles at conception. This design provides a principled way to identify causal effects of modifiable exposures when controlled experiments are infeasible or unethical. By integrating genetic epidemiology with instrumental variable methodology, MR has emerged as a key tool for strengthening causal claims in complex biomedical and epidemiological studies.

A commonly adopted framework in MR is the two-sample summary-data design, where genetic associations with an exposure and an outcome are estimated from two independent genome-wide association studies (GWAS)  \citep{pierce2013efficient}.
The exposure GWAS provides marginal SNP-exposure associations \( \widehat{\gamma}_j \), representing the association between variant \( Z_j \) and the exposure, while the outcome GWAS yields marginal SNP-outcome associations \( \widehat{\Gamma}_j \), representing the association between \( Z_j \) and the outcome. Under the standard IV assumptions, if the SNP \( Z_j \) is a valid instrumental variable, the underlying structural relationship satisfies \( \Gamma_j = \beta_0 \gamma_j.\) 
Here \( \beta_0 \) denotes the causal effect of the exposure on the outcome and is independent of the specific choice of \( Z_j \). Consequently, exploring the relationship between \( \widehat{\Gamma}_j \) on \( \widehat{\gamma}_j \) using appropriate models enables consistent estimation of \( \beta_0 \).

However, this approach hinges on a strong assumption of population \textit{homogeneity} across the two samples. In reality, exposure and outcome GWAS are often conducted in different populations and processed through distinct analytic pipelines, leading to discrepancies in the resulting population parameters. Certain SNPs may influence a trait only within particular subgroups. For example, gene--sex interactions  \citep{zhu2023amplification} and gene--environment interactions  \citep{manuck2014gene} can shape the genetic architecture of complex traits such as body mass index (BMI) through distinct biological pathways. Consequently, SNPs that are significant for a trait in one population may be insignificant in another, inducing population  \textit{heterogeneity} across the treatment and outcome GWAS.

A central difficulty with population \textit{heterogeneity} is the resulting parameter discrepancy between the GWAS estimates across samples. Specifically, the parameters $(\Gamma_j^{\mathrm{Tr}}, \gamma_j^{\mathrm{Tr}}, \beta_0)$ from the treatment GWAS need not coincide with $(\Gamma_j^{\mathrm{Ou}}, \gamma_j^{\mathrm{Ou}}, \beta_0)$ from the outcome GWAS. This discrepancy implies that the observed IV-outcome association $\widehat{\Gamma}_j^{\mathrm{Ou}}$ and IV-treatment association $\widehat{\gamma}_j^{\mathrm{Tr}}$ may no longer satisfy the linear structural IV relationship $\Gamma_j = \beta_0 \gamma_j$, thereby threatening identifiability of the causal parameter $\beta_0$. Consequently, the effect estimates might be biased and, in extreme cases, even reverse their direction, leading to conclusions opposite to the true causal relationship  \citep{li2025mind}. These risks undermine both the validity and reliability of MR findings, particularly when results are used to inform biomedical understanding or public health recommendations. Importantly, \cite{li2025mind}
also showed that some bias persists even when GWAS samples are drawn from populations of the same continental ancestry, a common practice in two-sample MR analyses that is often assumed to ensure comparability  \citep{pierce2013efficient, zhao2018statistical}. Addressing heterogeneity is therefore not only statistically essential but also critical for the credibility of MR studies.

\begin{figure}
\centering
\includegraphics[width=0.35\linewidth]{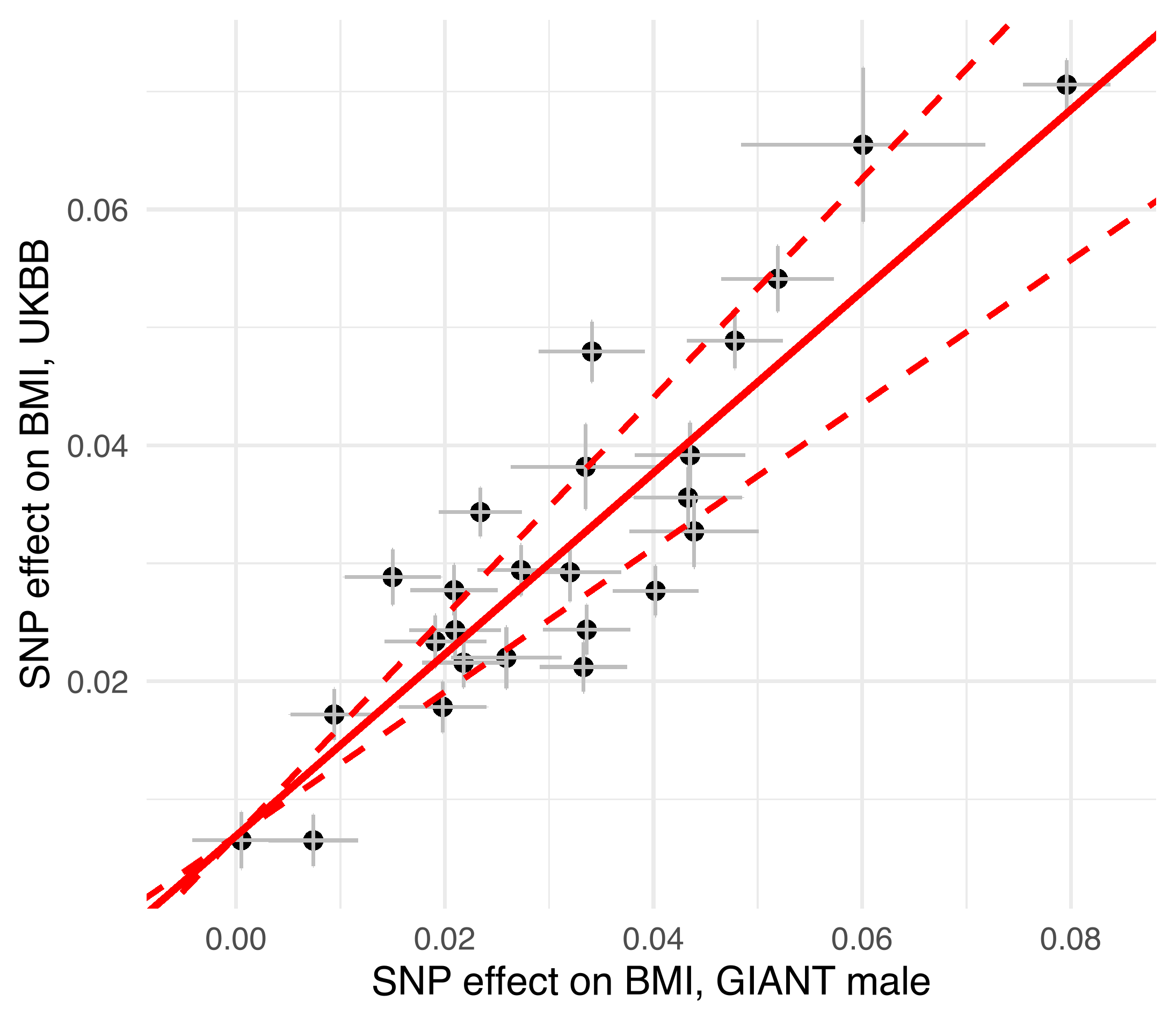}
\caption{Scatter plot of $\hat{\gamma}_j$ (UKBB data) versus $\hat{\gamma}_j$ ( GIANT male data) in the BMI--BMI example. 
Each point is accompanied by standard errors of $\hat{\gamma}_j^{\mathrm{Ou}}$ and $\hat{\gamma}_j^{\mathrm{Tr}}$ on the vertical and horizontal sides. 
{For presentation purposes only, we flipped the points with negative values on the $x$-axis so that all $\hat{\gamma}_j^{\mathrm{Tr}}$ are positive.} 
Solid lines are fitted using the MR--Egger method (slope $=0.769$). 
Dashed lines represent the 95\% confidence interval of the slope: $(0.610, 0.929)$. The estimated slope deviates substantially from $1$, indicating heterogeneity in $\gamma_j$ across the two populations.}
  \label{fig:bmibmimregger}
\end{figure}

To statistically assess heterogeneity across two samples, we introduce in Section~\ref{sec:heterogeneitytest} a test of the null hypothesis \(H_0: \gamma_j^{\mathrm{Tr}} = \gamma_j^{\mathrm{Ou}}\). In earlier work, \citet{zhao2018statistical} studied the causal effect of BMI on systolic blood pressure using MR, where the treatment data were BMI in males from the {\textbf{G}}enetic {\textbf{I}}nvestigation of {\textbf{AN}}thropometric {\textbf{T}}raits  \citep[GIANT]{locke2015genetic} consortium and the outcome data were SBP from the UK Biobank (UKBB). We applied our test to assess whether the genetic effects of BMI are homogeneous across the GIANT male and UKBB population. The test yielded a $p$-value of \(5 \times 10^{-4}\), providing strong evidence of heterogeneity between the treatment and outcome GWAS. We also used the MR-Egger estimator  \citep{bowden2015mendelian} to assess the causal effect of BMI (data from GIANT males) on BMI (data from UKBB), which gave a point estimate of $0.769$ ($95\%$ CI: $0.610, 0.929$). This provides strong evidence against the homogeneity assumption, under which the causal effect would equal one (see Figure~\ref{fig:bmibmimregger}). Such discrepancies are common; indeed, our empirical analyses also reveal substantial differences in SNP-BMI associations across GWAS.

An approach to avoid heterogeneity is to obtain both the SNP-treatment and SNP-outcome associations from the same population, often referred to as one-sample MR  \citep{burgess2023guidelines}. However, this approach suffers from dependency bias, since both sets of summary statistics are estimated from the same data  \citep{burgess2016bias}, leading to dependence between summary statistics and biased causal effect estimates. This limitation motivates the development of methods that 
mitigate such dependence while retaining robustness to heterogeneity across populations. 

We propose a novel procedure, the MR-Wald estimator, which simultaneously accounts for heterogeneity between $\widehat{\gamma}_j^{\mathrm{Tr}}$ and $\widehat{\gamma}_j^{\mathrm{Ou}}$ in two-sample MR, and the dependency between $\widehat{\gamma}_j^{\mathrm{Ou}}$ and $\widehat{\Gamma}_j^{\mathrm{Ou}}$ arising in one-sample MR. Our method extends the idea of the classical Wald estimator \citep{wald1940fitting} in the instrumental variable framework, which addresses unmeasured confounding using individual-level data. In contrast, we focus on summary statistics derived from two distinct samples. The key innovation is to treat $\widehat{\gamma}^{\mathrm{Tr}}_j$ as a hypothetical instrument (Figure \ref{fig:comparison}), thereby correcting the dependency bias between $\widehat{\gamma}^{\mathrm{Ou}}_j$ and $\widehat{\Gamma}^{\mathrm{Ou}}_j$.
The contributions of our  work are threefold: (i) the proposed estimator avoids imposing a parametric relationship between $\gamma_j^{\mathrm{Tr}}$ and $\gamma_j^{\mathrm{Ou}}$, yet still achieves the parametric rate of convergence;  
(ii) it attains improved efficiency over existing MR approaches in the homogeneous setting; and  
(iii) it introduces a new framework that employs robust loss functions to accommodate possibly invalid instruments arising from pleiotropy, the biological phenomenon in which a single genetic variant influences multiple, seemingly unrelated traits. In the presence of pleiotropy, the instrument $Z_j$ influences the outcome $Y$ through pathways not mediated by the exposure $X$, which violates the linear structural model $\Gamma_j = \beta_0 \gamma_j$.

\subsection{Prior works and our contribution}
There is a substantial and expanding body of literature addressing various challenges in MR, including weak instrument bias  \citep{ye2021debiasedmain, zhao2018statistical}, pleiotropy  \citep{
bowden2015mendelian, bowden2016consistent, hartwig2017robust}, and selection bias  \citep{gkatzionis2019contextualizing}. However, these works typically assume that the two sets of summary-level data are drawn from homogeneous populations.

Existing methods addressing heterogeneity in two-sample MR, such as \citet{zhao2019two}, primarily focus on correcting distributional discrepancies, for example differences in minor allele frequencies and linkage disequilibrium patterns. However, they still maintain the assumption of structural homogeneity, namely
\(\gamma_j^{\mathrm{Tr}} = \gamma_j^{\mathrm{Ou}}.
\)
In contrast, our proposed method accommodates both distributional and structural heterogeneity, allowing the SNP-exposure effects to differ across populations,
\(
\gamma_j^{\mathrm{Tr}} \neq \gamma_j^{\mathrm{Ou}}.
\)
This generalization is essential when causal effects are not invariant across populations, as illustrated by the observed heterogeneity in SNP-BMI associations. Furthermore, unlike the approach of \citet{zhao2019two}, which relies on an external reference dataset to estimate the covariance structure, our method does not require such information. It is fully operational using only two-sample summary-level data. \citet{iong2024latent} considered heterogeneity at the variant level, allowing different SNPs to follow different causal pathways in the biological process. However, their framework does not account for heterogeneity between the two samples.

The heterogeneity between $\gamma_j^{\mathrm{Tr}}$ and $\gamma_j^{\mathrm{Ou}}$ can be addressed in a one-sample MR framework using individual-level data and sample splitting  \citep{bibaut2024nonparametric}. Suppose individual-level data are available for the outcome $\{Y_i\}_{i=1}^n$, treatment $\{D_i\}_{i=1}^n$, and instruments $\{Z_{i,j}\}_{i=1}^n$ for $j = 1, \ldots, p$. One can estimate $\widehat{\gamma}_j$ using a first fold of the sample and $\widehat{\Gamma}_j$ using a second, non-overlapping fold. Sample splitting ensures independence between these two estimates while maintaining consistency with respect to the same underlying parameter, since both folds are drawn from the same population. However, this approach is infeasible in a summary data setting where individual-level data are unavailable. Conversely, our approach eliminates the need for splitting samples and can be directly applied to the current two-sample MR framework. 

Our method is distinguished from these aforementioned approaches in two key ways:  
(a) We explicitly account for structural heterogeneity by allowing \(\gamma_j^{\mathrm{Tr}} \ne \gamma_j^{\mathrm{Ou}}\), without imposing any parametric relationship between them;
(b) Our procedure requires only summary-level data, which makes the proposed method fully compatible with the classical two-sample MR framework. 

\subsection{Organization}
The remainder of the paper is organized as follows. Section~\ref{sec:setup} introduces the formal background and problem setup. In Section~\ref{sec:algorithm}, we present the proposed estimator and establish its theoretical properties. Section~\ref{sec:ple} extends our framework to allow for potentially invalid instruments arising from pleiotropic effects. In Section~\ref{sec:simulation}, we conduct extensive simulation studies, comparing our method with a range of existing approaches. The proposed estimator demonstrates the most favorable performance in both homogeneous and heterogeneous settings. Section~\ref{sec:realdata_new} illustrates the practical utility of our method through an application to summary-level data, assessing the causal effect of BMI on high-density lipoprotein (HDL) cholesterol across multiple populations. We conclude with a discussion in Section~\ref{sec:discussion}.

\section{Preliminary}
\label{sec:setup}

\subsection{Notation}
For a vector $\mathbf{x} \in \mathbb{R}^n$ and $q > 0$, we define the $\ell_q$ norm as  
\(
\|\mathbf{x}\|_q = \Big( \sum_{i=1}^n |x_i|^q \Big)^{1/q},
\) 
and the $\ell_\infty$ norm as $\|\mathbf{x}\|_\infty = \max_i |x_i|$.  For two positive sequences $\{a_j\}_{j \ge 1}$ and $\{b_j\}_{j \ge 1}$, we write $a_j = O(b_j)$ or $a_j \lesssim b_j$ if there exists a constant $C > 0$ such that $a_j \le C b_j$. We write $a_j = o(b_j)$ if $a_j/b_j \to 0$. Similarly, we write $a_j = \Omega(b_j)$ or $a_j \gtrsim b_j$ if $a_j/b_j \ge c$ for some constant $c > 0$. Finally, we use $a_j \asymp b_j$ if both $a_j = O(b_j)$ and $a_j = \Omega(b_j)$. 
For any parameter or quantity $a$, we write $a^{\mathrm{Ou}}$ and $a^{\mathrm{Tr}}$ to denote that it is from the outcome and treatment GWAS, respectively. Let $n_1, n_2$ be the sample sizes for treatment and outcome GWAS, respectively.

\subsection{MR with summary statistics}
Suppose that we are interested in estimating the causal effect of an exposure \( X \) on a health outcome \( Y \), using \( p \) genetic variants \( Z_1, \ldots, Z_p \) as instrumental variables. A commonly assumed structural equation model \citep{bowden2017framework} is given by:
\begin{equation}
  \label{eqn:IV}
  \begin{split}
   X &= \sum_{j=1}^p \gamma_j Z_j + \epsilon_x, \\
   Y &= X \beta_0 + \sum_{j=1}^p \alpha_j Z_j + \epsilon_y = \sum_{j=1}^p \Gamma_j Z_j + \widetilde{\epsilon}_y,
  \end{split}
\end{equation}
where $\epsilon_x$, $\epsilon_y$, and $\widetilde{\epsilon}_y = \epsilon_y + \beta_0 \epsilon_x$ denote random noise terms independent of the instruments $Z_j$, but may depend on $X$. Let $\beta_0$ being the parameter of interest. The error terms $\epsilon_x$ and $\epsilon_y$ may be dependent, reflecting the presence of endogeneity. Here, \( \Gamma_j = \beta_0 \gamma_j + \alpha_j \), where a nonzero \( \alpha_j \) indicates that the \( j \)-th instrument is invalid. In MR, genetic variants are typically assumed to be mutually independent after a pre-processing step known as LD clumping; that is, $Z_1,Z_2,\ldots Z_p $ are mutually independent.\citep{purcell2007plink,hemani2018mr,burgess2023guidelines}

In the summary statistics setting, researchers usually do not have access to individual-level data due to privacy concerns or the large scale of genetic data. Instead, the available data consist of summary statistics from GWAS: $\widehat{\gamma}_j^{\mathrm{Tr}}$, $\widehat{\gamma}_j^{\mathrm{Ou}}$, and $\widehat{\Gamma}_j^{\mathrm{Ou}}$, obtained from independent treatment and outcome samples, respectively. These are typically calculated from simple marginal linear regressions:
\[
\widehat{\gamma}_j^{\mathrm{Tr}} = \frac{\text{Cov}^{\mathrm{Tr}}_{n_1}(Z_j, X)}{\text{Var}^{\mathrm{Tr}}_{n_1}(Z_j, Z_j)}, \quad \widehat{\gamma}_j^{\mathrm{Ou}} = \frac{\text{Cov}^{\mathrm{Ou}}_{n_1}(Z_j, X)}{\text{Var}^{\mathrm{Ou}}_{n_1}(Z_j, Z_j)}, \quad\quad 
\widehat{\Gamma}_j^{\mathrm{Ou}} = \frac{\text{Cov}^{\mathrm{Ou}}_{n_2}(Z_j, Y)}{\text{Var}^{\mathrm{Ou}}_{n_2}(Z_j, Z_j)},
\]
where $n_1$ and $n_2$ are the sample sizes of the treatment and outcome summary datasets, respectively. 

MR analysis using summary data requires careful alignment of the summary statistics. When both quantities are measured from the \textit{same} outcome data, i.e., using $(\widehat{\Gamma}_j^{\mathrm{Ou}}, \widehat{\gamma}_j^{\mathrm{Ou}})_{j=1}^p$, the estimator of $\beta_0$ will generally be biased because $\widehat{\Gamma}_j^{\mathrm{Ou}}$ and $\widehat{\gamma}_j^{\mathrm{Ou}}$ are dependent. On the other hand, MR analysis based on summary data from \textit{different} sources, such as $(\widehat{\Gamma}_j^{\mathrm{Ou}}, \widehat{\gamma}_j^{\mathrm{Tr}})_{j=1}^p$, may also suffer from bias when the treatment and outcome samples differ in population structure.  In the following section, we introduce a heterogeneity test to assess and account for differences between the treatment and outcome samples before conducting MR inference.

\subsection{Heterogeneity test}
\label{sec:heterogeneitytest}

Under a heterogeneous design, the parameters \((\{\Gamma_j\}_{j=1}^p, \{\gamma_j\}_{j=1}^p, \beta_0)\) may differ between the treatment and outcome samples. In this setting, the relationship
\(
\Gamma_j^{\mathrm{Ou}} = \beta_0 \gamma_j^{\mathrm{Ou}} \neq \beta_0\gamma_j^{\mathrm{Tr}},
\)
no longer provides a valid link between the outcome GWAS summary statistics $\widehat{\Gamma}_j^{\mathrm{Ou}}$ and the treatment GWAS summary statistics $\widehat{\gamma}_j^{\mathrm{Tr}}$. As a result, the causal parameter \(\beta_0\) is not identified from heterogeneous data, and the classical two-sample MR estimator based on \((\widehat{\gamma}_j^{\mathrm{Tr}}, \widehat{\Gamma}_j^{\mathrm{Ou}})\) becomes inconsistent. Only in the ideal homogeneous setting, where \(\gamma_j^{\mathrm{Tr}} = \gamma_j^{\mathrm{Ou}}\) for all \(j\), can the two-sample MR procedure consistently recovers \(\beta_0\).

We first state the following test of the null hypothesis on homogeneity  
$H_0: \gamma_j^{\mathrm{Tr}} = \gamma_j^{\mathrm{Ou}}$ for all $j$. 
We define the test statistic
\(
T_j = ({\widehat{\gamma}_j^{\mathrm{Ou}} - \widehat{\gamma}_j^{\mathrm{Tr}}})
/{\sqrt{\sigma^2_{\gamma j,\mathrm {Ou}} + \sigma^2_{\gamma j,\mathrm{Tr}}}},
\)
where $\sigma_{\gamma j, \mathrm{Ou}}$ and $\sigma_{\gamma j, \mathrm{Tr}}$ denote the standard errors of 
$\widehat{\gamma}_j^{\mathrm{Ou}}$ and $\widehat{\gamma}_j^{\mathrm{Tr}}$, respectively. 
These standard errors are typically available from summary data \citep{burgess2013mendelian}. 
Under $H_0$, and assuming normality of 
$\widehat{\gamma}_j^{\mathrm{Tr}}$ and $\widehat{\gamma}_j^{\mathrm{Ou}}$, 
we have $T_j \sim \mathbb{N}(0,1)$. 
Thus, we may use $\sum_{j=1}^p T_j^2$ as a global test statistic.

\begin{proposition}[Asymptotic Null Distribution of the Test Statistic]
\label{prop:test}
Assume $\widehat{\gamma}_j^{\mathrm{Tr}}\sim \mathbb{N}(\gamma_j^{\mathrm{Tr}},\sigma_{\gamma j, \mathrm{Tr}}^2)$, $\widehat{\gamma}_j^{\mathrm{Ou}}\sim \mathbb{N}(\gamma_j^{\mathrm{Ou}},\sigma_{\gamma j, \mathrm{Ou}}^2)$ and assume that the $2p$ random variables $\{\widehat{\gamma}_j^{\mathrm{Tr}},\widehat{\gamma}_j^{\mathrm{Ou}}\}_{j=1}^p$ are mutually independent. 
Let \(T\) denote the test statistic defined by
\begin{equation}
    \label{eqn:heterotest}
    T =  \sum^p_{j=1}\frac{(\widehat{\gamma}_j^{\mathrm{Ou}}-\widehat\gamma_j^{\mathrm{Tr}})^2}{{\sigma^2_{\gamma j, \mathrm{Ou}}+\sigma^2_{\gamma j, \mathrm{Tr}}}}.
\end{equation}
Then, under the null hypothesis $H_0:\gamma_j^{\mathrm{Tr}} = \gamma_j^{\mathrm{Ou}}$ for all $j$,
\(
T\sim \chi^2(p).
\)
\end{proposition}

Proposition~\ref{prop:test} shows that, under mild regularity conditions, the test statistic follows a chi-square distribution with \(p\) degrees of freedom. The normality assumption, which is widely adopted in the MR literature \citep{zhao2018statistical, ye2021debiasedmain}, is slightly stronger than Assumption~A3, which will be stated and used later. We defer the explanation of the mutual independence assumption to Section~\ref{sec:Heteromodel}, where Assumption~A3 is introduced.

The test statistic defined in \eqref{eqn:heterotest} does not require a parametric model linking $\gamma_j^{\mathrm{Ou}}$ and $\gamma_j^{\mathrm{Tr}}$. Instead, it serves as a nonparametric tool for assessing whether there is systematic heterogeneity between the two sets of genetic associations. In particular, large values of \(T\) provide evidence against the null of structural homogeneity, thereby indicating population-specific differences in SNP-exposure marginal association. Later in Section~\ref{sec:realdata_new}, we show that this test provides strong evidence of heterogeneity across real datasets.

\subsection{Heterogeneity model}
\label{sec:Heteromodel}
We will use  $\widehat{\gamma}_j^{\mathrm{Ou}}$, the summary statistics of SNP-treatment marginal effects estimated from the outcome sample, to account for heterogeneity between $\gamma_j^{\mathrm{Ou}}$ and $\gamma_j^{\mathrm{Tr}}$. 
We impose the following assumptions throughout the manuscript.

\begin{itemize}

\item[A1.] \textbf{(IV assumption):}  
\(\Gamma_j^{\mathrm{Ou}} = \beta_0 \gamma_j^{\mathrm{Ou}} + \alpha_j\), where \(\alpha_j\) quantifies the violation of the IV assumption. 

\item[A2.] \textbf{(Observed summary statistics):}  
We observe three sets of summary statistics,
\(
(\widehat{\gamma}_j^{\mathrm{Tr}}, \sigma_{\gamma j,\mathrm{Tr}}^2)_{j=1}^{p},\) 
\(  (\widehat{\gamma}_j^{\mathrm{Ou}}, \sigma_{\gamma j,\mathrm{Ou}}^2)_{j=1}^{p}, \quad 
(\widehat{\Gamma}_j^{\mathrm{Ou}}, \sigma_{\Gamma j,\mathrm{Ou}}^2)_{j=1}^{p}.
\)
These satisfy 
\(
\mathbb{E}(\widehat{\gamma}_j^{\mathrm{Tr}}) = \gamma_j^{\mathrm{Tr}}, \; 
\mathbb{E}(\widehat{\gamma}_j^{\mathrm{Ou}}) = \gamma_j^{\mathrm{Ou}}, \; 
\mathbb{E}(\widehat{\Gamma}_j^{\mathrm{Ou}} \mid \alpha_j) = \Gamma_j^{\mathrm{Ou}},
\)
and
\(
\mathrm{Var}(\widehat{\gamma}_j^{\mathrm{Tr}}) = \sigma_{\gamma j,\mathrm{Tr}}^2, \; 
\mathrm{Var}(\widehat{\gamma}_j^{\mathrm{Ou}}) = \sigma_{\gamma j,\mathrm{Ou}}^2, \; 
\mathrm{Var}(\widehat{\Gamma}_j^{\mathrm{Ou}} \mid \alpha_j) = \sigma_{\Gamma j,\mathrm{Ou}}^2.
\)
\item[A3.] \textbf{(Dependency):}  
The \(2p\) random variables \(\widehat{\gamma}_j^{\mathrm{Tr}}\) and \(\widehat{\Gamma}_j^{\mathrm{Ou}}\) are mutually independent. Similarly, the \(2p\) random variables \(\widehat{\gamma}_j^{\mathrm{Tr}}\) and \(\widehat{\gamma}_j^{\mathrm{Ou}}\) are mutually independent. 

\end{itemize}

Condition A1 corresponds to the standard IV framework allowing for invalid IV, where $\alpha_j \neq 0$ indicates that the $j$th IV is invalid and exerts a direct effect on the outcome. A detailed discussion of invalid IVs  is provided in Section \ref{sec:ple}.
Condition A2 requires that the summary statistics are unbiased and that their variances are available from the summary data. 
This is reasonable because GWAS summary statistics are typically computed from hundreds of thousands of samples, 
yielding highly accurate variance estimates. Our numerical results do not use the actual variance of summary statistics, but we use its estimate from finite samples. 
Note that, unlike \citet{zhao2018statistical, ye2021debiasedmain}, we do not require a normality assumption. 
Condition A3 holds because $\widehat{\gamma}_j^{\mathrm{Tr}}$ and $\widehat{\gamma}_j^{\mathrm{Ou}}$ 
are estimated from independent samples, as are $\widehat{\gamma}_j^{\mathrm{Tr}}$ and $\widehat{\Gamma}_j^{\mathrm{Ou}}$. 
The independence between $\widehat{\gamma}_j^{\mathrm{Tr}}$ and $\widehat{\gamma}_i^{\mathrm{Tr}}$ (for $j \neq i$) 
holds approximately, 
as discussed in Section~2.2 of \citet{zhao2018statistical}.

\subsection{The Wald estimator}
\label{sec:wald}
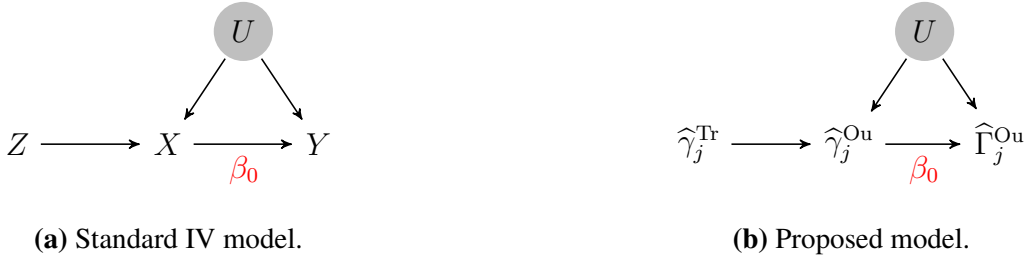
\begin{figure}[ht]
  \centering  
  \begin{subfigure}{0.45\textwidth}
  \begin{tikzpicture}[->,>=stealth',shorten >=1pt,auto,node distance=2.3cm,
   semithick, scale=0.50]
   \tikzstyle{every state}=[fill=none,draw=black,text=black]
  \node[shade] (U) at (2, 3){$U$};
  \node (Z) at (-4,0) {$Z$};
  \node (X) at (0,0) {$X$};
  \node (Y) at (4,0) {$Y$};
  \draw (U) -- (X);
  \draw (U) -- (Y);
  \draw[->] (X) -- (Y) node[midway, below] {\tred{$\beta_0$}};
  \draw (Z) -- (X);
  \end{tikzpicture}
  \centering
  \caption{Standard IV model.}
  \label{fig:classicIV}
  \end{subfigure}
  \hfill
  \begin{subfigure}{0.45\textwidth}
  \begin{tikzpicture}[->,>=stealth',shorten >=1pt,auto,node distance=2.3cm,
   semithick, scale=0.50]
   \tikzstyle{every state}=[fill=none,draw=black,text=black]
  \node[shade] (U) at (0, 3){$U$};
  \node (Z) at (-6,0) {$\widehat{\gamma}_j^{\mathrm{Tr}}$};
  \node (X) at (-2,0) {$\widehat{\gamma}_j^{\mathrm{Ou}}$};
  \node (Y) at (2,0) {$\widehat{\Gamma}_j^{\mathrm{Ou}}$};
  \draw (U) -- (X);
  \draw (U) -- (Y);
 \draw[->] (X) -- (Y) node[midway, below] {\tred{$\beta_0$}};
  \draw (Z) -- (X);
  \end{tikzpicture}
  \centering
  \caption{Proposed model.}
  \end{subfigure}
\caption{(a) Standard IV model with unmeasured confounding $U$; (b) proposed model, where $U$ captures dependence arising from using the same outcome GWAS data to estimate $\widehat{\gamma}_j^{\mathrm{Ou}}$ and $\widehat{\Gamma}_j^{\mathrm{Ou}}$.}
\label{fig:comparison}
\end{figure}

The Wald estimator provides a simple approach to estimating the causal effect of an exposure $X$ on an outcome $Y$ 
in the presence of unmeasured confounders \(U\). 
Let $Z$ be a valid instrument (Figure~\ref{fig:classicIV}), and consider the outcome model
\(
Y = \beta X + \gamma U + \pi Z + \epsilon_y,
\)
together with the standard IV assumptions: 
(i) \emph{exclusion}: $\pi = 0$; 
(ii) \emph{relevance}: $\mathrm{Cov}(X,Z) \neq 0$; 
and (iii) \emph{independence}: $\mathrm{Cov}(U,Z) = 0$. 
Under these conditions,
 \(
\hat{\beta}_{\text{Wald}} = {\mathrm{Cov}(Y,Z)}/{\mathrm{Cov}(X,Z)}
\) 
 consistently estimates $\beta$. 
Intuitively, it can be interpreted as the ratio of the association between $Z$ and $Y$ 
to that between $Z$ and $X$.
This interpretation motivates the construction of our proposed estimator, which takes the same form as the ratio of two quantities. 

The Wald estimator provides a consistent estimate the causal effect of \(X\) on \(Y\)  without assuming a parametric relationship between $X$ and $Z$. This property is also linked to the proposed estimator's model-free feature in the context of two-sample MR.

\section{Mendelian randomization with heterogeneous data}
\label{sec:algorithm}
To motivate our estimator in the heterogeneous setting ($\gamma_j^{\mathrm{Tr}} \neq \gamma_j^{\mathrm{Ou}}$), we consider the linear working model, 
\(
\gamma_j^{\mathrm{Ou}} = b\, \gamma_j^{\mathrm{Tr}},
\)
which provides a framework for linking the independent samples ${\widehat{\gamma}_j^{\mathrm{Tr}}}$ with $({\widehat{\gamma}_j^{\mathrm{Ou}}, \widehat{\Gamma}_j^{\mathrm{Ou}}})$. In Section~\ref{sec:thm}, we show that the validity of the proposed method does not rely on the correct specification of this working model. Under this working model, the following relationships hold:
\begin{align}
\label{eqn:MR1}
  & \{\widehat{\gamma}_j^{\mathrm{Ou}}, \widehat{\gamma}_j^{\mathrm{Tr}}\}_{j=1}^p, 
  \quad \text{with } \gamma_j^{\mathrm{Ou}}= b\,\gamma_j^{\mathrm{Tr}},\\
  \label{eqn:MR2}
   & \{\widehat{\Gamma}_j^{\mathrm{Ou}}, \widehat{\gamma}_j^{\mathrm{Tr}}\}_{j=1}^p, 
   \quad \text{with } \Gamma_j^{\mathrm{Ou}}= \beta_0 \;\gamma_j^{\mathrm{Ou}} = b\, \beta_0\, \gamma_j^{\mathrm{Tr}}.
\end{align}
Equation \eqref{eqn:MR1} corresponds to a standard MR model\citep{davey2014mendelian} for independent samples
\(\{\widehat{\gamma}_j^{\mathrm{Ou}}\}_{j=1}^p\) and \(\{\widehat{\gamma}_j^{\mathrm{Tr}}\}_{j=1}^p\), 
without pleiotropic effects. The parameter \(b\) can be estimated using the inverse-variance weighted (IVW) estimator  \citep{burgess2013mendelian},
\[
\widehat{b} = \bigl\{(\widehat{\boldsymbol{ \gamma}}^{\mathrm{Tr}})^{\T} W^{(1)} \widehat{\boldsymbol \gamma}^{\mathrm{Tr}}\bigr\}^{-1} 
(\widehat{\bm \gamma}^{\mathrm{Tr}})^{\T} W^{(1)} \widehat{\bm \gamma}^{\mathrm{Ou}},
\quad W^{(1)} = \operatorname{diag}\!\left(1/\sigma_{\gamma 1,\mathrm{Ou}}^2, \ldots, 1/\sigma_{\gamma p,\mathrm{Ou}}^2\right).
\]
Analogously, equation \eqref{eqn:MR2} corresponds to the standard MR model for independent samples \(\{\widehat{\Gamma}_j^{\mathrm{Ou}}\}_{j=1}^p\) and \(\{\widehat{\gamma}_j^{\mathrm{Tr}}\}_{j=1}^p\) with the structure equation \(\Gamma_j^{\mathrm{Ou}} = \beta_0\gamma_j^{\mathrm{Ou}}=  b \beta_0 \gamma_j^{\mathrm{Tr}}\). The product \(b\beta\) can be estimated by
\[
\widehat{b\beta} = \bigl\{(\widehat{\bm \gamma}^{\mathrm{Tr}})^{\T} W^{(2)} \widehat{\bm \gamma}^{\mathrm{Tr}}\bigr\}^{-1} 
(\widehat{\bm \gamma}^{\mathrm{Tr}})^{\T} W^{(2)} \widehat{\bm \Gamma}^{\mathrm{Ou}},
\quad W^{(2)} = \operatorname{diag}\!\left(1/\sigma_{\Gamma 1,\mathrm{Ou}}^2, \ldots, 1/\sigma_{\Gamma p,\mathrm{Ou}}^2\right).
\]

We define the MR-Wald estimator for the causal effect,
\begin{equation}
\label{eqn:mr_walds}
\widehat{\beta}^{\mathrm{MR\text{-}Wald}} 
= \frac{\widehat{b\beta}}{\widehat{b}}
= \frac{\{(\widehat{\bm \gamma}^{\mathrm{Tr}})^{\T} W^{(2)} \widehat{\bm \gamma}^{\mathrm{Tr}}\}^{-1} 
(\widehat{\bm \gamma}^{\mathrm{Tr}})^{\T} W^{(2)} \widehat{\bm \Gamma}^{\mathrm{Ou}}}{\{(\widehat{\bm \gamma}^{\mathrm{Tr}})^{\T} W^{(1)} \widehat{\bm \gamma}^{\mathrm{Tr}}\}^{-1} 
(\widehat{\bm \gamma}^{\mathrm{Tr}})^{\T} W^{(1)} \widehat{\bm \gamma}^{\mathrm{Ou}}}.
\end{equation}

Both $\widehat{b}$ and $\widehat{b\beta}$ are biased for their population counterparts due to the measurement error, in the sense that $\widehat{\gamma}_j^{\mathrm{Tr}} \neq \gamma_j^{\mathrm{Tr}}$. 
Under suitable regularity conditions, this bias can be corrected using the debiasing procedure of \citet{ye2021debiasedmain}. 
Nevertheless, as shown in Section~\ref{sec:thm}, the MR-Wald estimator  \(\widehat{\beta}^{\mathrm{MR\text{-}Wald}}\) is consistent and asymptotically normal under mild regularity conditions, without relying on a debiasing procedure. This property arises because the biases in the numerator and denominator cancel each other out.

\begin{remark} 
Equation~\eqref{eqn:mr_walds} can also be expressed as
\[
\widehat{\beta}^{\mathrm{MR\text{-}Wald}} 
= \frac{\text{effect of }\; \widehat{\gamma}_j^{\mathrm{Tr}} \text{ on } \widehat{\Gamma}_j^{\mathrm{Ou}}}
{\text{effect of }\; \widehat{\gamma}_j^{\mathrm{Tr}} \text{ on } \widehat{\gamma}_j^{\mathrm{Ou}}},
\]
which mirrors the classical Wald estimator we discussed in Section \ref{sec:wald},
\[
\widehat{\beta}^{\mathrm{Wald}} 
= \frac{\text{effect of }Z \text{ on }Y}{\text{effect of }Z \text{ on }X}.
\]
Refer to Figure \ref{fig:comparison} for a depiction and comparison of these two estimators. This similarity highlights the model-free nature of the estimator. The MR-Wald estimator follows the same principle as the classical Wald estimator and therefore does not require a linear or parametric link between \(\gamma^{\mathrm{Ou}}\) and \(\gamma^{\mathrm{Tr}}\). In the same spirit, the classical Wald estimator remains valid in the IV framework even when the relationship between \(X\) and \(Z\) is unknown. In the next section, a formal proof of consistency and asymptotic normality of \eqref{eqn:mr_walds} without assuming a parametric form for the relationship \({\gamma}^{\mathrm{Ou}} \sim {\gamma}^{\mathrm{Tr}}\) is given.
\end{remark}

\subsection{Theoretical guarantees}
\label{sec:thm}

In this section, we establish the theoretical properties of the proposed estimator in \eqref{eqn:mr_walds}. We show that it is consistent and asymptotically normal as $p$ tends to infinity in heterogeneous two-sample settings. Moreover, in the special case of homogeneous summary statistics, our estimator attains a smaller asymptotic variance compared to existing methods known to us.

We begin by listing the regularity conditions that we require for the theoretical results, followed by a discussion of intuition and implications.

\begin{assumption}{(Assumptions for the MR-Wald estimator)}
  \begin{itemize}
   \item [B1.] (Relationship among variances) Define residual $U_j^{\mathrm{Ou}} = (\widehat{\Gamma}^{\mathrm{Ou}}_j-{\Gamma}^{\mathrm{Ou}}_j) - \beta_0(\widehat{\gamma}^{\mathrm{Ou}}_j-{\gamma}^{\mathrm{Ou}}_j)$ and let $\sigma_{Uj}^2 = \text{Var}(U_j^{\mathrm{Ou}})$, assumes  
there exist positive constant $C_1$ and $C_2$ such that $C_1\leq {\sigma_{\gamma j, \mathrm{Ou}}^2}/{\sigma_{Uj}^2},{\sigma_{\gamma j, \mathrm{Ou}}^2}/{\sigma_{\gamma j, \mathrm{Tr}}^2}\leq C_2$. 
Furthermore, we assume there exist a constant $k$ such that $\sigma_{\Gamma j, \mathrm{Ou}}^2 = k\sigma_{\gamma j, \mathrm{Ou}}^2$, where $\sigma_{\Gamma j, \mathrm{Ou}}^2 = \text{Var}(\widehat{\Gamma}^{\mathrm{Ou}}_j)$, $\sigma_{\gamma j, \mathrm{Ou}}^2 =  \text{Var}(\widehat{\gamma}^{\mathrm{Ou}}_j)$, and $\sigma_{\gamma j, \mathrm{Tr}}^2 = \text{Var}(\widehat{\gamma}_j^{\mathrm{Tr}}).$
   \item [B2.](Average IV strength) Define the IV strength for treatment data, outcome data, and the IV-coherence (and their average): 
\begin{align*}
&\kappa_j^{\mathrm{Tr}} =\left(\frac{\gamma_j^{\mathrm{Tr}}}{\sigma_{\gamma j, \mathrm{Tr}}}\right)^2,\quad\quad \kappa_j^{\mathrm{Ou}} =\left(\frac{\gamma_j^{\mathrm{Ou}}}{\sigma_{\gamma j, \mathrm{Tr}}}\right)^2,\quad\quad\kappa_j^{Co} =\sqrt{\kappa_j^{\mathrm{Tr}}\kappa_j^{\mathrm{Ou}}};\\
&
\kappa^{\mathrm{Tr}} = \frac{1}{p}\sum^p_{j=1}\kappa^{\mathrm{Tr}}_j,\;\;\quad\quad \kappa^{\mathrm{Ou}} = \frac{1}{p}\sum^p_{j=1}\kappa^{\mathrm{Ou}}_j,\;\;\;\quad\quad\kappa^{Co} = \frac{1}{p}\sum^p_{j=1}\kappa^{Co}_j.
\end{align*}

Assume that the following order assumptions are satisfied: $(\kappa^{\mathrm{Tr}})^2p,(\kappa^{\mathrm{Ou}})^2p \rightarrow\infty$, and $\sqrt{\kappa^{\mathrm{Tr}}\kappa^{\mathrm{Ou}}}\lesssim\kappa^{Co}$.
\item [B3.] (Bounded 4th moments)  There exists a constant $C_3 \in (0,\infty)$ such that, for all $j$,
\[
\mathbb{E}\!\left[\bigl(\widehat{\gamma}_j^{\mathrm{Tr}}-\mathbb{E}\widehat{\gamma}_j^{\mathrm{Tr}}\bigr)^4\right]
\le
C_3\,\sigma_{\gamma j,\mathrm{Tr}}^{4},
\qquad
\mathbb{E}\!\left[\bigl(U_j^{\mathrm{Ou}}-\mathbb{E}U_j^{\mathrm{Ou}}\bigr)^4\right]
\le
C_3\,\sigma_{Uj}^{4}.
\]
  \end{itemize}
\end{assumption}

We now discuss the validity and intuition behind these assumptions. Condition B1 characterizes the relationships between the variances of summary statistics. Based on direct variance calculations under model \eqref{eqn:IV} assuming $\alpha_j=0$, we have:
\begin{equation*}
\begin{split}
  &\frac{\sigma_{\gamma j, \mathrm{Ou}}^2 }{\sigma_{Uj}^2} \approx \frac{\text{Var}(X^{\mathrm{Ou}}) -(\gamma_j^{\mathrm{Ou}})^2\text{Var}(Z_j^{\mathrm{Ou}}) }{\text{Var}(\epsilon_y^{\mathrm{Ou}})}\approx\frac{\text{Var}(X^{\mathrm{Ou}})}{\text{Var}(\epsilon_y^{\mathrm{Ou}})},\\
  &\frac{\sigma_{\gamma j, \mathrm{Ou}}^2 }{\sigma_{\gamma j, \mathrm{Tr}}^2} \approx\frac{n_1}{n_2} \cdot \frac{\text{Var}(X^{\mathrm{Ou}}) -(\gamma_j^{\mathrm{Ou}})^2\text{Var}(Z_j^{\mathrm{Ou}}) }{\text{Var}(X^{\mathrm{Tr}}) -(\gamma_j^{\mathrm{Tr}})^2\text{Var}(Z_j^{\mathrm{Tr}}) } \cdot \frac{\text{Var}(Z_j^{\mathrm{Tr}})}{\text{Var}(Z_j^{\mathrm{Ou}})}\approx\frac{n_1}{n_2} \cdot \frac{\text{Var}(X^{\mathrm{Ou}})}{\text{Var}(X^{\mathrm{Tr}})} \cdot \frac{\text{Var}(Z_j^{\mathrm{Tr}})}{\text{Var}(Z_j^{\mathrm{Ou}})},\\
  &\frac{\sigma_{\gamma j, \mathrm{Ou}}^2 }{\sigma_{\Gamma j, \mathrm{Ou}}^2} \approx \frac{\text{Var}(X^{\mathrm{Ou}}) -(\gamma_j^{\mathrm{Ou}})^2\text{Var}(Z_j^{\mathrm{Ou}}) }{\text{Var}(Y^{\mathrm{Ou}}) -(\Gamma_j^{\mathrm{Ou}})^2\text{Var}(Z_j^{\mathrm{Ou}})} \approx \frac{\text{Var}(X^{\mathrm{Ou}})}{\text{Var}(Y^{\mathrm{Ou}})}.
\end{split}  
\end{equation*}

The second approximation in all equations relies on the fact that, in genetic studies, a single SNP typically explains only a negligible proportion of the total variance in the exposure and outcome variables. These equations suggest that Condition B1 is a reasonable assumption when the sample sizes of the treatment and outcome summary data grow at similar rates.





Condition B2 requires that the average IV strength and the average IV coherence be of a certain order. We do not require the average IV strength to go to infinity, as the classical IVW estimator requires \citep{burgess2013mendelian}. Assumptions of this type, which allow the average IV strength to remain bounded, have also appeared in the MR literature.\citep{ye2021debiasedmain,zhao2018statistical}. The bound for IV coherence is required as $\widehat{b} \asymp \kappa^{Co}/\sqrt{\kappa^{\mathrm{Tr}}\kappa^{\mathrm{Ou}}}$ and we need $\widehat{b}$ to be bounded away from $0$. We have the following proposition showing that this assumption holds under weak requirements for $|\gamma_j^{\mathrm{Ou}}|$ and $|\gamma_j^{\mathrm{Tr}}|$:

\begin{proposition}
\label{prop:conditionB2}
  Assume that $(\kappa^{\mathrm{Tr}})^2p\rightarrow \infty$ and  $ |\gamma_j^{\mathrm{Ou}}| \asymp |\gamma_j^{\mathrm{Tr}}|$, we have 
  $$
  (\kappa^{\mathrm{Ou}})^2p \rightarrow\infty, \text{and}\;\; \kappa^{Co}\gtrsim\sqrt{\kappa^{\mathrm{Tr}}\kappa^{\mathrm{Ou}}}.$$
\end{proposition}
Condition B3 requires that the kurtosis statistics for $\widehat{\gamma}_j$ and $U_j^{\text{Ou}}$ be uniformly bounded across the $p$ instruments. This essentially means that their tails are not too heavy. 

We now present our first theoretical result, which establishes that the proposed MR-Wald estimator \eqref{eqn:mr_walds} possesses desirable theoretical properties.

\begin{theorem}[Properties of the MR-Wald Estimator]
\label{thm:walds}
Suppose Assumptions A1--A4 and B1--B2 hold with \(\alpha_j = 0\) (no pleiotropy). Then the MR-Wald estimator \(\widehat{\beta}^{\mathrm{MR\text{-}Wald}}\) satisfies:  
\begin{itemize}
  \item[a.] \textbf{Consistency:}
  \(\;\;
  \widehat{\beta}^{\mathrm{MR\text{-}Wald}} \xrightarrow{p} \beta_0.
  \)
  
  \item[b.] \textbf{Asymptotic normality:} Under Assumption B3, 
  \(\;\;
  V_1^{-1} \left( \widehat{\beta}^{\mathrm{MR\text{-}Wald}} - \beta_0 \right) \xrightarrow{d} \mathcal{N}(0, 1),
  \)
  where
  \[
  V_1^2 = \frac{\sum_{j=1}^p {((\gamma_j^{\mathrm{Tr}})^2+\sigma_{\gamma j, \mathrm{Tr}}^2)\sigma^2_{Uj}}{\sigma^{-4}_{\gamma j,\mathrm{Ou}}}}{\left(\sum_{j=1}^p \gamma^{\mathrm{Tr}}_j \gamma^{\mathrm{Ou}}_j{\sigma_{\gamma j, \mathrm{Ou}}^{-2}}\right)^{2}}.
  \]

  \item[c.] \textbf{Efficiency:} Under Assumption B3, if $\sigma^2_{Uj} \leq \sigma^2_{\Gamma j, \mathrm{Ou}}$ and \(\gamma_j^{\mathrm{Tr}} = \gamma_j^{\mathrm{Ou}}\) (homogeneity), then \(\widehat{\beta}^{\mathrm{MR\text{-}Wald}}\) is more efficient than the debiased IVW estimator \(\widehat{\beta}^{\lambda = 0, \mathrm{dIVW}}\) of \citet{ye2021debiasedmain}.
\end{itemize}
\end{theorem}

Theorem \ref{thm:walds} establishes the asymptotic properties of the estimator defined in \eqref{eqn:mr_walds}. Under standard regularity conditions, the estimator is consistent and asymptotically normal, and it achieves a smaller asymptotic variance than the existing state-of-the-art method. However, the asymptotic variance involves the nuisance parameter $\sigma_{Uj}^2$, which is not estimable from the observed data distribution. Consequently, we estimate the variance $\widehat{V}_1$ using a bootstrap procedure in simulation and real data analysis.

\begin{remark}
The third result relies on the key assumption that $\sigma^2_{Uj} \leq \sigma^2_{\Gamma j, \mathrm{Ou}},$ which is a sufficient condition for efficiency.  Using some basic algebra, we can show that
\[
\frac{\sigma^2_{Uj}}{\sigma^2_{\Gamma j, \mathrm{Ou}}} = \frac{\operatorname{Var}(Y^{\mathrm{Ou}} - X^{\mathrm{Ou}} \beta_0)}{\operatorname{Var}(Y^{\mathrm{Ou}} - \Gamma_j^{\mathrm{Ou}} Z_j^{\mathrm{Ou}})}.
\]

The key assumption essentially implies that including the causal term in the outcome model reduces the residual variance more effectively than including a single variable \( Z_j^{\text{Ou}} \). Given that individual SNPs typically account for only a negligible proportion of the variance in complex traits, we believe this is a reasonable assumption. 
However, this condition may not hold in all scenarios. In particular, 
\(
\operatorname{Var}(Y^{\mathrm{Ou}} - \Gamma_j^{\mathrm{Ou}} Z_j^{\mathrm{Ou}}) \leq \operatorname{Var}(Y^{\mathrm{Ou}}),
\)
and if confounding is strong, it is possible that
\(
\operatorname{Var}(Y^{\mathrm{Ou}}) \leq \operatorname{Var}(Y^{\mathrm{Ou}} - X^{\mathrm{Ou}} \beta_0).
\)
This suggests that the sufficient condition may be violated when the effect of the confounders is substantial.


The dIVW method incorporates a screening step based on a hard-thresholding rule, using a $z$-score threshold $\lambda \geq 0$, to select instruments; see equations (4), (12), and (13) in \citet{ye2021debiasedmain}.  To keep our notation simple and the presentation concise, we do not apply this screening step. Instead, we take screening as a preprocessing procedure for data analysis.  In our analysis, we compare the variance of $\widehat{\beta}^{\text{MR-Wald}}$ (our proposed estimator) with that of $\widehat{\beta}^{\lambda = 0, \text{dIVW}}$ (their estimator without screening). If we adopt the same screening procedure, it would be possible to obtain a more general result, which we leave for future work.
\end{remark}

\section{Pleiotropy}
\label{sec:ple}
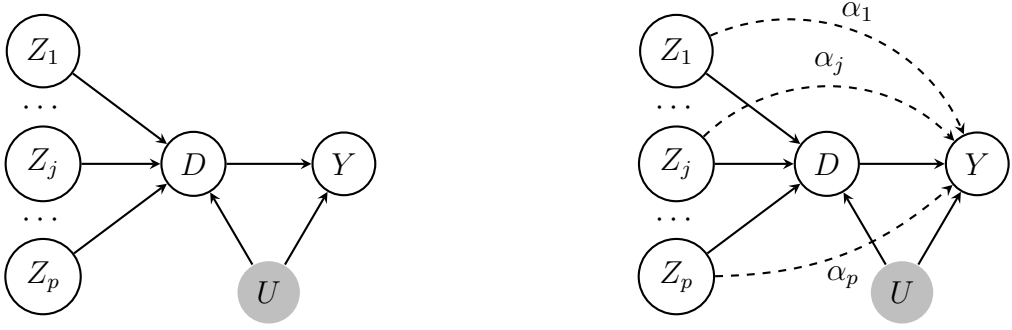
\begin{figure}[ht]
  \centering  
  \begin{subfigure}{0.49\textwidth}
  \begin{tikzpicture}[->,>=stealth,thick,node distance=2.5cm]
    \node (Y)  at (2,0) [circle, draw] {$Y$};
    \node (U)  at (1,-1.71) [circle, fill=gray!50] {$U$};
    \draw (U) -- (Y);
    \node (D)  at (0,0) [circle, draw] {$D$};
    \node (Z)  at (-2,0) [circle, draw] {$Z_j$};
    \node (Z1)  at (-2,1.5) [circle, draw] {$Z_1$};
    \node at (-2,0.75) {$\cdots$};
    \node at (-2,-0.75) {$\cdots$};
    \node (Zp)  at (-2,-1.5) [circle, draw] {$Z_p$};

    \draw[->] (D) -- (Y) node[midway, above] {\tred{$ $}};
    \draw (Z) -- (D);
    \draw (Z1) -- (D);
    \draw (Zp) -- (D);
    \draw (U) -- (D);
    \end{tikzpicture}
    \centering
  \caption{MR without pleiotropy, where $\Gamma_j^\mathrm{Ou} = \gamma_j^\mathrm{Ou}\beta_0$.}

  \end{subfigure}
  \hfill
  \begin{subfigure}{0.49\textwidth}
  \begin{tikzpicture}[->,>=stealth,thick,node distance=2.5cm]
    \node (Y)  at (2,0) [circle, draw] {$Y$};
    \node (U)  at (1,-1.71) [circle, fill=gray!50] {$U$};
    \draw (U) -- (Y);
    \node (D)  at (0,0) [circle, draw] {$D$};
    \node (Z)  at (-2,0) [circle, draw] {$Z_j$};
    \draw[->] (D) -- (Y) node[midway, above] {\tred{$ $}};
    \draw (Z) -- (D);
    \draw (U) -- (D);
    \node (Z1)  at (-2,1.5) [circle, draw] {$Z_1$};
    \node at (-2,0.75) {$\cdots$};
    \node at (-2,-0.75) {$\cdots$};
    \node (Zp)  at (-2,-1.5) [circle, draw] {$Z_p$};
        \draw (Z1) -- (D);
    \draw (Zp) -- (D);
    \draw[->,dashed] (Z) to [bend left=45] node[midway, above] {$\alpha_j$} (Y);
    \draw[->,dashed] (Z1) to [bend left=45] node[midway, above] {$\alpha_1$} (Y);
    \draw[->,dashed] (Zp) to [bend right=20] node[midway, below] {$\alpha_p$} (Y);
  \end{tikzpicture}
      \centering
  \caption{MR with pleiotropy, where $\Gamma_j^\mathrm{Ou} = \gamma_j^\mathrm{Ou}\beta_0+\alpha_j$.}
  \end{subfigure}
  \caption{Comparison of MR models with and without Pleiotropy effects.}
\label{fig:comparison_ple}
\end{figure}

We now discuss the pleiotropy effect, a biological phenomenon in which a single genetic variant influences multiple phenotypic traits. Pleiotropy violates the valid IV assumption because certain genetic variants may exert a direct effect on the outcome of interest rather than acting solely through the exposure. We use the parameter $\alpha_j$ to characterize this violation. See Figure \ref{fig:comparison_ple} for a graphical comparison of MR models without and with pleiotropy. We consider the following three pleiotropy models.


\begin{enumerate}[label=P\arabic*]
  \item \label{P1} \textbf{Balanced pleiotropy:} 
  Assume that $\alpha_j \overset{\text{i.i.d.}}{\sim} \mathcal{N}(0, \tau_0^2)$.
  \item \label{P2}\textbf{Idiosyncratic pleiotropy:}  
  Assume there exists a vector $\alpha^* = (\alpha^*_1,\ldots,\alpha^*_p)$ and an index set $\mathcal{A} \subset \{1, 2, \ldots, p\}$ such that
  \(
  \alpha_j \overset{\text{i.i.d.}}{\sim} \mathcal{N}(\alpha_j^*, \tau_0^2),
  \)
   where 
   $\alpha^*_j = 0$, if $j\in \mathcal{A}$, and $\alpha^*_j = \mu_j\not=0$, if $j\notin \mathcal{A}$.
  \item \label{P3}\textbf{Directional pleiotropy:}  
  \(
  \alpha_j \overset{\text{i.i.d.}}{\sim} \mathcal{N}(\mu, \tau_0^2),\;\text{with } \mu \neq 0.
  \)
\end{enumerate}

In these models, the pleiotropic effect follows either a centered normal distribution, a contaminated centered normal distribution, or a shifted normal distribution. Correlated pleiotropy, in which \(\alpha_j\) may depend on \(\gamma_j\), is discussed in the Section \ref{sec:discussion} and is not covered by our method. In the following section, we discuss strategies to address each of these pleiotropic effects in turn.

\subsection {Balanced pleiotropy}
We first investigate the effect of balanced pleiotropy, where the pleiotropic effects $\alpha_j$ follow a mean-zero normal distribution (model~\ref{P1}). The following proposition establishes that the estimators in equations~\eqref{eqn:mr_walds} 
remain valid under Assumption~P1.

\begin{proposition}
\label{prop:balanced_p}
Suppose Assumption~P1 holds and that $\tau_0^2 \lesssim \max_j \sigma_{\Gamma j, \mathrm{Ou}}^2$. 
  Under Assumptions~A1--A4 and B1--B3, the estimator in equation~\eqref{eqn:mr_walds} is consistent and asymptotically normal.
\end{proposition}

The condition $\tau_0^2 \lesssim \max_j \sigma_{\Gamma j,\mathrm{Ou}}^2$ 
essentially requires that balanced pleiotropic effects should not be excessively large, 
but instead remain comparable to the measurement error in $\widehat{\Gamma}_j^{\mathrm{Ou}}$.
This assumption is standard in the MR literature for balanced pleiotropy \citep{zhao2018statistical, ye2021debiasedmain}.

\subsection{Idiosyncratic pleiotropy}

We now investigate the effect of idiosyncratic pleiotropy, where the pleiotropic effects $\alpha_j$ follow a contaminated normal distribution (model~\ref{P2}). The estimator proposed in Section~\ref{sec:algorithm} is based on the IVW MR estimator, which is known to be biased under idiosyncratic pleiotropy. Thus, under Assumption P2, the MR-Wald estimator \eqref{eqn:mr_walds} 
also exhibits bias.

Existing MR procedures for handling idiosyncratic pleiotropy mainly rely on two strategies:
(a) down-weighting outliers using a robust loss function \citep{zhao2018statistical}; and 
 (b) remove the outlier following an outlier selection method \citep{rees2019robust}. These approaches are typically developed under parametric assumptions such as
\(\Gamma_j^{\mathrm{Ou}} = \beta_0 \gamma_j^{\mathrm{Tr}} + \alpha_j\), which correspond to an idealized homogeneous setting.
However, they do not readily extend to heterogeneous designs, where a parametric relationship between \(\Gamma_j^{\mathrm{Ou}}\) and \(\gamma_j^{\mathrm{Tr}}\) is difficult to specify.

To address bias induced by idiosyncratic pleiotropy, we generalize the MR-Wald estimator by replacing its least-squares components with robust regression counterparts.
In equation~\eqref{eqn:mr_walds}, \(\widehat{b\beta }\) and \(\widehat{b}\) are obtained by the weighted least squares, minimizing \(\sum w_j(\widehat{\Gamma}_j^{\text{Ou}} - \beta \widehat{\gamma}_j^{\text{Tr}})^2\) and \(\sum w_j(\widehat{\gamma}_j^{\text{Ou}} - \beta \widehat{\gamma}_j^{\text{Tr}})^2\), respectively. The idiosyncratic pleiotropy will introduce outliers in the $\widehat{\Gamma}_j^{\mathrm{Ou}}$, making the original MR-Wald estimator biased. To improve robustness against outliers, we replace the squared loss with a robust loss function $\rho(\cdot)$, such as the quantile loss, Huber's loss, or Tukey's bi-weight loss. Specifically, we obtain \(\widetilde{b\beta }\) and \(\widetilde{b}\) by solving the weighted robust regression problems (i) \(\sum w_j \rho(\widehat{\Gamma}_j^{\text{Ou}} - \beta \widehat{\gamma}_j^{\text{Tr}})\) and  (ii) \(\sum w_j \rho(\widehat{\gamma}_j^{\text{Ou}} - \beta \widehat{\gamma}_j^{\text{Tr}})\), respectively.
As we show in the proof of Theorem~\ref{thm:idio_sycratic}, 
robust regression is essential to estimate $b$, even when no outliers are present in regression (ii) involving $\widehat{\gamma}_j^{\mathrm{Ou}}$ and $\widehat{\gamma}_j^{\mathrm{Tr}}$. 
This is because the solution to a misspecified robust regression generally differs from that of a
misspecified ordinary least squares regression. 
Consequently, applying robust loss in one regression but the least squares in the other would result in a biased estimator formed by their ratio.

Based on the previous observations, our proposed estimator is
\begin{equation}
\label{eqn:MR-Wald-idio}
\begin{split}\widetilde{b}&= \argmin_{\beta\in \mathbb{R}} \widehat{g}_1(\beta),\;\text{where}\; \widehat{g}_1(\beta) =  \frac{1}{p} \sum_{j=1}^p w_j^{(1)}\rho(\widehat{\gamma}_j^{\text{Ou}} - \widehat{\gamma}_j^{\text{Tr}} \beta);\\
\widetilde{b\beta} &= \argmin_{\beta\in \mathbb{R}} \;\widehat{g}_2(\beta),\; \text{where}\;\widehat{g}_2(\beta) = \frac{1}{p} \sum_{j=1}^p w_j^{(2)} \rho(\widehat{\Gamma}_j^{\text{Ou}} - \widehat{\gamma}_j^{\text{Tr}} \beta ); \\
&\widetilde{\beta}^{MR-Wald} = \widetilde{b\beta}/\widetilde{b}.
\end{split}
\end{equation}
In both our theoretical and numerical analyses, we specify the weights as  
\begin{equation}
  \label{eqn:weights}
  w_j^{(1)} = \frac{\min_j \sigma_{\Gamma j,\mathrm{Ou}}}{\sigma_{\Gamma j,\mathrm{Ou}}},\quad 
  w_j^{(2)} = \frac{\min_j \sigma_{\gamma j,\mathrm{Ou}}}{\sigma_{\gamma j,\mathrm{Ou}}},
\end{equation}
where \(\sigma_{\Gamma j,\mathrm{Ou}}\) and \(\sigma_{\gamma j,\mathrm{Ou}}\) denote the standard deviations of \(\widehat{\Gamma}_j^{\text{Ou}}\) and \(\widehat{\gamma}_j^{\text{Ou}}\), respectively. The scaling factors \(\min_j \sigma_{\Gamma j,\mathrm{Ou}}\) and \(\min_j \sigma_{\gamma j,\mathrm{Ou}}\) are introduced to ensure that the weights remain uniformly bounded away from infinity.

We now present the theoretical result for \eqref{eqn:MR-Wald-idio}. 
Before doing so, we state the assumptions required to establish the asymptotic result.

\begin{assumption}
  (Idiosyncratic pleiotropy)
\begin{itemize}  
\item [C1.] Let \( n = \min({n_1}, n_{2})\). Assume there exist constants \( c_\sigma > 0 \) and \( c'_\sigma > 0 \) such that
\[
\frac{c_\sigma}{n} \leq \sigma^2_{\gamma j,\mathrm{Ou}}, \sigma^2_{\Gamma j, \mathrm{Ou}}, \sigma^2_{\gamma j,\mathrm{Tr}} \leq \frac{c'_\sigma}{n} \quad  \text{for all } 1 \leq j \leq p.
\]
Furthermore, assumes $\tau_0^2 \lesssim \max_j \sigma_{\Gamma j, \mathrm{Ou}}^2$, and that there exists a constant $k$ such that $\sigma_{\Gamma j, \mathrm{Ou}}^2 = k \sigma_{\gamma j, \mathrm{Ou}}^2$, where $\sigma_{\Gamma j, \mathrm{Ou}}^2 = \text{Var}(\widehat{\Gamma}^{\mathrm{Ou}}_j)$, $\sigma_{\gamma j, \mathrm{Ou}}^2 =  \text{Var}(\widehat{\gamma}^{\mathrm{Ou}}_j)$, and $\sigma_{\gamma j, \mathrm{Tr}}^2 = \text{Var}(\widehat{\gamma}_j^{\mathrm{Tr}}).$
\item [C2.] Let \( g_1(\beta) := \mathbb{E}[\widehat{g}_1(\beta)] \). Assume that \( g_1 \) has a unique minimizer, denoted by \( \beta_1^* \) and \( \beta_1^*\not=0\).

\item [C3.] $ ||\gamma^{\mathrm{Tr}}||_\infty= o(\sqrt{p})$, $||\gamma^{\mathrm{Ou}}||_\infty= o(\sqrt{p})$,  $||\alpha^*||_1 = o(p)$, and $n\rightarrow  
\infty$, $p\rightarrow  
\infty$.
\item [C4.] Assume that $\widehat{\gamma}_j^{\mathrm{Ou}}$ has absolutely continuous distribution functions $F_j^{(1)}$, with continuous
densities $f_j^{(1)}$ that are uniformly bounded away from 0 and $\infty$ at the points $\widehat{\gamma}_j^{\mathrm{Tr}}\beta_1^*$; Assumes $\widehat{\Gamma}_j^{\mathrm{Ou}}$  has absolutely continuous distribution functions $F_j^{(2)}$, with continuous
densities $f_j^{(2)}$ that are uniformly bounded away from 0 and  $\infty$  at the points $\widehat{\gamma}_j^{\mathrm{Tr}}\beta_1^*\beta_0$.
\item [C5.] Define the following quantity:
$$
J_1 =\frac{1}{p}\sum^p_{j=1}w_j^{(1)}f_j^{(1)}(\widehat{\gamma}_j^{\mathrm{Tr}}\beta_1^*)(\widehat{\gamma}_j^{\mathrm{Tr}})^2,\;\;J_2 =\frac{1}{p}\sum^p_{j=1}w_j^{(2)}f_j^{(2)}(\widehat{\gamma}_j^{\mathrm{Tr}}\beta_1^*\beta_0)(\widehat{\gamma}_j^{\mathrm{Tr}})^2.
$$
We assume: 
$
J_1 \xrightarrow{p.} \bar{J}_1:= \mathbb{E}(J_1),\quad J_2 \xrightarrow{p.} \bar{J}_2:= \mathbb{E}(J_2),
$
and $\bar{J}_1, \bar{J}_2$ bounded away from zero.
\end{itemize}
\noindent
\end{assumption}
Condition~C1 strengthens B1 by requiring the variances of the summary statistics to be of order $n^{-1}$.
Under \eqref{eqn:IV} we have $\operatorname{Var}(\widehat{\gamma}_j^{\mathrm{Tr}})\asymp n_1^{-1}$ and 
$\operatorname{Var}(\widehat{\Gamma}_j^{\mathrm{Ou}})\asymp n_2^{-1}$, so when $n_1\asymp n_2$ this is of order $1/n$.
A similar condition has appeared in the literature \citep[Assumption~3]{zhao2018statistical}. 
Condition~C2 requires that the population-level optimization problem admit a unique minimizer $\beta_1^*$ with $\beta_1^* \neq 0$. 
Since this minimizer appears in the denominator of our formulation, it must be bounded away from zero. 
This condition parallels B2, which requires $\sqrt{\kappa^{\mathrm{Tr}}\kappa^{\mathrm{Ou}}} \lesssim \kappa^{\mathrm{Co}}$. 
Note that $b \asymp \kappa^{\mathrm{Co}}/\sqrt{\kappa^{\mathrm{Tr}}\kappa^{\mathrm{Ou}}}$ also appears in the denominator. 
Condition~C3 restricts the growth rate. 
It allows the largest elements of $\gamma^{\mathrm{Tr}}$ and $\gamma^{\mathrm{Ou}}$ to diverge, but at a slower rate than $\sqrt{p}$. 
The magnitude of pleiotropic effects is also permitted to diverge under this condition. 
Conditions~C4 and C5 are standard regularity assumptions from robust statistics, adapted to our setting; 
see also \citet[Theorem~4.1]{koenker2005quantile}.







Based on these assumptions, the following theoretical result shows that when combined with a robust loss, the Wald-MR estimator with a robust loss remains consistent and asymptotically normal in the presence of idiosyncratic pleiotropy.  

\begin{theorem}[Properties of the MR-Wald Estimator with Robust Loss]
\label{thm:idio_sycratic}
Suppose model~P2 holds and consider the median loss function 
\(
\rho(t) = |t|.
\)
Then the estimator defined in~\eqref{eqn:MR-Wald-idio} satisfies the following:  

\begin{itemize}
  \item[a.] \textbf{Consistency.} Under Assumptions C1--C3,
  \(
  \widetilde{\beta}^{\mathrm{MR\text{-}Wald}} \xrightarrow{p} \beta_0.
  \)
  
  \item[b.] \textbf{Asymptotic normality.} Under Assumptions C1--C5,
  \(
  \frac{\sqrt{p}}{V_3}\big(\widetilde{\beta}^{\mathrm{MR\text{-}Wald}} - \beta_0\big) \xrightarrow{d} \mathcal{N}(0,1),
  \)
  where
  \[
  V_3^2 = \frac{1}{p(\beta_1^*)^2}\left(
    \beta_0^2 \bar{J}_1^{-2}\sum_{j=1}^p \zeta^{(1)}_j
    + \bar{J}_2^{-2}\sum_{j=1}^p \zeta^{(2)}_j
    - 2\beta_0 \bar{J}_1^{-1}\bar{J}_2^{-1}\sum_{j=1}^p \zeta^{(12)}_j
  \right), \text{with}
  \]
\[
\zeta^{(1)}_j = \mathbb{E}\Big[\{w_j^{(1)}\widehat{\gamma}_j^{\mathrm{Tr}}
\big(\tfrac{1}{2} - \mathbb{I}(\widehat{\gamma}_j^{\mathrm{Ou}} \leq \widehat{\gamma}_j^{\mathrm{Tr}}\beta_1^*)\big)\}^2\Big],\quad 
\zeta^{(2)}_j = \mathbb{E}\Big[\{w_j^{(2)}\widehat{\gamma}_j^{\mathrm{Tr}}
\big(\tfrac{1}{2} - \mathbb{I}(\widehat{\Gamma}_j^{\mathrm{Ou}} \leq \widehat{\gamma}_j^{\mathrm{Tr}}\beta_1^*\beta_0)\big)\}^2\Big],
\]
\[
\zeta^{(12)}_j = \mathbb{E}\Big[\{w_j^{(1)}\widehat{\gamma}_j^{\mathrm{Tr}}
\big(\tfrac{1}{2} - \mathbb{I}(\widehat{\gamma}_j^{\mathrm{Ou}} \leq \widehat{\gamma}_j^{\mathrm{Tr}}\beta_1^*)\big)\}
\{w_j^{(2)}\widehat{\gamma}_j^{\mathrm{Tr}}
\big(\tfrac{1}{2} - \mathbb{I}(\widehat{\Gamma}_j^{\mathrm{Ou}} \leq \widehat{\gamma}_j^{\mathrm{Tr}}\beta_1^*\beta_0)\big)\}\Big].
\]
\end{itemize}
\end{theorem}

In Theorem~\ref{thm:idio_sycratic}, the estimator is constructed using the median loss, 
a widely used choice in the robust statistics literature. 
Alternative robust loss functions, such as Huber's loss or Tukey's biweight loss, may also be employed.  

Theorem~\ref{thm:idio_sycratic} is particularly appealing, as it shows that even in realistic settings with heterogeneous samples, idiosyncratic pleiotropy, and weak instruments, the estimator remains consistent and asymptotically normal. 
The asymptotic variance involves quantities such as the densities of certain variables, 
which are difficult to estimate directly. 
To address this, we employ a bootstrap procedure that resamples the summary statistics with replacement 
and computes the variance across the resulting bootstrap samples.


\subsection{Directional pleiotropy}
We now consider \textit{directional pleiotropy} \citep{bowden2015mendelian,burgess2017interpreting}, 
where the pleiotropic effects $\alpha_j$ are assumed to follow a normal distribution with nonzero mean (model~\ref{P3}).  

Under Assumption~P3, the previously proposed estimators are biased because of the nonzero mean of the pleiotropic effects. 
To address this, we extend the MR-Wald estimator via a two-step procedure. 
First, we estimate $b\beta$ using a method robust to directional pleiotropy, such as the MR-Egger method \citep{bowden2015mendelian}, 
denoted by $\widetilde{b\beta}$, by regressing $\widehat{\Gamma}_j^{\mathrm{Ou}}$ on $\widehat{\gamma}_j^{\mathrm{Tr}}$. 
Second, we estimate $b$ using the same loss function that is robust to directional pleiotropy. 
The resulting estimator accommodates directional pleiotropy. 
Technical details and corresponding simulation results are provided in Section~\ref{sec:app_directional_pleiotropy} and \ref{sec:supp_numerical} of the Supplementary Material.

\section{Numerical examples}
\label{sec:simulation}

We consider a simulation setting with \( p = 200 \) SNPs and a sample size of \( n = 10{,}000 \), 
following a data-generating mechanism similar to \citet{burgess2016bias}. 
The simulation involves two independent samples: one for the treatment GWAS and the other for the outcome GWAS.

\paragraph{Treatment GWAS.} 
For each individual \( i \in \{1, \dots, n\} \) and SNP \( j \in \{1, \dots, p\} \), 
the genotype is generated as \( Z_{i,j}^{(2)} \overset{\mathrm{iid}}{\sim}\text{Binomial}(2, 0.3) \). 
The treatment variable is given by
\(
D^{(2)}_i = \sum_{j=1}^{p} \gamma_j^{\mathrm{Tr}} Z_{i,j}^{(2)} + U_i + \epsilon_{d,i},
\)
where \( \gamma_j^{\mathrm{Tr}} \overset{\mathrm{iid}}{\sim} \text{Unif}(0.05, 0.10) \), and \( U_i, \epsilon_{d,i} \overset{\text{iid}}{\sim} \mathcal{N}(0, 1) \). 
The marginal SNP-treatment associations \( \widehat{\gamma}_j^{\mathrm{Tr}} \) are estimated by regressing \( D^{(2)} \) on \( Z_j^{(2)} \).

\paragraph{Outcome GWAS.} 
Genotypes are independently generated as \( Z_{i,j}^{(1)} \overset{\mathrm{iid}}{\sim} \text{Binomial}(2, 0.3) \). 
The treatment variable is defined by
\(
D^{(1)}_i = \sum_{j=1}^{p} \gamma_j^{\mathrm{Ou}} Z_{i,j}^{(1)} + U_i + \epsilon_{d,i},
\)
where \( \gamma_j^{\mathrm{Ou}} = g(\gamma_j^{\mathrm{Tr}}) \) for a given function \( g(\cdot) \), and \( U_i, \epsilon_{d,i} \overset{\mathrm{iid}}{\sim} \mathcal{N}(0,1) \). 
The outcome variable is generated as
\(
Y^{(1)}_i = 0.5 D^{(1)}_i + U_i + \epsilon_{y,i} + \sum_{j=1}^{p} \alpha_j Z_{i,j}^{(1)},
\)
where \( \epsilon_{y,i} \overset{\mathrm{iid}}{\sim}   \mathcal{N}(0,1) \), and the pleiotropic effects satisfy \( \alpha_j \sim \mathcal{N}(\alpha_j^*, \tau^2_0) \). 
The marginal SNP-outcome associations \( \widehat{\Gamma}_j^{\mathrm{Ou}} \) and SNP-treatment associations \( \widehat{\gamma}_j^{\mathrm{Ou}} \) 
are estimated by regressing \( Y^{(1)} \) and \( D^{(1)} \), respectively, on \( Z_j^{(1)} \).

Under this data-generating mechanism, the true parameter satisfies
\[
\Gamma_j^{\mathrm{Ou}} = \beta_0 \gamma_j^{\mathrm{Ou}} + \alpha_j, 
\quad \alpha_j \sim \mathcal{N}(\alpha_j^*, \tau_0^2).
\]
We examine four pleiotropy scenarios:  
(i) \emph{no pleiotropy}, with $\alpha_j^* = 0$ and $\tau_0 = 0$ for all $j$;  
(ii) \emph{balanced pleiotropy}, with $\alpha_j^* = 0$ and $\tau_0 = 0.02$;  
(iii) \emph{idiosyncratic pleiotropy (single SNP)}, with $\alpha_j^* = 0.1$ for the SNP corresponding to $j = \arg\max_j \gamma_j^{\mathrm{Tr}}$ and $\tau_0 = 0.02$;  
(iv) \emph{idiosyncratic pleiotropy (multiple SNPs)}, with $\alpha_j^* = 0.1$ for five randomly chosen SNPs and $\tau_0 = 0.02$.

We compare the following six methods: \begin{itemize} \item[E1:] \textbf{MR-Wald (proposed):} \(\widehat{\beta} = \widehat{\beta b} / \widehat{b}\), where \(\widehat{\beta b}\) is obtained from the weighted least squares regression of \(\widehat{\Gamma}_j^{\mathrm{Ou}}\) on \(\widehat{\gamma}_j^{\mathrm{Tr}}\), and \(\widehat{b}\) from the weighted least squares regression of \(\widehat{\gamma}_j^{\mathrm{Ou}}\) on \(\widehat{\gamma}_j^{\mathrm{Tr}}\). \item[E2:] \textbf{MR-Wald-R (proposed):} \(\widehat{\beta} = \widetilde{\beta b} / \widetilde{b}\), where \(\widetilde{\beta b}\) is obtained from the weighted median regression of \(\widehat{\Gamma}_j^{\mathrm{Ou}}\) on \(\widehat{\gamma}_j^{\mathrm{Tr}}\), and \(\widetilde{b}\) from the weighted median regression of \(\widehat{\gamma}_j^{\mathrm{Ou}}\) on \(\widehat{\gamma}_j^{\mathrm{Tr}}\). \item[E3:] \textbf{Weighted Median \citep{bowden2016consistent}:} defined as the weighted median of the ratio estimates \( \widehat{\beta}_j = {\widehat{\Gamma}^{\mathrm{Ou}}_j}/{\widehat{\gamma}^{\mathrm{Tr}}_j}, \quad j = 1, \dots, p, \) with weights \(w_j = 1 / \mathrm{Var}(\widehat{\beta}_j)\). \item[E4:] \textbf{MR-Egger \citep{burgess2017interpreting}:} estimating \((\beta, \alpha)\) by minimizing \( \sum_{j=1}^p w_j \bigl(\widehat{\Gamma}_j - \widehat{\gamma}_j^{\mathrm{Tr}} \beta - \alpha \bigr)^2, \) which corresponds to the IVW estimator with an intercept term. \item[E5:] \textbf{RAPS \citep{zhao2018statistical}:} a robust adjusted profile likelihood estimator constructed under the assumption that \((\widehat{\Gamma}_j, \widehat{\gamma}_j^{\mathrm{Tr}})\) follows a bivariate normal distribution. \item[E6:] \textbf{DIVW \citep{ye2021debiasedmain}:} a de-biased IVW estimator that corrects for weak instrument bias in the standard IVW method. \end{itemize}

\begin{figure}
    \centering
    \includegraphics[width=0.5\linewidth]{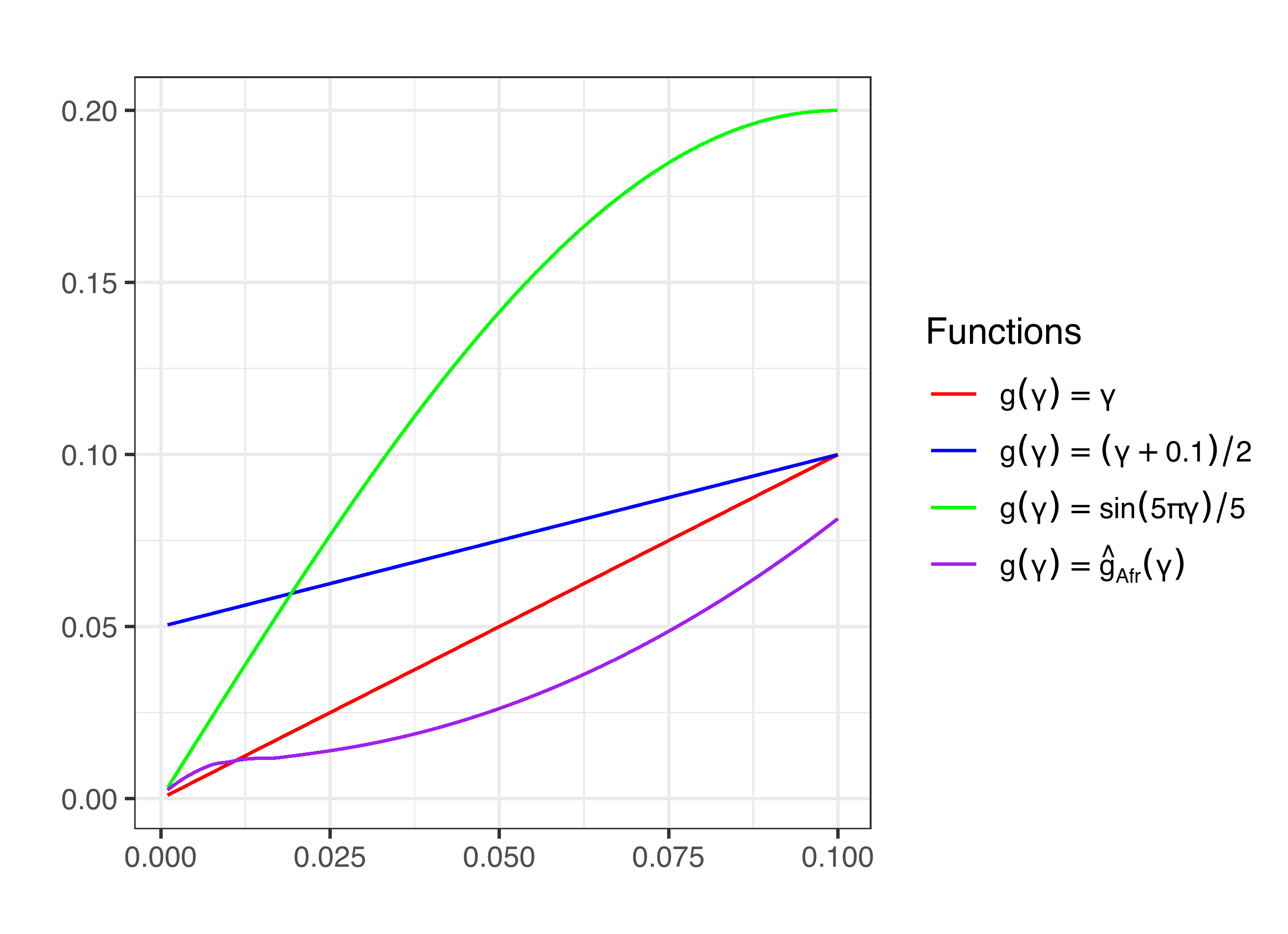}
    \caption{Functions considered in the simulation study.}
    \label{fig:gfun_graphical}
\end{figure}

To allow heterogeneity between $\gamma_j^{\mathrm{Tr}}$ and $\gamma_j^{\mathrm{Ou}}$, we set $\gamma_j^{\mathrm{Ou}} = g(\gamma_j^{\mathrm{Tr}})$ 
and consider four specifications for $g$: the identity $g(\gamma) = \gamma$, 
a scaled shift $g(\gamma) = (\gamma+0.1)/2$, a nonlinear transformation $g(\gamma) = \sin(5\pi\gamma)/5$, 
and a nonparametric estimator $\widehat{g}_{\mathrm{Afr}}$ derived from BMI-SNP associations in European and African samples 
(see Section~\ref{sec:realdata_new}). 
Figure~\ref{fig:gfun_graphical} illustrates these functions.

\subsection{Results and discussions}

Table~\ref{tab:simresultta} reports the simulation results for scenarios (i) no pleiotropy and (ii) balanced pleiotropy. 
In the homogeneous setting where $\gamma_j^\mathrm{Tr} = \gamma_j^\mathrm{Ou}$, 
the MR-Egger and Weighted Median estimators exhibit substantial bias and fail to achieve nominal coverage, 
reflecting their lack of adjustment for weak IV bias. 
RAPS and DIVW yield reliable inference but incur larger RMSEs than MR-Wald and comparable RMSEs to MR-Wald-R. 
In the heterogeneous setting where $\gamma_j^\mathrm{Tr} \neq \gamma_j^\mathrm{Ou}$, 
only the proposed MR-Wald and MR-Wald-R estimators deliver valid inference and exhibit negligible bias, 
as they explicitly account for heterogeneity across samples. 
MR-Wald attains the smallest RMSE and is the most efficient estimator in all cases, 
while MR-Wald-R is less efficient, consistent with the usual efficiency loss of the median regression relative to the OLS for Gaussian data.

In the presence of idiosyncratic pleiotropy, as shown in Table~\ref{tab:simresulttab}, 
MR-Wald-R achieves coverage at the nominal 95\% level, whereas competing methods do not. 
Approaches such as MR-Egger, Weighted Median, and DIVW fail to deliver satisfactory coverage 
because they do not account for idiosyncratic pleiotropy, leading to coverage rates below 95\% even under homogeneity. 
RAPS employs a robust loss function to mitigate the influence of invalid IVs, 
making it less biased than MR-Egger, DIVW, and Weighted Median in the homogeneous setting where $\gamma_j^\mathrm{Tr} = \gamma_j^\mathrm{Ou}$. 
However, when the number of invalid IVs increases to five, as in scenario (iv), the use of robust loss there is insufficient: even under homogeneity, RAPS fails to maintain nominal coverage. RAPS also performs poorly under heterogeneity because it does not account for differences between the treatment and outcome GWAS. MR-Wald maintains good coverage in scenario (iii).  This is because it constructs confidence intervals via the bootstrap: when only a single invalid SNP is present, many bootstrap samples exclude the invalid IV, yielding a distribution close to valid and maintaining coverage. However, once the number of invalid IVs increases to five, as in scenario (iv), 
the likelihood that bootstrap samples include invalid instruments rises substantially, causing coverage to fall well below the nominal $95\%$ level.

\begin{table}
\centering
\caption{Simulation results for various estimators under Scenarios (iii) and (iv). Reported summary statistics include the bias, root-mean-square error (RMSE), confidence interval (CI) length, and CI coverage rate (nominal level 95 \%).  We report $(\text{Bias}  /\beta_0)\times100\%$, $(\text{RMSE}/\beta_0)\times100\%$, $(\text{CI length}/\beta_0)\times100\%$, and CI coverage rate. Cell colors indicate performance, with yellow denoting poor and green denoting favorable results.}
\vspace{-1.5em}

\begin{flushleft}
\begin{tikzpicture}[baseline=0ex]
  \begin{axis}[
    hide axis,
    colorbar horizontal,
    point meta min=0,
    point meta max=1,
    colormap={mycolormap}{
      rgb255(0cm)=(255,255,0);
      rgb255(0.33cm)=(204,204,0);
      rgb255(0.66cm)=(0,153,0);
      rgb255(1cm)=(0,255,0)
    },
    colorbar style={
      height=0.12cm,
      width=5cm,
      xtick={0,1},
      xticklabels={Poor,Favorable},
      ticklabel style={font=\scriptsize}
    }]
  \addplot [draw=none] coordinates {(0,0)};
  \end{axis}
\end{tikzpicture}
\end{flushleft}
\label{tab:simresultta}
\vspace{-4em}
\scalebox{1}{
\begin{tabular}{ccABFDc@{\hskip 0.1in}ABFD}
\cmidrule(r){1-6} \cmidrule(l){8-11}
$g(\gamma)$&Method&\multicolumn{4}{c}{Scenario (i)}  &  & \multicolumn{4}{c}{Scenario (ii)}  \\ 
\cmidrule(r){1-6} \cmidrule(l){8-11}
\cmidrule(r){1-6} \cmidrule(l){8-11}
&&\multicolumn{1}{c}{Bias} & \multicolumn{1}{l}{RMSE} & \multicolumn{1}{l}{\hspace{-2em}CI length} & \multicolumn{1}{l}{CI} & &\multicolumn{1}{c}{Bias} & \multicolumn{1}{l}{RMSE} & \multicolumn{1}{l}{\hspace{-2em}CI length} & \multicolumn{1}{l}{CI} \\
&MR-Wald&0.1 & 4.3 & 16.7 & 94.2&&0.1 & 5.8 & 22.7 & 94.4\\
&MR-Wald-R&0.2 & 6.7 & 26.6 & 93.9&&0.0 & 8.0 & 33.9 &95.5\\
$\gamma$&W.Median&-24.7 & 25.7 & 32.5 & 11.5&&-28.3 & 29.4 & 34.1 & 9.5\\
&MR-Egger&-70 & 71.6 & 60.4 & 0.5&&-70.1 & 72.6 & 71.9 & 3.4\\
&RAPS&-0.2 & 6.6 & 25.5 & 94.9&&0.3 & 7.6 & 29.8 & 95.0\\
&DIVW&0.3 & 6.5 & 24.7 & 94.5&&0.4 & 7.4 & 24.8 & 91.4\\
\cmidrule(r){1-6} \cmidrule(l){8-11}
\cmidrule(r){1-5} \cmidrule(l){7-10}
&MR-Wald&0.1 & 3.7 & 14.5 & 94.2&&0.1 & 5.1 & 19.8 & 94.4\\
&MR-Wald-R&0.2 & 5.7 & 23.6 & 94.8&&0.0 & 7.2 & 30.0 & 95.3\\
$\frac{\gamma+0.1}{2}$&W.Median&-15.2 & 16.7 & 33.2 & 58.5&&-19 & 20.7 & 34.6 & 42.6\\
&MR-Egger&-81.8 & 83.3 & 60.3 & 0.0&&-81.9 & 84.2 & 71.6 & 1.2\\
&RAPS&14.9 & 16.4 & 26.4 & 40.7&&15.3 & 17.2 & 30.7 & 50.6\\
&DIVW&15.2 & 16.7 & 25.4 & 36.4&&15.4 & 17.2 & 25.6 & 35.5\\
\cmidrule(r){1-6} \cmidrule(l){8-11}
\cmidrule(r){1-6} \cmidrule(l){8-11}
&MR-Wald&0.0 & 1.8 & 7.0 & 94.4&&0.1 & 2.5 & 9.6 & 94.4\\
&MR-Wald-R& 0.0 & 3.3 & 13.8 & 95.2&&-0.2 & 4.0 & 16.7 & 95.7\\
$ \frac{\sin(5\pi\gamma)}{5}$&W.Median&75.9 & 76.4 & 41.1 & 0.0&&70.4 & 71.1 & 42.2 & 0.0\\
&MR-Egger&-59.2 & 62.4 & 67.1 & 11.2&&-59.2 & 63.4 & 77.8 & 19.4\\
&RAPS&136.7 & 137.0 & 35.0 & 0.0&&137.5 & 137.9 & 38.7 & 0.0\\
&DIVW&137.3 & 137.6 & 33.7 & 0.0&&137.7 & 138.0 & 33.8 & 0.0\\
\cmidrule(r){1-6} \cmidrule(l){8-11}
\cmidrule(r){1-6} \cmidrule(l){8-11}
&MR-Wald&0.1 & 4.6 & 17.8 & 94.2&&0.1 & 6.2 & 24.2 & 94.4\\
&MR-Wald-R&0.1 & 7.0 & 28.1 & 94.1&&0.0 & 8.5 & 36.0 & 95.2\\
$ \widehat{g}_{Afr}(\gamma)$&W.Median&-29.1 & 29.9 & 32.3 & 3.6&&-32.6 & 33.5 & 33.9 & 3.2\\
&MR-Egger&-69.9 & 71.6 & 60.2 & 0.4&&-70.1 & 72.6 & 71.8 & 3.4\\
&RAPS&-6.2 & 9.0 & 25.3 & 82.0&&-5.8 & 9.5 & 29.6 & 87.6\\
&DIVW&-5.8 & 8.7 & 24.4 & 82.9&&-5.6 & 9.3 & 24.6 & 80.8\\
\cmidrule(r){1-6} \cmidrule(l){8-11}

\end{tabular}
}
\end{table}

\begin{table}
\centering
\caption{Simulation results for various estimators under Scenarios (iii) and (iv). Reported summary statistics include the bias, root-mean-square error (RMSE), confidence interval (CI) length, and CI coverage rate (nominal level 95 \%).  We report $(\text{Bias}  /\beta_0)\times100\%$, $(\text{RMSE}/\beta_0)\times100\%$, $(\text{CI length}/\beta_0)\times100\%$, and CI coverage rate. Cell colors indicate performance, with yellow denoting poor and green denoting favorable results.}
\vspace{-1.5em}

\begin{flushleft}
\begin{tikzpicture}[baseline=0ex]
  \begin{axis}[
    hide axis,
    colorbar horizontal,
    point meta min=0,
    point meta max=1,
    colormap={mycolormap}{
      rgb255(0cm)=(255,255,0);
      rgb255(0.33cm)=(204,204,0);
      rgb255(0.66cm)=(0,153,0);
      rgb255(1cm)=(0,255,0)
    },
    colorbar style={
      height=0.12cm,
      width=5cm,
      xtick={0,1},
      xticklabels={Poor,Favorable},
      ticklabel style={font=\scriptsize}
    }]
  \addplot [draw=none] coordinates {(0,0)};
  \end{axis}
\end{tikzpicture}
\end{flushleft}
\vspace{-4em}
\label{tab:simresulttab}
\scalebox{1}{
\begin{tabular}{ccABFDc@{\hskip 0.1in}ABFD}
\cmidrule(r){1-6} \cmidrule(l){8-11}
$g(\gamma)$&Method&\multicolumn{4}{c}{Scenario (iii)}  &  & \multicolumn{4}{c}{Scenario (iv)}  \\ 
\cmidrule(r){1-6} \cmidrule(l){8-11}
\cmidrule(r){1-6} \cmidrule(l){8-11}
&&\multicolumn{1}{c}{Bias} & \multicolumn{1}{l}{RMSE} & \multicolumn{1}{l}{\hspace{-2em}CI length} & \multicolumn{1}{l}{CI} & &\multicolumn{1}{c}{Bias} & \multicolumn{1}{l}{RMSE} & \multicolumn{1}{l}{\hspace{-2em}CI length} & \multicolumn{1}{l}{CI} \\
&MR-Wald&1.9 & 6.1 & 23.7 & 94.4&&7.0 & 9.1 & 25.8 & 83.0\\
&MR-Wald-R&0.8 & 8.0 & 34.3 & 95.4&&3.2 & 8.8 & 35.1 & 93.8\\
$\gamma$&W.Median&-27.9 & 29.1 & 34.2 & 10.1&&-27.0 & 28.2 & 34.7 & 12.9\\
&MR-Egger&-66.9 & 69.6 & 73.5 & 6.0&&-67.4 & 70.4 & 78.7 & 8.8\\
&RAPS&1.2 & 7.8 & 30.1 & 94.5&&4.1 & 8.9 & 31.2 & 91.9\\
&DIVW&2.2 & 7.7 & 24.9 & 90.0&&7.3 & 10.5 & 25.2 & 75.6\\
 \cmidrule(r){1-6} \cmidrule(l){8-11}
\cmidrule(r){1-6} \cmidrule(l){8-11}
&MR-Wald&1.6 & 5.3 & 20.6 & 94.4&&6.1 & 7.9 & 22.5 & 82.9\\
&MR-Wald-R&0.7 & 7.2 & 30.3 & 95.2&&2.9 & 7.9 & 31.2 & 94.8\\
$\frac{\gamma+0.1}{2}$&W.Median&-18.4 & 20.2 & 34.8 & 45.3&&-17.6 & 19.5 & 35.3 & 49.8\\
&MR-Egger&-78.7 & 81.1 & 73.2 & 1.7&&-79.2 & 81.9 & 78.4 & 2.4\\
&RAPS&16.3 & 18.2 & 31.0 & 45.5&&19.3 & 20.9 & 32.1 & 33.6\\
&DIVW&17.2 & 18.8 & 25.7 & 27.4&&22.3 & 23.6 & 26.0 & 9.9\\
 \cmidrule(r){1-6} \cmidrule(l){8-11}
\cmidrule(r){1-6} \cmidrule(l){8-11}
&MR-Wald&0.8 & 2.6 & 10.0 & 94.5&&3.0 & 3.9 & 10.9 & 83.2\\
&MR-Wald-R&0.4 & 4.0 & 16.9 & 95.9&&1.6 & 4.4 & 17.4 & 95.0\\
$\frac{\sin(5\pi\gamma)}{5}$&W.Median&71.0 & 71.7 & 42.4 & 0.0&&71.7 & 72.4 & 43.0 & 0.0\\
&MR-Egger&-56.0 & 60.5 & 79.3 & 23.3&&-56.5 & 61.3 & 84.4 & 28.2\\
&RAPS&138.9 & 139.3 & 39.0 & 0.0&&142.5 & 142.9 & 40.4 & 0.0\\
&DIVW&139.4 & 139.8 & 34.0 & 0.0&&144.6 & 144.9 & 34.4 & 0.0\\
 \cmidrule(r){1-6} \cmidrule(l){8-11}
\cmidrule(r){1-6} \cmidrule(l){8-11}
&MR-Wald&2.0 & 6.5 & 25.3 & 94.4&&7.4 & 9.7 & 27.5 & 83.0\\
&MR-Wald-R&0.8 & 8.6 & 36.3 & 94.5&&3.3 & 9.3 & 37.2 & 94.3\\
$\widehat{g}_{Afr}$&W.Median&-32.2 & 33.2 & 34.0 & 3.6&&-31.2 & 32.3 & 34.5 & 4.8\\
&MR-Egger&-66.9 & 69.5 & 73.4 & 6.1&&-67.4 & 70.3 & 78.6 & 8.5\\
&RAPS&-4.9 & 9.1 & 29.9 & 89.6&&-2.1 & 8.1 & 31.0 & 93.8\\
&DIVW&-3.9 & 8.3 & 24.7 & 85.1&&1.2 & 7.6 & 24.9 & 90.8\\
\cmidrule(r){1-6} \cmidrule(l){8-11}
\end{tabular}
}
\end{table}

\section{Real data analysis}
\label{sec:realdata_new}
To evaluate the performance of various MR methods in a real-world scenario, we examine the causal effect of BMI on HDL. 
BMI, a widely used measure of overall adiposity, is strongly associated with cardiometabolic health. 
HDL, often referred to as the "good cholesterol," plays a protective role in lipid metabolism and cardiovascular disease. 
Observational studies consistently document an inverse association between BMI and HDL \citep{rashid2007effect}, 
but such associations may be confounded by lifestyle and environmental factors, making causal interpretation difficult. 
MR provides a principled framework to address this issue by leveraging genetic variants as instrumental variables, thereby mitigating bias from unmeasured confounding and reverse causation.

As the instrument-selection dataset for identifying relevant genetic variants, we use the participants from the GIANT consortium \citep{locke2015genetic}, comprising $322{,}154$ individuals of European ancestry. 
SNPs are selected as instrumental variables based on two conventional significance thresholds, with $p$-value cutoffs of $5 \times 10^{-8}$ and $1 \times 10^{-4}$, following the recommendations of \citet{zhao2018statistical}. 
To reduce correlation among instruments, we perform linkage disequilibrium clumping with a threshold $r^2 \leq 0.0001$ and a window size of $10{,}000$ kb, yielding $69$ and $265$ SNPs, respectively.

For the treatment dataset, we use SNP--BMI summary statistics from four ancestrally diverse populations: Japanese individuals ($n=158{,}284$) \citep{kanai2018genetic}, Hispanic or Latin American individuals ($n=56{,}161$) \citep{fernandez2022ancestral}, African individuals ($n=118{,}993$) \citep{verma2024diversity}, and South Asian individuals ($n=8{,}658$) \citep{karczewski2024pan}. 
The outcome dataset consists of SNP--HDL associations from European participants in the UK Biobank ($n=407{,}609$) \citep{mbatchou2021computationally}. 
For consistency, SNP--BMI summary statistics ($\widehat{\gamma}_j^{\mathrm{Ou}}$) from the same study are also included.

We conducted the heterogeneity test described in Section~\ref{sec:heterogeneitytest}, 
comparing BMI--SNP effect estimates in European participants with those from South Asian, African, Hispanic, and Japanese populations. 
The results, reported in Table~\ref{tab:testresult}, show that most test statistics are highly significant, with $p$-values well below the 0.05 threshold, providing strong evidence of heterogeneity between the treatment and outcome GWAS. 
The null hypothesis of homogeneity is not rejected for the South Asian versus European comparison, likely due to the relatively small South Asian sample size. 
Nonetheless, the estimates based on South Asian data still differ substantially from those obtained in other populations, underscoring the presence of meaningful heterogeneity across ancestry groups in our analysis.

\begin{table}[ht]
\centering
\caption{P-values from heterogeneity tests comparing SNP-BMI associations between European and non-European ancestries}
\label{tab:testresult}
\begin{tabular}{lllll}
\toprule
Source & South Asian & Latin American & Japanese & African \\
\hline
Pvalue(Threshold 1) & 0.29 & $7.3\times 10^{-10}$& 0 &0 \\
Pvalue(Threshold 2) & 0.69 & $3.6\times10^{-10}$   &  0  & 0  \\
\bottomrule
\end{tabular}
\end{table}

\begin{table}[ht]
\centering
\caption{Comparison of results in the BMI-HDL example with heterogeneous treatment GWAS sources. Reported values are point estimates, with confidence intervals shown in brackets}
\label{tab:realdata}
\resizebox{1\textwidth}{!}{
\begin{tabular}{ccccccccc}
\toprule
&\multicolumn{7}{c}{Threshold = $5\times10^{-8}$, $p = 69\quad$}\\
\midrule
Source & MR-Egger & W.Median & IVW & DIVW & RAPS & MR-Wald & MR-Wald-R\\
\midrule
\multicolumn{1}{c}{\multirow{2}{*}{South Asian }}& 0.062& -0.096 & -0.088 & -0.171 & -0.681 & -0.177 & -0.262\\
\multicolumn{1}{c}{} & (-0.062,0.187) & (-0.156,-0.036) & (-0.18,0.004) & (-0.305,-0.037) & (-0.907,-0.456) & (-0.338,-0.015) & (-0.369,-0.156)\\
\multicolumn{1}{c}{\multirow{2}{*}{African }} & 0.066 & -0.115 & -0.122 & -0.157 & -0.659 & -0.194& -0.207\\
\multicolumn{1}{c}{}& (-0.102,0.234) & (-0.164,-0.067) & (-0.247,0.003) & (-0.264,-0.050) & (-0.737,-0.582) & (-0.316,-0.072) & (-0.311,-0.102)
\\\multicolumn{1}{c}{\multirow{2}{*}{Japanese }} & -0.055& -0.208 & -0.254 & -0.255 & -0.416 & -0.251 & -0.283\\
\multicolumn{1}{c}{} &(-0.252,0.142) & (-0.271,-0.144) & (-0.372,-0.136) & (-0.363,-0.147) & (-0.459,-0.373) & (-0.360,-0.142) & (-0.389,-0.176)\\
\multicolumn{1}{c}{\multirow{2}{*}{Latin American }} & 0.002 & -0.161& -0.206 & -0.236& -0.554 & -0.246 & -0.231\\
\multicolumn{1}{c}{} &(-0.228,0.232) & (-0.211,-0.110) &(-0.329,-0.084) & (-0.353,-0.119) & (-0.631,-0.477) & (-0.360,-0.131) & (-0.305,-0.156)\\
\midrule
&\multicolumn{7}{c}{{Threshold = $1\times10^{-4}$}, $p = 265$}  \\
\midrule
\multicolumn{1}{c}{\multirow{2}{*}{South Asian }}& -0.088 & -0.062 & -0.087& -0.263 & -1.555 & -0.202 & -0.237\\
\multicolumn{1}{c}{} & (-0.169,-0.006) & (-0.098,-0.026) & (-0.141,-0.033) & (-0.437,-0.089) & (-2.140,-0.969) & (-0.326,-0.077) & (-0.306,-0.168)\\
\multicolumn{1}{c}{\multirow{2}{*}{African}} & -0.032& -0.116 & -0.109 & -0.137 & -1.395 & -0.178 & -0.237\\
\multicolumn{1}{c}{} & (-0.130,0.067) & (-0.153,-0.078) & (-0.179,-0.039) & (-0.209,-0.064) & (-1.584,-1.206) & (-0.277,-0.079) & (-0.324,-0.151)\\
\multicolumn{1}{c}{\multirow{2}{*}{Japanese}} & -0.153 & -0.183 & -0.244& -0.270& -0.663 & -0.252& -0.251\\
\multicolumn{1}{c}{}  & (-0.279,-0.026) & (-0.232,-0.134) & (-0.322,-0.167) & (-0.352,-0.189) & (-0.716,-0.611) & (-0.323,-0.181) & (-0.310,-0.191)\\
\multicolumn{1}{c}{\multirow{2}{*}{Latin American}}  & -0.165& -0.160& -0.201& -0.248& -1.041& -0.249 & -0.248\\
 \multicolumn{1}{c}{}& (-0.277,-0.052) & (-0.199,-0.121) & (-0.270,-0.132) & (-0.328,-0.168) & (-1.177,-0.904) & (-0.327,-0.171) & (-0.312,-0.183)\\
\bottomrule
\end{tabular}
}
\end{table}


\begin{figure}[ht]
\centering
\begin{subfigure}{0.48\textwidth}
  \centering
  \includegraphics[width=0.8\linewidth]{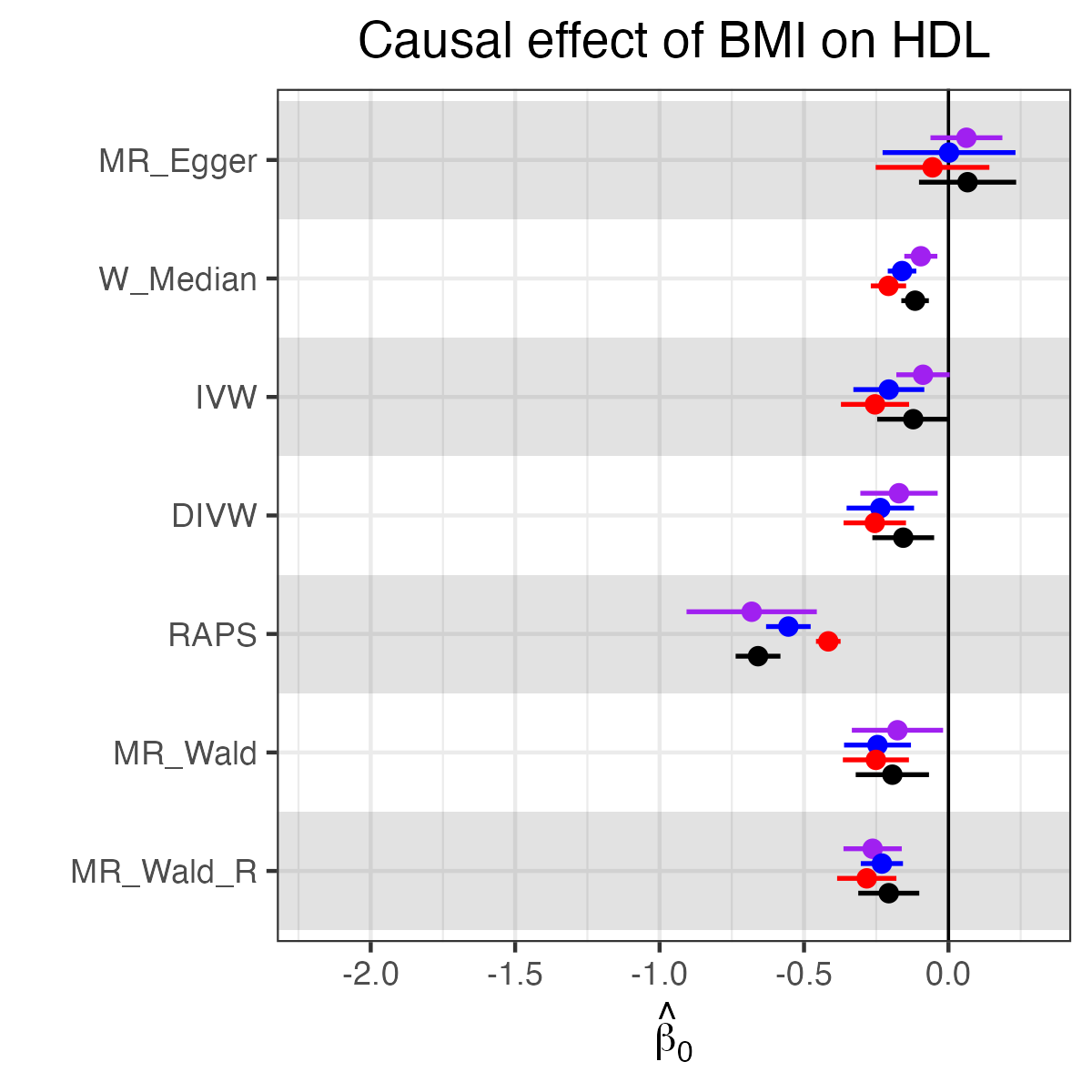}
  \caption{Threshold = $5\times 10^{-8}$.}
  \label{fig:realdataa}
  \label{fig:realdatasubfig1}
\end{subfigure}
\hfill
\begin{subfigure}{0.48\textwidth}
  \centering
  \includegraphics[width=0.8\linewidth]{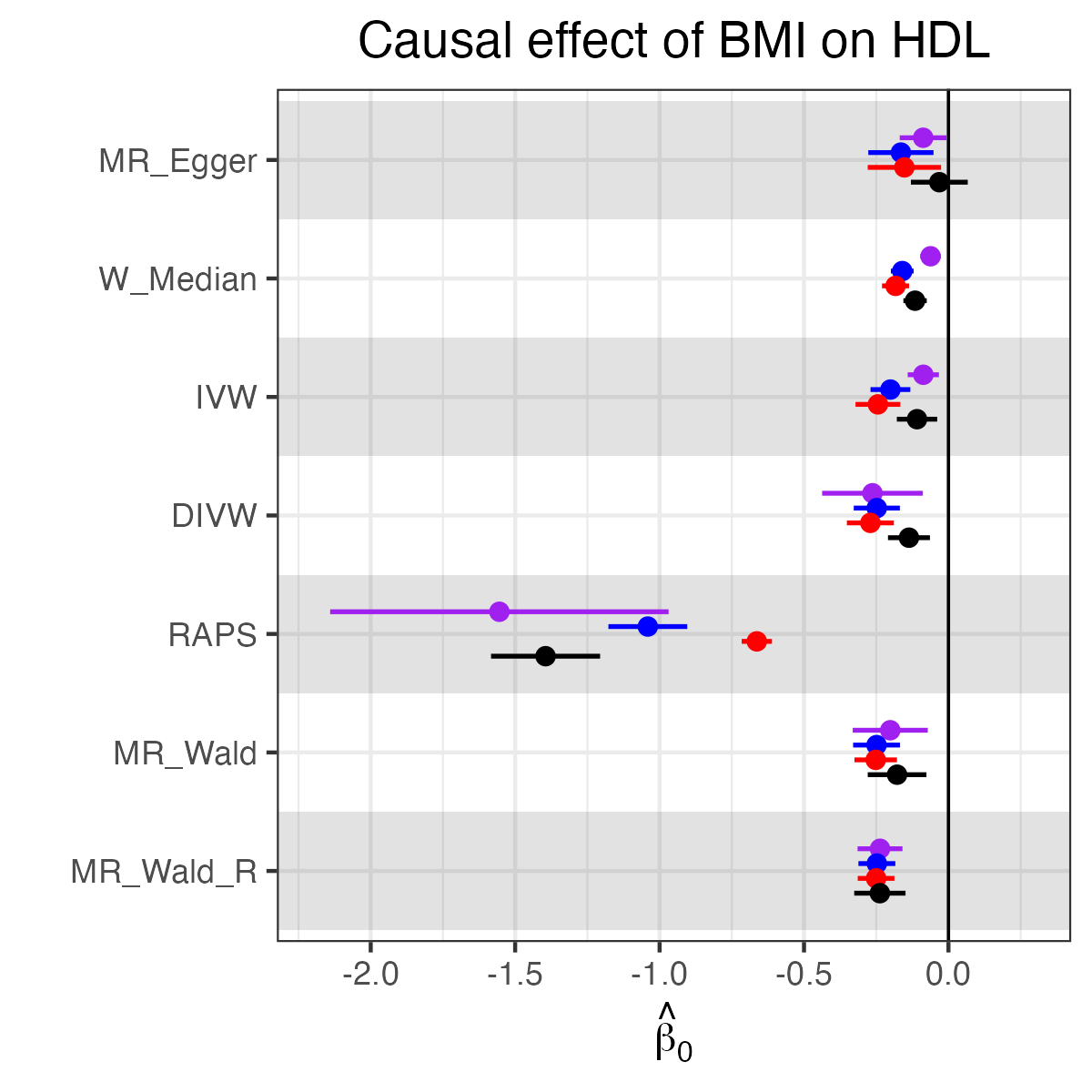}
  \caption{Threshold = $1\times 10^{-4}$.}
  \label{fig:realdatab}
\end{subfigure}
\caption{Estimated causal effect of BMI on HDL using different thresholds and estimators. Treatment data are shown for South Asian (purple), Latin American (blue), Japanese (red), and African (black) populations.}
\label{fig:realdataresult}
\end{figure}

\subsection{MR results and interpretations}

The primary parameter of interest is the causal effect of BMI on HDL cholesterol in the European population, 
as the outcome GWAS data are drawn from European participants in the UK Biobank. 
To assess robustness, we apply our method using treatment summary statistics from multiple sources 
and benchmark its performance against existing MR estimators. 
Figure~\ref{fig:realdataresult} and Table~\ref{tab:realdata} summarize the estimated causal effects 
across four ancestrally diverse populations, obtained using seven MR methods under two instrument selection thresholds ($5\times 10^{-8}$ and $1\times 10^{-4}$).

The estimates obtained with the stringent threshold of $5 \times 10^{-8}$ are shown in Figure~\ref{fig:realdataa}. 
Most estimators yield negative causal effect estimates, consistent with current scientific understanding of the BMI--HDL relationship. 
The MR-Egger estimator produces positive point estimates in South Asian and African populations, though these are not statistically significant. 
The weighted median method shows inconsistency across populations: for example, the estimate from the African population ($-0.116$) does not lie within the confidence interval for the Japanese population ($-0.271,-0.145$).   
The IVW estimator fails to adjust for weak instrument bias, which may explain why its confidence intervals include zero for both South Asian and African populations. 
The debiased IVW estimator corrects this bias and yields more significant results than IVW. 
However, confidence intervals from both IVW and debiased IVW in the African population do not align with those from the Japanese population. 
For instance, the IVW point estimate for the African population ($-0.121$) lies outside the confidence interval obtained from the Japanese population ($-0.372,-0.136$). 
Similarly, RAPS produces divergent estimates across populations. 
In contrast, our proposed estimators (MR-Wald and MR-Wald-R) provide consistent results across populations, as they explicitly account for cross-sample heterogeneity.  

When the threshold is relaxed to $1 \times 10^{-4}$, the results are shown in Figure~\ref{fig:realdatab}. Compared with Figure~\ref{fig:realdataa}, most estimators produce narrower confidence intervals, reflecting efficiency gains from including additional SNPs that contribute signal to causal effect estimation.  Nonetheless, inconsistency remains a major issue for several methods. 
In particular, the confidence intervals for the weighted median, IVW, debiased IVW, and RAPS estimators show limited overlap between African and Japanese, or African and Hispanic populations. 
By contrast, our proposed estimators again yield consistent results across all four populations. 
The MR-Wald-R estimator is especially stable, with point estimates close to $-0.24$ and confidence intervals that align across populations. 
This robustness likely stems from its ability to accommodate heterogeneity between the exposure and outcome GWAS, mitigate weak instrument bias, and guard against idiosyncratic pleiotropy.

\section{Discussion}
\label{sec:discussion}

We propose a new estimator for MR that does not require a parametric model linking the SNP--exposure effects across two samples. A key contribution of our work is the development of a new debiased version of the inverse-variance weighted estimator \citep{burgess2013mendelian}. Even when the heterogeneous model is linear (i.e., \(\gamma^\mathrm{Ou} = b\gamma^\mathrm{Tr}\)), both the numerator and denominator in equation~\eqref{eqn:mr_walds} are biased for \(b\) and \(b\beta\), respectively. However, these biases cancel in the ratio, enabling consistent estimation of the causal effect. Our method improves efficiency relative to existing approaches and provides a flexible, robust framework for modeling diverse pleiotropic structures. This conceptual advancement offers a new perspective on addressing heterogeneity and pleiotropy in two-sample summary-data MR studies.

Our work considered various pleiotropy models. These models do not accommodate \emph{correlated pleiotropy} \citep{morrison2020mendelian}, where the magnitude of the direct effect $\alpha_j$ depends on the instrument strength $\gamma_j^\mathrm{Ou}$; addressing such dependence requires additional structural assumptions and is left for future work.

Our estimator achieves robustness to heterogeneity between the two samples, 
$\widehat{\Gamma}_j^{\mathrm{Ou}}$ and $\widehat{\gamma}_j^{\mathrm{Tr}}$, 
by incorporating information from the SNP--treatment association estimated in the outcome sample, 
$\widehat{\gamma}_j^{\mathrm{Ou}}$. A natural question is whether similar results can be obtained 
using the SNP--outcome association estimated in the treatment sample, 
$\widehat{\Gamma}_j^{\mathrm{Tr}}$. An even broader question is whether 
greater efficiency and robustness can be achieved when both 
$\widehat{\Gamma}_j^{\mathrm{Tr}}$ and $\widehat{\gamma}_j^{\mathrm{Ou}}$ 
are available. These extensions lie beyond the scope of this paper and are left for future research.

\putbib
\end{bibunit}

\appendix

\setcounter{section}{0}
\renewcommand{\thesection}{S.\arabic{section}}

\setcounter{equation}{0}
\renewcommand{\theequation}{S\arabic{equation}}

\setcounter{figure}{0}
\renewcommand{\thefigure}{S\arabic{figure}}

\setcounter{table}{0}
\renewcommand{\thetable}{S\arabic{table}}

\def\spacingset#1{\renewcommand{\baselinestretch}%
{#1}\small\normalsize} \spacingset{1}
\if1\blind
{
\vspace{2cm}
  {\bf \large{\centering Supporting materials for ``A Robust Framework for Two-Sample Mendelian Randomization under Population Heterogeneity"}}
  \vspace{11pt}
  \date{}
} \fi

\if0\blind
{
  \bigskip
  \bigskip
  \bigskip
  \begin{center}
  {\LARGE\bf Title}
\end{center}
  \medskip
} \fi

\vspace{-0.8cm}
\begin{abstract}
 In Section \ref{sec:proof_sec3}, we provide proofs of the main results stated in the paper. Section \ref{sec:app_pleiotropy} presents the additional results for pleiotropy that complement Section \ref{sec:ple} of the manuscript.  Section \ref{sec:supp_numerical} contains additional numerical result for the directional pleiotropy.
\end{abstract}

\spacingset{1.45} 

\begin{bibunit}
\defaultbibliographystyle{asa}
\defaultbibliography{ref}
\section{Theoretical details}
\label{sec:proof_sec3}
We provide the proofs of the propositions and theorems presented in Section~\ref{sec:algorithm}.

\subsection{Proof of Proposition~\ref{prop:test}}

Note that 
\[
\widehat{\gamma}_j^{\mathrm{Ou}} \sim N\!\left(\gamma_j^{\mathrm{Ou}}, \sigma^{2}_{\gamma j, \mathrm{Ou}}\right), 
\qquad 
\widehat{\gamma}_j^{\mathrm{Tr}} \sim N\!\left(\gamma_j^{\mathrm{Tr}}, \sigma^{2}_{\gamma j, \mathrm{Tr}}\right),
\text{and}\quad \widehat{\gamma}_j^{\mathrm{Ou}} \ind \widehat{\gamma}_j^{\mathrm{Tr}}.\]

Under the null hypothesis, we have
\[
T_j = 
\frac{\widehat{\gamma}_j^{\mathrm{Ou}} - \widehat{\gamma}_j^{\mathrm{Tr}}}
{\sqrt{\sigma^{2}_{\gamma j, \mathrm{Ou}} + \sigma^{2}_{\gamma j, \mathrm{Tr}}}}
\sim N(0,1),
\]
and $T_1,\ldots,T_p$ are independent random variables. Therefore,
\[
T = \sum_{j=1}^p T_j^2
\]
follows a chi-square distribution with \(p\) degrees of freedom.
\subsection{Proof of Proposition~\ref{prop:conditionB2}}

For the first part, we note that
\begin{align*}
\kappa_j^{\text{Ou}} = \frac{g^2(\gamma_j^{\text{Tr}})}{\sigma_{\gamma j,\text{Tr}}^2}
\geq C_1^2 \cdot \frac{(\gamma_j^{\text{Tr}})^2}{\sigma_{\gamma j,\text{Tr}}^2} 
= C_1^2 \kappa_j^{\text{Tr}},
\end{align*}
which implies that $\kappa^{\text{Ou}} \gtrsim \kappa^{\text{Tr}}$. Consequently, if $(\kappa^{\text{Tr}})^2 p \to \infty$, then $(\kappa^{\text{Ou}})^2 p \to \infty$ as well.

For the second part, we observe that
\begin{align*}
C_2^2 \kappa_j^{\text{Tr}} \geq \kappa_j^{\text{Ou}} \geq C_1^2 \kappa_j^{\text{Tr}}.
\end{align*}
This leads to
\begin{align*}
\kappa_j^{\text{Co}} = \sqrt{\kappa_j^{\text{Ou}} \kappa_j^{\text{Tr}}}
\geq \max \left\{ C_1 \kappa_j^{\text{Tr}}, \, \frac{\kappa_j^{\text{Ou}}}{C_2} \right\}.
\end{align*}
Taking the maximum over $j$ and using the inequality above, we obtain
\begin{align*}
\kappa^{\text{Co}} \geq \max \left\{ C_1 \kappa^{\text{Tr}}, \, \frac{\kappa^{\text{Ou}}}{C_2} \right\}.
\end{align*}
It follows that
\begin{align*}
\kappa^{\text{Co}} \geq \sqrt{\frac{C_1}{C_2}} \cdot \sqrt{\kappa^{\text{Tr}} \kappa^{\text{Ou}}} \sim \sqrt{\kappa^{\text{Tr}} \kappa^{\text{Ou}}},
\end{align*}
which completes the proof.

\subsection{Proof of Theorem \ref{thm:walds} (a)}

Under Assumption B1, $W^{(1)} = kW^{(2)}$, thus the Wald estimator \eqref{eqn:mr_walds} can be simplified as

\begin{equation}
\label{eqn:simple_mrwalds}
  \widehat{\beta}^{MR-Wald}  = 
\frac{(\widehat{\bm \gamma}^{\mathrm{Tr}})^{\T} W^{(1)} \widehat{\bm \Gamma}^{\mathrm{Ou}}}{(\widehat{\bm \gamma}^{\mathrm{Tr}})^{\T} W^{(1)} \widehat{\bm \gamma}^{\mathrm{Ou}}} = \frac{\sum^p_{j=1}\frac{\widehat{ \gamma}^{\mathrm{Tr}}_j\widehat{ \Gamma}^{\mathrm{Ou}}_j}{\sigma_{\gamma j, \mathrm{Ou}}^2}}{\sum^p_{j=1}\frac{\widehat{ \gamma}^{\mathrm{Tr}}_j\widehat{ \gamma}^{\mathrm{Ou}}_j}{\sigma_{\gamma j, \mathrm{Ou}}^2}}
\end{equation}

We handle the numerator and denominator separately.
For the denominator, whose mean is $\sum^p_{j=1}\frac{\gamma_j^{\mathrm{Tr}}\gamma_j^{\mathrm{Ou}}}{\sigma_{\gamma j, \mathrm{Ou}}^2}$, and whose variance is $\sum^p_{j=1}\left\{\frac{(\gamma_j^{\mathrm{Tr}})^2}{ \sigma_{\gamma j, \mathrm{Ou}}^2}\right.$ $\left.+ \frac{(\gamma_j^{\mathrm{Ou}})^2 \sigma_{\gamma j, \mathrm{Tr}}^2}{\sigma_{\gamma j, \mathrm{Ou}}^4} + \frac{\sigma_{\gamma j, \mathrm{Tr}}^2}{\sigma_{\gamma j, \mathrm{Ou}}^2}\right\}$. 
We now show that the denominator is close to its mean, in the sense that

\begin{equation}
\label{eqn:denomeratorgoesto1}
\frac{\sum^p_{j=1}\frac{\widehat{ \gamma}^{\mathrm{Tr}}_j\widehat{ \gamma}^{\mathrm{Ou}}_j}{\sigma_{\gamma j, \mathrm{Ou}}^2}}{\sum^p_{j=1}\frac{\gamma_j^{\mathrm{Tr}}\gamma_j^{\mathrm{Ou}}}{\sigma_{\gamma j, \mathrm{Ou}}^2}}\xrightarrow{p.}1  
\end{equation}

We note that for the mean of the denominator, we have 
\begin{equation}
\label{eqn:meandomominat}
  \mathbb{E}\left({\sum^p_{j=1}\frac{\widehat{ \gamma}^{\mathrm{Tr}}_j\widehat{ \gamma}^{\mathrm{Ou}}_j}{\sigma_{\gamma j, \mathrm{Ou}}^2}}\right) = {\sum^p_{j=1}\frac{{ \gamma}^{\mathrm{Tr}}_j{ \gamma}^{\mathrm{Ou}}_j}{\sigma_{\gamma j, \mathrm{Ou}}^2}} \geq \sum^p_{j=1}\frac{\gamma_j^{\mathrm{Tr}}\gamma_j^{\mathrm{Ou}}}{C_2\sigma_{\gamma j, \mathrm{Tr}}^2} \gtrsim \kappa^{Co} p,
\end{equation}

The first inequality is given by Assumption B1, and the second inequality is given by Assumption B2. We also note that, for the variance of the denominator, we have

\begin{equation}
\label{eqn:vardomominat}
\begin{split}
\text{Var}\left({\sum^p_{j=1}\frac{\widehat{ \gamma}^{\mathrm{Tr}}_j\widehat{ \gamma}^{\mathrm{Ou}}_j}{\sigma_{\gamma j, \mathrm{Ou}}^2}}\right) 
&= \sum^p_{j=1}\left\{\frac{(\gamma_j^{\mathrm{Tr}})^2}{ \sigma_{\gamma j, \mathrm{Ou}}^2} + \frac{(\gamma_j^{\mathrm{Ou}})^2 \sigma_{\gamma j, \mathrm{Tr}}^2}{\sigma_{\gamma j, \mathrm{Ou}}^4} + \frac{\sigma_{\gamma j, \mathrm{Tr}}^2}{\sigma_{\gamma j, \mathrm{Ou}}^2}\right\}\\
&\leq \sum^p_{j=1}\left\{\frac{(\gamma_j^{\mathrm{Tr}})^2}{C_1 \sigma_{\gamma j, \mathrm{Tr}}^2} + \frac{(\gamma_j^{\mathrm{Ou}})^2}{C_1\sigma_{\gamma j, \mathrm{Tr}}^2} + \frac{1}{C_1}\right\}\\
&\lesssim (\kappa^{\mathrm{Tr}}+\kappa^{\mathrm{Ou}}+1)p,
\end{split}
\end{equation}
where the first inequality also follows from Assumption B2. Combining equations \eqref{eqn:meandomominat} and \eqref{eqn:vardomominat}, we finally have

\begin{equation}
\label{eqn:demominator/vargoesto1}
\text{Var}\left(\frac{\sum^p_{j=1}\frac{\widehat{ \gamma}^{\mathrm{Tr}}_j\widehat{ \gamma}^{\mathrm{Ou}}_j}{\sigma_{\gamma j, \mathrm{Ou}}^2}}{\sum^p_{j=1}\frac{\gamma_j^{\mathrm{Tr}}\gamma_j^{\mathrm{Ou}}}{\sigma_{\gamma j, \mathrm{Ou}}^2}}\right)\lesssim\frac{\kappa^{\mathrm{Tr}}+\kappa^{\mathrm{Ou}}+1}{(\kappa^{Co})^2 p} \lesssim \frac{1}{\kappa^{\mathrm{Tr}} p} + \frac{1}{\kappa^{\mathrm{Ou}} p} + \frac{1}{(\kappa^{Co})^2 p}  = o(1),
\end{equation}
as $\kappa^{\mathrm{Tr}} p = \sqrt{(\kappa^{\mathrm{Tr}})^2 p}\sqrt{p}\geq \sqrt{(\kappa^{\mathrm{Tr}})^2 p}\rightarrow\infty $, $\kappa^{\mathrm{Ou}} p = \sqrt{(\kappa^{\mathrm{Ou}})^2 p}\sqrt{p}\geq\sqrt{(\kappa^{\mathrm{Ou}})^2 p}\rightarrow\infty $, and $(\kappa^{Co})^2 p \gtrsim \sqrt{(\kappa^{\mathrm{Tr}})^2 p}\sqrt{(\kappa^{\mathrm{Ou}})^2 p}\rightarrow\infty$, provided condition B2. Equation \eqref{eqn:demominator/vargoesto1} is sufficient to prove \eqref{eqn:denomeratorgoesto1}.

We now turn to the numerator, whose variance is
\begin{align*}
&\sum^p_{j=1}\left\{\frac{(\gamma_j^{\mathrm{Tr}})^2\sigma_{\Gamma j, \mathrm{Ou}}^2}{ \sigma_{\gamma j, \mathrm{Ou}}^4} + \frac{(\Gamma_j^{\mathrm{Ou}})^2 \sigma_{\gamma j, \mathrm{Tr}}^2}{\sigma_{\gamma j, \mathrm{Ou}}^4} + \frac{\sigma_{\gamma j, \mathrm{Tr}}^2\sigma_{\Gamma j, \mathrm{Ou}}^2}{\sigma_{\gamma j, \mathrm{Ou}}^4}\right\} \\
\leq &\sum^p_{j=1}\left\{\frac{(\gamma_j^{\mathrm{Tr}})^2t}{ C_1\sigma_{\gamma j, \mathrm{Tr}}^2} + \frac{\beta_0^2(\gamma_j^{\mathrm{Ou}})^2}{C_2^2\sigma_{\gamma j, \mathrm{Tr}}^2} + \frac{t}{C_1}\right\}\\
\lesssim &(\kappa^{\mathrm{Tr}}+\kappa^{\mathrm{Ou}}+1)p,  
\end{align*}

where the first inequality is provided by assumption B1, which, together with \eqref{eqn:meandomominat}, implies
\begin{equation}
  \label{eqn:numerrator/meanofdenominator}
  \text{Var}\left(\frac{\sum^p_{j=1}\frac{\widehat{ \gamma}^{\mathrm{Tr}}_j\widehat{ \Gamma}^{\mathrm{Ou}}_j}{\sigma_{\gamma j, \mathrm{Ou}}^2}}{\sum^p_{j=1}\frac{\gamma_j^{\mathrm{Tr}}\gamma_j^{\mathrm{Ou}}}{\sigma_{\gamma j, \mathrm{Ou}}^2}}\right)\lesssim \frac{\kappa^{\mathrm{Tr}}+\kappa^{\mathrm{Ou}}+1}{(\kappa^{Co})^2p} = o(1),
\end{equation}
provided Assumption B2.

Equation \eqref{eqn:numerrator/meanofdenominator} suggests that $\frac{\sum^p_{j=1}\frac{\widehat{ \gamma}^{\mathrm{Tr}}_j\widehat{ \Gamma}^{\mathrm{Ou}}_j}{\sigma_{\gamma j, \mathrm{Ou}}^2}}{\sum^p_{j=1}\frac{\gamma_j^{\mathrm{Tr}}\gamma_j^{\mathrm{Ou}}}{\sigma_{\gamma j, \mathrm{Ou}}^2}}$ will converge in probability to its mean:

\begin{equation}
\label{eqn:convergeofnumerator}
  \frac{\sum^p_{j=1}\frac{\widehat{ \gamma}^{\mathrm{Tr}}_j\widehat{ \Gamma}^{\mathrm{Ou}}_j}{\sigma_{\gamma j, \mathrm{Ou}}^2}}{\sum^p_{j=1}\frac{\gamma_j^{\mathrm{Tr}}\gamma_j^{\mathrm{Ou}}}{\sigma_{\gamma j, \mathrm{Ou}}^2}} \xrightarrow{p.} \frac{\sum^p_{j=1}\frac{\gamma_j^{\mathrm{Tr}} \Gamma_j^{\mathrm{Ou}}}{\sigma_{\gamma j, \mathrm{Ou}}^2}}{\sum^p_{j=1}\frac{\gamma_j^{\mathrm{Tr}}\gamma_j^{\mathrm{Ou}}}{\sigma_{\gamma j, \mathrm{Ou}}^2}} = \frac{\sum^p_{j=1}\frac{\gamma_j^{\mathrm{Tr}}\gamma_j^{\mathrm{Ou}}\beta_0}{\sigma_{\gamma j, \mathrm{Ou}}^2}}{\sum^p_{j=1}\frac{\gamma_j^{\mathrm{Tr}}\gamma_j^{\mathrm{Ou}}}{\sigma_{\gamma j, \mathrm{Ou}}^2}} = \beta_0.
\end{equation}

Combining equations \eqref{eqn:denomeratorgoesto1} and \eqref{eqn:convergeofnumerator}, we finally arrive at

$$
 \widehat{\beta}^{MR-Wald}  = 
\frac{\sum^p_{j=1}\frac{\widehat{ \gamma}^{\mathrm{Tr}}_j\widehat{ \Gamma}^{\mathrm{Ou}}_j}{\sigma_{\gamma j, \mathrm{Ou}}^2}}{\sum^p_{j=1}\frac{\widehat{ \gamma}^{\mathrm{Tr}}_j\widehat{ \gamma}^{\mathrm{Ou}}_j}{\sigma_{\gamma j, \mathrm{Ou}}^2}} = 
\frac{\left(\sum^p_{j=1}\frac{\widehat{ \gamma}^{\mathrm{Tr}}_j\widehat{ \Gamma}^{\mathrm{Ou}}_j}{\sigma_{\gamma j, \mathrm{Ou}}^2}\right)/\left(\sum^p_{j=1}\frac{ \gamma^{\mathrm{Tr}}_j \gamma^{\mathrm{Ou}}_j }{\sigma_{\gamma j, \mathrm{Ou}}^2}\right)}{\left(\sum^p_{j=1}\frac{\widehat{ \gamma}^{\mathrm{Tr}}_j \widehat{ \gamma}^{\mathrm{Ou}}_j}{\sigma_{\gamma j, \mathrm{Ou}}^2}\right)/\left(\sum^p_{j=1}\frac{ \gamma^{\mathrm{Tr}}_j \gamma^{\mathrm{Ou}}_j }{\sigma_{\gamma j, \mathrm{Ou}}^2}\right)} \xrightarrow{p.} \beta_0,
$$

by the Slutsky's Theorem. This completes the proof for the first part of Theorem \ref{thm:walds}.

\subsection{Proof of Theorem \ref{thm:walds} (b)}

We first decompose the $\widehat{\beta}^{MR-Wald}-\beta_0$, and then process the numerator of the target quantity:

\begin{equation}
\label{eqn:mrwalddecompose}
  \widehat{\beta}^{MR-Wald} -\beta_0 
 = \frac{\sum^p_{j=1}\frac{\widehat{ \gamma}^{\mathrm{Tr}}_j\widehat{ \Gamma}^{\mathrm{Ou}}_j}{\sigma_{\gamma j, \mathrm{Ou}}^2}}{\sum^p_{j=1}\frac{\widehat{ \gamma}^{\mathrm{Tr}}_j\widehat{ \gamma}^{\mathrm{Ou}}_j}{\sigma_{\gamma j, \mathrm{Ou}}^2}} -\beta_0
 =
 \frac{\sum^p_{j=1}\frac{\widehat{ \gamma}^{\mathrm{Tr}}_j\widehat{ \Gamma}^{\mathrm{Ou}}_j-\beta_0\widehat{\gamma}_j^{\mathrm{Tr}}\widehat{\gamma}_j^{\mathrm{Ou}}}{\sigma_{\gamma j, \mathrm{Ou}}^2}}{\sum^p_{j=1}\frac{\widehat{ \gamma}^{\mathrm{Tr}}_j\widehat{ \gamma}^{\mathrm{Ou}}_j}{\sigma_{\gamma j, \mathrm{Ou}}^2}}
 =  \frac{\sum^p_{j=1}\frac{\widehat{ \gamma}^{\mathrm{Tr}}_jU_j^{\mathrm{Ou}}}{\sigma_{\gamma j, \mathrm{Ou}}^2}}{\sum^p_{j=1}\frac{\widehat{ \gamma}^{\mathrm{Tr}}_j\widehat{ \gamma}^{\mathrm{Ou}}_j}{\sigma_{\gamma j, \mathrm{Ou}}^2}},
\end{equation}



The last equation holds because \( U_j^{\mathrm{Ou}} = (\widehat{\Gamma}_j^{\mathrm{Ou}} - \Gamma_j^{\mathrm{Ou}}) - \beta_0(\widehat{\gamma}_j^{\mathrm{Ou}} - \gamma_j^{\mathrm{Ou}}) = \widehat{\Gamma}_j^{\mathrm{Ou}} - \beta_0 \widehat{\gamma}_j^{\mathrm{Ou}} \), since \( \Gamma_j^{\mathrm{Ou}} - \beta_0 \gamma_j^{\mathrm{Ou}} = \alpha_j = 0 \).
Let's consider the numerator of \eqref{eqn:mrwalddecompose}, who has the form

\begin{equation}
  \label{eqn:MR_walds_linderburg_numerator}
  \sum^p_{j=1}\frac{\widehat{ \gamma}^{\mathrm{Tr}}_jU_j^{\mathrm{Ou}}}{\sigma_{\gamma j, \mathrm{Ou}}^2} = \sum_j^p L_j,
\end{equation}
Each term $L_j$ has mean 0, variance $\sigma_j^2 = \frac{((\gamma_j^{\mathrm{Tr}})^2+\sigma_{\gamma j, \mathrm{Tr}}^2)\sigma^2_{Uj}}{\sigma^4_{\gamma j, \mathrm{Ou}}}$. Let $s_p^2 = \sum^p_{j=1}\sigma_j^2$. We now verify the Linderburg's condition for \eqref{eqn:MR_walds_linderburg_numerator}.

We first write the Lindeberg condition for $\{L_j\}^p_{j=1}$:
\begin{equation}
\label{eqn:Lindburg_statement}
\begin{split}
  \frac{1}{s_p^2}\sum_{j=1}^p\mathbb{E}(L_j^2\cdot\bm{1}_{|L_j|>\epsilon s_p}) 
  = \frac{1}{s_p^2}\sum_{j=1}^p \frac{1}{\sigma_{\gamma j, \mathrm{Ou}}^4} \mathbb{E}((\widehat{\gamma}_j^{\mathrm{Tr}}U_j^{\mathrm{Ou}})^2\cdot\bm{1}_{|L_j|>\epsilon s_p})
\end{split}
\end{equation}

To verify that \eqref{eqn:Lindburg_statement} goes to 0 for $\forall \epsilon > 0$, we first provide a bound for the following quantity:
\begin{equation}
\label{eqn:bound1}
\begin{split}
\mathbb{E}((\widehat{\gamma}_j^{\mathrm{Tr}}U_j^{\mathrm{Ou}})^2\cdot\bm{1}_{|L_j|>\epsilon s_p})
&\leq \sqrt{\mathbb{E}\left((\widehat{\gamma}_j^{\mathrm{Tr}}U_j^{\mathrm{Ou}})^4\right)}\sqrt{\mathbb{P}({|L_j|>\epsilon s_p})} \\
& \leq \sqrt{\mathbb{E}\left((\widehat{\gamma}_j^{\mathrm{Tr}}U_j^{\mathrm{Ou}})^4\right)}\sqrt{\frac{\mathbb{E}(L_j^2)}{\epsilon^2 s_p^2}} \\
& = \frac{\sigma_j}{\epsilon s_p}\sqrt{\mathbb{E}\left((\widehat{\gamma}_j^{\mathrm{Tr}}U_j^{\mathrm{Ou}})^4\right)},
\end{split}
\end{equation}

The first line follows from the Cauchy-Schwarz inequality, and the second line follows from Markov's inequality. Using equation \eqref{eqn:bound1}, the Lindeberg condition \eqref{eqn:Lindburg_statement} will be bounded by 
\begin{equation}
\label{eqn:bound2}
\begin{split}
  LSH &\lesssim  \frac{1}{s_p^3}\sum_{j=1}^p{\sigma_{j}}\cdot \frac{\sqrt{\mathbb{E}\left((\widehat{\gamma}_j^{\mathrm{Tr}}U_j^{\mathrm{Ou}})^4\right)}}{\sigma_{\gamma j, \mathrm{Ou}}^4} \\
  &\leq\frac{1}{s_p^3}\sqrt{\sum_{j=1}^p{\sigma_{j}^2}}\sqrt{\sum^p_{j=1}\frac{{\mathbb{E}\left((\widehat{\gamma}_j^{\mathrm{Tr}}U_j^{\mathrm{Ou}})^4\right)}}{\sigma_{\gamma j, \mathrm{Ou}}^8}} \\
  &= \frac{1}{s_p^2}\sqrt{\sum^p_{j=1}\frac{\mathbb{E}\left((\widehat{\gamma}_j^{\mathrm{Tr}}U_j^{\mathrm{Ou}})^4\right)}{\sigma_{\gamma j, \mathrm{Ou}}^8}},
\end{split}
\end{equation}

where the second inequality follows from H\"older's inequality. We now discuss its upper bound and thus derive the rate of \eqref{eqn:Lindburg_statement}. From Assumptions B1, B2 and B3, $s_p^2$ can be lower bounded by
\begin{equation}
  \label{eqn:bound3}
  s_p^2 =\sum_j^p \frac{((\gamma_j^{\mathrm{Tr}})^2+\sigma_{\gamma j, \mathrm{Tr}}^2)\sigma^2_{Uj}}{\sigma^4_{\gamma j, \mathrm{Ou}}} 
  \geq \sum_j^p \frac{((\gamma_j^{\mathrm{Tr}})^2+\sigma_{\gamma j, \mathrm{Tr}}^2)\frac{1}{C_2}\sigma^2_{\gamma j, \mathrm{Ou}}}{\sigma^2_{\gamma j, \mathrm{Ou}}C_2\sigma^2_{\gamma j, \mathrm{Tr}}} 
  \gtrsim \sum^p_{j=1}(\kappa_j^{\mathrm{Tr}}+1) = (\kappa^{\mathrm{Tr}} +1)p.
\end{equation}

We now bound the other term under the square root in equation \eqref{eqn:bound2}. Define new random variables \(M_j = \widehat{\gamma}_j^{\mathrm{Tr}} - {\gamma}_j^{\mathrm{Tr}}\), and let \(m_j^{(k)}\) denotes the \(k\)th moments of \(M_j\). We then have the following bound:
\begin{equation}
  \label{eqn:bound4}
  \begin{split}\sqrt{\sum^p_{j=1}\frac{\mathbb{E}\left((\widehat{\gamma}_j^{\mathrm{Tr}}U_j^{\mathrm{Ou}})^4\right)}{\sigma_{\gamma j, \mathrm{Ou}}^8}} 
  &=\sqrt{\sum^p_{j=1}\frac{\mathbb{E}\left((M_j+{\gamma}_j^{\mathrm{Tr}})^4(U_j^{\mathrm{Ou}})^4\right)}{\sigma_{\gamma j, \mathrm{Ou}}^8}}\\
  &= \sqrt{\sum^p_{j=1}\frac{ \left( m_j^{(4)}+ 4(\gamma_j^{\mathrm{Tr}}) m_j^{(3)} + 6(\gamma_j^{\mathrm{Tr}})^2 m_j^{(2)} + (\gamma_j^{\mathrm{Tr}})^4\right)
}{\sigma_{\gamma j, \mathrm{Ou}}^4} \frac{\mathbb{E}((U_j^{\mathrm{Ou}})^4)}{\sigma_{\gamma j, \mathrm{Ou}}^4}}\\
&\lesssim \sqrt{\sum_j^p\frac{ \left( m_j^{(4)}+ 4(\gamma_j^{\mathrm{Tr}}) m_j^{(3)} + 6(\gamma_j^{\mathrm{Tr}})^2 m_j^{(2)} + (\gamma_j^{\mathrm{Tr}})^4\right)
}{\sigma_{\gamma j, \mathrm{Tr}}^4}\frac{\mathbb{E}((U_j^{\mathrm{Ou}})^4)}{\sigma_{Uj}^4}}\\
& \lesssim   
\sqrt{\sum^p_{j=1}\left\{\frac{m^{(4)}_j}{(m_j^{(2)})^2}+\frac{\gamma_j^{\mathrm{Tr}}}{(m_j^{(2)})^{1/2}}\frac{m_j^{(3)}}{(m_j^{(2)})^{3/2}}+\frac{(\gamma_j^{\mathrm{Tr}})^2}{m_j^{(2)}}+\frac{(\gamma_j^{\mathrm{Tr}})^4}{(m_j^{(2)})^2}\right\}}\\
&\lesssim\sqrt{\sum^p_{j=1}\left\{1+\frac{\gamma_j^{\mathrm{Tr}}}{(m_j^{(2)})^{1/2}}+\frac{(\gamma_j^{\mathrm{Tr}})^2}{m_j^{(2)}}+\frac{(\gamma_j^{\mathrm{Tr}})^4}{(m_j^{(2)})^2}\right\}}\\
&=\sqrt{\sum_{j=1}^p\left(1 + \sqrt{\kappa_j^{\mathrm{Tr}}} + \kappa_j^{\mathrm{Tr}} +(\kappa_j^{\mathrm{Tr}})^2\right)}\\
&\lesssim
\sqrt{\sum_{j=1}^p\left(1  + \kappa_j ^{\mathrm{Tr}}+\kappa_j^{\mathrm{Tr}}(\max_j\kappa_j^{\mathrm{Tr}})\right)}=
  \sqrt{p+p\kappa^{\mathrm{Tr}} + p\kappa^{\mathrm{Tr}}(\max_j\kappa_j^{\mathrm{Tr}})}.
  \end{split}
\end{equation}
The first line follows the decomposition using $M_j$, the second line follows the tedious calculation, the third line use the assumption B3(bounded forth moments),  B1(bound among variance) and indendence among two sample, the fourth line follows equivalent transformation up to a constant, and noticing that $\sigma_{\gamma j, \mathrm{Tr}}^2 = m_j^{(2)}$,
The fifth line follows from condition B3 and an application of H\"older's inequality:
$m_j^{(3)} \leq \sqrt{m_j^{(2)} m_j^{(4)}} \lesssim (m_j^{(2)})^{3/2}.
$The sixth line follows from the definition of \(\kappa_j\); the sixth line uses the inequality \(2\sqrt{a} \leq 1 + a\); and the final line follows from the definition of \(\kappa\).

Combining equations \eqref{eqn:bound2}, \eqref{eqn:bound3}, and \eqref{eqn:bound4}, the Lindeberg condition is bounded by
\begin{equation}
  \label{eqn:boundlinderburg_final}
  LSH \lesssim \frac{\sqrt{p+p\kappa^{\mathrm{Tr}} + p\kappa^{\mathrm{Tr}}(\max_j\kappa_j^{\mathrm{Tr}})}}{p (\kappa^{\mathrm{Tr}}+1)} \lesssim\sqrt{\frac{1}{p(\kappa^{\mathrm{Tr}})^2}+\frac{1}{p\kappa^{\mathrm{Tr}}}+\frac{\max_j\kappa_j^{\mathrm{Tr}}}{p\kappa^{\mathrm{Tr}}}} = o(1),
\end{equation}
provided that both $p(\kappa^{\mathrm{Tr}})^2 \to \infty$ and $p\kappa^{\mathrm{Tr}} = \sqrt{p}\sqrt{p(\kappa^{\mathrm{Tr}})^2} \to \infty$. So far, we have verified Lindeberg's condition, which implies 

\begin{equation}
\label{eqn:thm1final1}
  \frac{1}{s_p}\sum^p_{j=1}\frac{\widehat{ \gamma}^{\mathrm{Tr}}_jU_j^{\mathrm{Ou}}}{\sigma_{\gamma j, \mathrm{Ou}}^2} \xrightarrow{d.}\mathcal{N}(0,1).
\end{equation}

Using the pointwise convergence result: $\left(\sum^p_{j=1}\frac{\widehat{ \gamma}^{\mathrm{Tr}}_j \widehat{ \gamma}^{\mathrm{Ou}}_j}{\sigma_{\gamma j, \mathrm{Ou}}^2}\right)/\left(\sum^p_{j=1}\frac{ \gamma^{\mathrm{Tr}}_j \gamma^{\mathrm{Ou}}_j }{\sigma_{\gamma j, \mathrm{Ou}}^2}\right)\xrightarrow{p.} 1 $ and let $V_1^{-1} = \frac{\left(\sum^p_{j=1}\frac{ \gamma^{\mathrm{Tr}}_j \gamma^{\mathrm{Ou}}_j }{\sigma_{\gamma j, \mathrm{Ou}}^2}\right)}{s_p}$, we arrive

$$
V_1^{-1}(\widehat{\beta}^{MR-Wald} - \beta_0) = \frac{\frac{1}{s_p}\sum^p_{j=1}\frac{\widehat{ \gamma}^{\mathrm{Tr}}_jU_j^{\mathrm{Ou}}}{\sigma_{\gamma j, \mathrm{Ou}}^2} }{\left(\sum^p_{j=1}\frac{\widehat{ \gamma}^{\mathrm{Tr}}_j \widehat{ \gamma}^{\mathrm{Ou}}_j}{\sigma_{\gamma j, \mathrm{Ou}}^2}\right)/\left(\sum^p_{j=1}\frac{ \gamma^{\mathrm{Tr}}_j \gamma^{\mathrm{Ou}}_j }{\sigma_{\gamma j, \mathrm{Ou}}^2}\right)}\xrightarrow{d.}\mathcal{N}(0,1),$$
which proved the final result.

\subsection{Proof of Theorem \ref{thm:walds} (c)}
Assuming $\gamma^{\mathrm{Tr}}_j = \gamma^{\mathrm{Ou}}_j$, from the previous subsection we know that  
\[
V^{-1}(\widehat{\beta}^{\mathrm{Mr\text{-}wald}} - \beta_0) \xrightarrow{d} \mathcal{N}(0,1),
\]
where
\[
V^2 = \frac{s_p^2}{\left(\sum_{j=1}^p \frac{(\gamma^{\mathrm{Tr}}_j)^2}{\sigma_{\gamma j, \mathrm{Ou}}^2}\right)^2} 
= \frac{\sum_{j=1}^p \frac{(\gamma_j^{\mathrm{Tr}})^2 + \sigma_{\gamma j, \mathrm{Tr}}^2}{\sigma_{\gamma j, \mathrm{Ou}}^2}\frac{ \sigma_{Uj}^2}{\sigma_{\gamma j, \mathrm{Ou}}^2}}
{\left(\sum_{j=1}^p \frac{(\gamma_j^{\mathrm{Tr}})^2}{\sigma_{\gamma j, \mathrm{Ou}}^2}\right)^2}
=\frac{\sum_{j=1}^p \frac{(\gamma_j^{\mathrm{Tr}})^2 + \sigma_{\gamma j, \mathrm{Tr}}^2}{\sigma_{\Gamma j, \mathrm{Ou}}^2}\frac{ \sigma_{Uj}^2}{\sigma_{\Gamma j, \mathrm{Ou}}^2}}
{\left(\sum_{j=1}^p \frac{(\gamma_j^{\mathrm{Tr}})^2}{\sigma_{\Gamma j, \mathrm{Ou}}^2}\right)^2}.
\]
The third equality follows the assumption B1 that $\sigma_{\Gamma j, \mathrm{Ou}}^2 = k\sigma_{\gamma j, \mathrm{Ou}}^2$.

From Theorem 4.1 of \citet{ye2021debiased}, we also know that  
\[
\widetilde{V}^{-1}(\widehat{\beta}^{0,\mathrm{dIVW}} - \beta_0) \xrightarrow{d} \mathcal{N}(0,1),
\]
where
\[
\widetilde{V}^2 = \frac{\sum_{j=1}^{p} \left( 
\frac{(\gamma_j^{\mathrm{Tr}})^2}{\sigma_{\Gamma j, \mathrm{Ou}}^2}  
+ \frac{\sigma_{\gamma j, \mathrm{Tr}}^2}{\sigma_{\Gamma j, \mathrm{Ou}}^2}  
+ \beta_0^2 \frac{\sigma_{\gamma j, \mathrm{Tr}}^2}{\sigma_{\Gamma j, \mathrm{Ou}}^2}  
\left( \frac{(\gamma_j^{\mathrm{Tr}})^2}{\sigma_{\Gamma j, \mathrm{Ou}}^2}  
+ \frac{2\sigma_{\gamma j, \mathrm{Tr}}^2}{\sigma_{\Gamma j, \mathrm{Ou}}^2}  \right) 
\right)}
{\left(\sum_{j=1}^p \frac{(\gamma_j^{\mathrm{Tr}})^2}{\sigma_{\Gamma j, \mathrm{Ou}}^2}  \right)^2}.
\]

Comparing the expressions for $V^2$ and $\widetilde{V}^2$, it suffices to show that $\sigma_{Uj}^2 \leq \sigma_{\Gamma j, \mathrm{Ou}}^2$ in order to conclude that $V^2 \leq \widetilde{V}^2$. This has been guaranteed by the assumption of the claim.  

\section{Additional results for pleiotropy}
\label{sec:app_pleiotropy}
We present additional theoretical results and numerical results accounting for invalid IV and pleiotropy effect we have discussed in \ref{sec:ple}. We will discuss 

\subsection{Additional result for balanced pleiotropy}
\label{sec:plei_sub}
We provide proof for Proposition \ref{prop:balanced_p} in this section. The proof is very similar to the one we provided in Section \ref{sec:proof_sec3}, the only difference is noticing that $\text{Var}(\widehat{\Gamma}_j^{\mathrm{Ou}}) = \tau_0^2+\sigma_{\Gamma j, \mathrm{Ou}}^2$. We only prove the first part of Proposition \ref{prop:balanced_p}, and we omit the details of the second part for brevity.

\subsubsection{Proof of Proposition~\ref{prop:balanced_p} (Part 1, consistentcy)}
Under Assumption B2, $W^{(1)} = kW^{(2)}$, thus the Wald estimator \eqref{eqn:mr_walds} can be simplified as

\begin{equation}
\label{eqn:simple_mrwalds_bp}
  \widehat{\beta}^{MR-Wald}  = 
\frac{(\widehat{\bm \gamma}^{\mathrm{Tr}})^{\T} W^{(1)} \widehat{\bm \Gamma}^{\mathrm{Ou}}}{(\widehat{\bm \gamma}^{\mathrm{Tr}})^{\T} W^{(1)} \widehat{\bm \gamma}^{\mathrm{Ou}}} = \frac{\sum^p_{j=1}\frac{\widehat{ \gamma}^{\mathrm{Tr}}_j\widehat{ \Gamma}^{\mathrm{Ou}}_j}{\sigma_{\gamma j, \mathrm{Ou}}^2}}{\sum^p_{j=1}\frac{\widehat{ \gamma}^{\mathrm{Tr}}_j\widehat{ \gamma}^{\mathrm{Ou}}_j}{\sigma_{\gamma j, \mathrm{Ou}}^2}}
\end{equation}

We handle the numerator and denominator of equation \eqref{eqn:simple_mrwalds_bp} separately.

For the denominator of equation \eqref{eqn:simple_mrwalds_bp}, the following result has been established in \eqref{eqn:denomeratorgoesto1}.

\begin{equation*}
\frac{\sum^p_{j=1}\frac{\widehat{ \gamma}^{\mathrm{Tr}}_j\widehat{ \gamma}^{\mathrm{Ou}}_j}{\sigma_{\gamma j, \mathrm{Ou}}^2}}{\sum^p_{j=1}\frac{\gamma_j^{\mathrm{Tr}}\gamma_j^{\mathrm{Ou}}}{\sigma_{\gamma j, \mathrm{Ou}}^2}}\xrightarrow{p.}1  
\end{equation*}
We now turn to the numerator of equation~\eqref{eqn:simple_mrwalds_bp}. 
We first note that the variance and expectation of $\widehat{\Gamma}_j^{\mathrm{Ou}}$ are given by
\[
\mathbb{E}(\widehat{\Gamma}_j^{\mathrm{Ou}}) 
= \mathbb{E}\!\left( \mathbb{E}(\widehat{\Gamma}_j^{\mathrm{Ou}} \mid \alpha_j) \right) 
= \mathbb{E}(\beta_0 \gamma_j^{\mathrm{Ou}} + \alpha_j) 
= \beta_0 \gamma_j^{\mathrm{Ou}},
\]
\[
\text{Var}(\widehat{\Gamma}_j^{\mathrm{Ou}}) 
= \text{Var}\!\left( \mathbb{E}(\widehat{\Gamma}_j^{\mathrm{Ou}} \mid \alpha_j) \right) 
+ \mathbb{E}\!\left( \text{Var}(\widehat{\Gamma}_j^{\mathrm{Ou}} \mid \alpha_j) \right) 
= \text{Var}(\beta_0 \gamma_j^{\mathrm{Ou}} + \alpha_j) + \mathbb{E}(\sigma_{\Gamma j, \mathrm{Ou}}^2) 
= \tau_0^2 + \sigma_{\Gamma j, \mathrm{Ou}}^2.
\]

Using these results, the variance of the numerator of equation~\eqref{eqn:simple_mrwalds_bp} is given by
\begin{align*}
&\sum_{j=1}^p \left\{ 
\frac{(\gamma_j^{\mathrm{Tr}})^2 (\sigma_{\Gamma j, \mathrm{Ou}}^2 + \tau_0^2)}{\sigma_{\gamma j, \mathrm{Ou}}^4} 
+ \frac{(\Gamma_j^{\mathrm{Ou}})^2 \, \sigma_{\gamma j, \mathrm{Tr}}^2}{\sigma_{\gamma j, \mathrm{Ou}}^4} 
+ \frac{\sigma_{\gamma j, \mathrm{Tr}}^2 (\sigma_{\Gamma j, \mathrm{Ou}}^2 + \tau_0^2)}{\sigma_{\gamma j, \mathrm{Ou}}^4} \right\} \\
\leq\;& \sum_{j=1}^p \left\{ 
\frac{(\gamma_j^{\mathrm{Tr}})^2 t}{C_1 \sigma_{\gamma j, \mathrm{Tr}}^2} 
+ \frac{\beta_0^2 (\gamma_j^{\mathrm{Ou}})^2}{C_2^2 \sigma_{\gamma j, \mathrm{Tr}}^2} 
+ \frac{t}{C_1} \right\} \\
\lesssim\;& (\kappa^{\mathrm{Tr}} + \kappa^{\mathrm{Ou}} + 1)\, p.
\end{align*}

where the first inequality follows from Assumption~B1 and the condition $\tau_0^2 \lesssim \max_j \sigma_{\Gamma j, \mathrm{Ou}}^2$. Together with equation~\eqref{eqn:meandomominat}, this implies
\begin{equation}
\label{eqn:numerrator/meanofdenominator_bp}
\mathrm{Var}\left( \frac{ \sum_{j=1}^p \frac{\widehat{\gamma}_j^{\mathrm{Tr}} \widehat{\Gamma}_j^{\mathrm{Ou}}}{\sigma_{\gamma j, \mathrm{Ou}}^2} }{ \sum_{j=1}^p \frac{\gamma_j^{\mathrm{Tr}} \gamma_j^{\mathrm{Ou}}}{\sigma_{\gamma j, \mathrm{Ou}}^2} } \right) \lesssim \frac{\kappa^{\mathrm{Tr}} + \kappa^{\mathrm{Ou}} + 1}{ (\kappa^{\mathrm{Co}})^2 p } = o(1),
\end{equation}
under Assumption~B2.

Equation \eqref{eqn:numerrator/meanofdenominator_bp} suggests that $\frac{\sum^p_{j=1}\frac{\widehat{ \gamma}^{\mathrm{Tr}}_j\widehat{ \Gamma}^{\mathrm{Ou}}_j}{\sigma_{\gamma j, \mathrm{Ou}}^2}}{\sum^p_{j=1}\frac{\gamma_j^{\mathrm{Tr}}\gamma_j^{\mathrm{Ou}}}{\sigma_{\gamma j, \mathrm{Ou}}^2}}$ will converge in probability to its mean:

\begin{equation}
\label{eqn:convergeofnumerator_bp}
  \frac{\sum^p_{j=1}\frac{\widehat{ \gamma}^{\mathrm{Tr}}_j\widehat{ \Gamma}^{\mathrm{Ou}}_j}{\sigma_{\gamma j, \mathrm{Ou}}^2}}{\sum^p_{j=1}\frac{\gamma_j^{\mathrm{Tr}}\gamma_j^{\mathrm{Ou}}}{\sigma_{\gamma j, \mathrm{Ou}}^2}} \xrightarrow{p.} = \frac{\sum^p_{j=1}\frac{\gamma_j^{\mathrm{Tr}}\gamma_j^{\mathrm{Ou}}\beta_0}{\sigma_{\gamma j, \mathrm{Ou}}^2}}{\sum^p_{j=1}\frac{\gamma_j^{\mathrm{Tr}}\gamma_j^{\mathrm{Ou}}}{\sigma_{\gamma j, \mathrm{Ou}}^2}} = \beta_0.
\end{equation}

Combining equations \eqref{eqn:denomeratorgoesto1} and \eqref{eqn:convergeofnumerator_bp}, we finally arrive at

$$
 \widehat{\beta}^{MR-Wald}  = 
\frac{\sum^p_{j=1}\frac{\widehat{ \gamma}^{\mathrm{Tr}}_j\widehat{ \Gamma}^{\mathrm{Ou}}_j}{\sigma_{\gamma j, \mathrm{Ou}}^2}}{\sum^p_{j=1}\frac{\widehat{ \gamma}^{\mathrm{Tr}}_j\widehat{ \gamma}^{\mathrm{Ou}}_j}{\sigma_{\gamma j, \mathrm{Ou}}^2}} = 
\frac{\left(\sum^p_{j=1}\frac{\widehat{ \gamma}^{\mathrm{Tr}}_j\widehat{ \Gamma}^{\mathrm{Ou}}_j}{\sigma_{\gamma j, \mathrm{Ou}}^2}\right)/\left(\sum^p_{j=1}\frac{ \gamma^{\mathrm{Tr}}_j \gamma^{\mathrm{Ou}}_j }{\sigma_{\gamma j, \mathrm{Ou}}^2}\right)}{\left(\sum^p_{j=1}\frac{\widehat{ \gamma}^{\mathrm{Tr}}_j \widehat{ \gamma}^{\mathrm{Ou}}_j}{\sigma_{\gamma j, \mathrm{Ou}}^2}\right)/\left(\sum^p_{j=1}\frac{ \gamma^{\mathrm{Tr}}_j \gamma^{\mathrm{Ou}}_j }{\sigma_{\gamma j, \mathrm{Ou}}^2}\right)} \xrightarrow{p.} \beta_0,
$$

by Slutsky's Theorem. This completes the proof of the consistency result in the first part of Proposition~\ref{prop:balanced_p}.

\subsubsection{Proof of Proposition~\ref{prop:balanced_p} (Part 2, asymptotic normality)}

We first decompose the $\widehat{\beta}^{MR-Wald}-\beta_0$, and then process the numerator of the target quantity:

\begin{equation}
\label{eqn:mrwalddecompose_bp}
  \widehat{\beta}^{MR-Wald} -\beta_0 
 = \frac{\sum^p_{j=1}\frac{\widehat{ \gamma}^{\mathrm{Tr}}_j\widehat{ \Gamma}^{\mathrm{Ou}}_j}{\sigma_{\gamma j, \mathrm{Ou}}^2}}{\sum^p_{j=1}\frac{\widehat{ \gamma}^{\mathrm{Tr}}_j\widehat{ \gamma}^{\mathrm{Ou}}_j}{\sigma_{\gamma j, \mathrm{Ou}}^2}} -\beta_0
 =
 \frac{\sum^p_{j=1}\frac{\widehat{ \gamma}^{\mathrm{Tr}}_j\widehat{ \Gamma}^{\mathrm{Ou}}_j-\beta_0\widehat{\gamma}_j^{\mathrm{Tr}}\widehat{\gamma}_j^{\mathrm{Ou}}}{\sigma_{\gamma j, \mathrm{Ou}}^2}}{\sum^p_{j=1}\frac{\widehat{ \gamma}^{\mathrm{Tr}}_j\widehat{ \gamma}^{\mathrm{Ou}}_j}{\sigma_{\gamma j, \mathrm{Ou}}^2}}
 =  \frac{\sum^p_{j=1}\frac{\widehat{ \gamma}^{\mathrm{Tr}}_j(U_j^{\mathrm{Ou}}+\alpha_j)}{\sigma_{\gamma j, \mathrm{Ou}}^2}}{\sum^p_{j=1}\frac{\widehat{ \gamma}^{\mathrm{Tr}}_j\widehat{ \gamma}^{\mathrm{Ou}}_j}{\sigma_{\gamma j, \mathrm{Ou}}^2}},
\end{equation}

The last equation holds because \( U_j^{\mathrm{Ou}} = (\widehat{\Gamma}_j^{\mathrm{Ou}} - \Gamma_j^{\mathrm{Ou}}) - \beta_0(\widehat{\gamma}_j^{\mathrm{Ou}} - \gamma_j^{\mathrm{Ou}})\) and \(\alpha_j =\Gamma_j^{\mathrm{Ou}} - \beta_0 \gamma_j^{\mathrm{Ou}}  \).
Let's consider the numerator of \eqref{eqn:mrwalddecompose_bp}, who has the form

\begin{equation}
  \label{eqn:MR_walds_linderburg_numerator_bp}
  \sum^p_{j=1}\frac{\widehat{ \gamma}^{\mathrm{Tr}}_j(U_j^{\mathrm{Ou}}+\alpha_j)}{\sigma_{\gamma j, \mathrm{Ou}}^2} = \sum_j^p L_j,
\end{equation}
Each term $L_j$ has mean 0, variance $\sigma_j^2 = {((\gamma_j^{\mathrm{Tr}})^2+\sigma_{\gamma j, \mathrm{Tr}}^2)(\sigma^2_{Uj}+\tau_0^2)}/{\sigma^4_{\gamma j, \mathrm{Ou}}}$. Let $s_p^2 = \sum^p_{j=1}\sigma_j^2$. We now verify the Linderburg's condition for \eqref{eqn:MR_walds_linderburg_numerator_bp}.

We first write the Lindeberg condition for $\{L_j\}^p_{j=1}$:
\begin{equation}
\label{eqn:Lindburg_statement_bp}
\begin{split}
  \frac{1}{s_p^2}\sum_{j=1}^p\mathbb{E}(L_j^2\cdot\bm{1}_{|L_j|>\epsilon s_p}) 
  = \frac{1}{s_p^2}\sum_{j=1}^p \frac{1}{\sigma_{\gamma j, \mathrm{Ou}}^4} \mathbb{E}((\widehat{\gamma}_j^{\mathrm{Tr}}(U_j^{\mathrm{Ou}}+\alpha_j))^2\cdot\bm{1}_{|L_j|>\epsilon s_p})
\end{split}
\end{equation}

To verify that \eqref{eqn:Lindburg_statement_bp} goes to 0 for $\forall \epsilon > 0$, we first provide a bound for the following quantity:
\begin{equation}
\label{eqn:bound1_bp}
\begin{split}
\mathbb{E}((\widehat{\gamma}_j^{\mathrm{Tr}}(U_j^{\mathrm{Ou}}+\alpha_j))^2\cdot\bm{1}_{|L_j|>\epsilon s_p})
&\leq \sqrt{\mathbb{E}\left((\widehat{\gamma}_j^{\mathrm{Tr}}(U_j^{\mathrm{Ou}}+\alpha_j))^4\right)}\sqrt{\mathbb{P}({|L_j|>\epsilon s_p})} \\
& \leq \sqrt{\mathbb{E}\left((\widehat{\gamma}_j^{\mathrm{Tr}}(U_j^{\mathrm{Ou}}+\alpha_j))^4\right)}\sqrt{\frac{\mathbb{E}(L_j^2)}{\epsilon^2 s_p^2}} \\
& = \frac{\sigma_j}{\epsilon s_p}\sqrt{\mathbb{E}\left((\widehat{\gamma}_j^{\mathrm{Tr}}(U_j^{\mathrm{Ou}}+\alpha_j))^4\right)},
\end{split}
\end{equation}

The first line follows from the Cauchy-Schwarz inequality, and the second line follows from Markov's inequality. Using equation \eqref{eqn:bound1}, the Lindeberg condition \eqref{eqn:Lindburg_statement} will be bounded by 
\begin{equation}
\label{eqn:bound2_bp}
\begin{split}
  LSH &\lesssim  \frac{1}{s_p^3}\sum_{j=1}^p{\sigma_{j}}\cdot \frac{\sqrt{\mathbb{E}\left((\widehat{\gamma}_j^{\mathrm{Tr}}(U_j^{\mathrm{Ou}}+\alpha_j))^4\right)}}{\sigma_{\gamma j, \mathrm{Ou}}^4} \\
  &\leq\frac{1}{s_p^3}\sqrt{\sum_{j=1}^p{\sigma_{j}^2}}\sqrt{\sum^p_{j=1}\frac{{\mathbb{E}\left((\widehat{\gamma}_j^{\mathrm{Tr}}(U_j^{\mathrm{Ou}}+\alpha_j))^4\right)}}{\sigma_{\gamma j, \mathrm{Ou}}^8}} \\
  &= \frac{1}{s_p^2}\sqrt{\sum^p_{j=1}\frac{\mathbb{E}\left((\widehat{\gamma}_j^{\mathrm{Tr}}(U_j^{\mathrm{Ou}}+\alpha_j))^4\right)}{\sigma_{\gamma j, \mathrm{Ou}}^8}},
\end{split}
\end{equation}

where the second inequality follows from H\"older's inequality. We now discuss its upper bound and thus derive the rate of \eqref{eqn:Lindburg_statement}. From Assumptions B1, B2 and B3, $s_p^2$ can be lower bounded by
\begin{equation}
  \label{eqn:bound3_bp}
  s_p^2 =\sum_j^p \frac{((\gamma_j^{\mathrm{Tr}})^2+\sigma_{\gamma j, \mathrm{Tr}}^2)(\sigma^2_{Uj}+\tau_0^2)}{\sigma^4_{\gamma j, \mathrm{Ou}}} 
  \geq \sum_j^p \frac{((\gamma_j^{\mathrm{Tr}})^2+\sigma_{\gamma j, \mathrm{Tr}}^2)\frac{1}{C_2}\sigma^2_{\gamma j, \mathrm{Ou}}}{\sigma^2_{\gamma j, \mathrm{Ou}}C_2\sigma^2_{\gamma j, \mathrm{Tr}}} 
  \gtrsim \sum^p_{j=1}(\kappa_j^{\mathrm{Tr}}+1) = (\kappa^{\mathrm{Tr}} +1)p.
\end{equation}
The first inequality holds since $\tau_0^2 \geq 0$ and Assumption B1 holds. 

We now bound the other term under the square root in equation \eqref{eqn:bound2_bp}. Define new random variables \(M_j = \widehat{\gamma}_j^{\mathrm{Tr}} - {\gamma}_j^{\mathrm{Tr}}\), and let \(m_j^{(k)}\) denotes the \(k\)th moments of \(M_j\). We then have the following bound:
\begin{equation}
  \label{eqn:bound4_bp}
  \begin{split}
&\sqrt{\sum^p_{j=1}\frac{\mathbb{E}\left((\widehat{\gamma}_j^{\mathrm{Tr}}(U_j^{\mathrm{Ou}}+\alpha_j))^4\right)}{\sigma_{\gamma j, \mathrm{Ou}}^8}} \\  =&\sqrt{\sum^p_{j=1}\frac{\mathbb{E}\left((M_j+{\gamma}_j^{\mathrm{Tr}})^4((U_j^{\mathrm{Ou}}+\alpha_j))^4\right)}{\sigma_{\gamma j, \mathrm{Ou}}^8}}\\
=& \sqrt{\sum^p_{j=1}\frac{ \left( m_j^{(4)}+ 4(\gamma_j^{\mathrm{Tr}}) m_j^{(3)} + 6(\gamma_j^{\mathrm{Tr}})^2 m_j^{(2)} + (\gamma_j^{\mathrm{Tr}})^4\right)
}{\sigma_{\gamma j, \mathrm{Ou}}^4} \frac{\mathbb{E}(((U_j^{\mathrm{Ou}}+\alpha_j))^4)}{\sigma_{\gamma j, \mathrm{Ou}}^4}}\\
\lesssim& \sqrt{\sum_j^p\frac{ \left( m_j^{(4)}+ 4(\gamma_j^{\mathrm{Tr}}) m_j^{(3)} + 6(\gamma_j^{\mathrm{Tr}})^2 m_j^{(2)} + (\gamma_j^{\mathrm{Tr}})^4\right)
}{\sigma_{\gamma j, \mathrm{Tr}}^4}\frac{\mathbb{E}(((U_j^{\mathrm{Ou}}+\alpha_j))^4)}{\sigma_{Uj}^4}}\\
\lesssim&
\sqrt{\sum^p_{j=1}\left\{\frac{m^{(4)}_j}{(m_j^{(2)})^2}+\frac{\gamma_j^{\mathrm{Tr}}}{(m_j^{(2)})^{1/2}}\frac{m_j^{(3)}}{(m_j^{(2)})^{3/2}}+\frac{(\gamma_j^{\mathrm{Tr}})^2}{m_j^{(2)}}+\frac{(\gamma_j^{\mathrm{Tr}})^4}{(m_j^{(2)})^2}\right\}}\\
\lesssim&\sqrt{\sum^p_{j=1}\left\{1+\frac{\gamma_j^{\mathrm{Tr}}}{(m_j^{(2)})^{1/2}}+\frac{(\gamma_j^{\mathrm{Tr}})^2}{m_j^{(2)}}+\frac{(\gamma_j^{\mathrm{Tr}})^4}{(m_j^{(2)})^2}\right\}}\\
=&\sqrt{\sum_{j=1}^p\left(1 + \sqrt{\kappa_j^{\mathrm{Tr}}} + \kappa_j^{\mathrm{Tr}} +(\kappa_j^{\mathrm{Tr}})^2\right)}\\
\lesssim&
\sqrt{\sum_{j=1}^p\left(1  + \kappa_j ^{\mathrm{Tr}}+\kappa_j^{\mathrm{Tr}}(\max_j\kappa_j^{\mathrm{Tr}})\right)}=
\sqrt{p+p\kappa^{\mathrm{Tr}} + p\kappa^{\mathrm{Tr}}(\max_j\kappa_j^{\mathrm{Tr}})}.
\end{split}
\end{equation}
The first line follows from the decomposition using \( M_j \); the second line follows from a straightforward but tedious calculation. The third line uses Assumption B3 (bounded fourth moments), Assumption B1 (bounded variances), the fact that \( \alpha_j \) are i.i.d.\ normal (so the fourth moment is also bounded), and the independence between the two samples. The fourth line follows from an equivalent transformation up to a constant, noting that \( \sigma_{\gamma j,\text{Tr}}^2 = m_j^{(2)} \). The fifth line follows from Assumption B3 and an application of H\"older's inequality:
\[
m_j^{(3)} \leq \sqrt{m_j^{(2)} m_j^{(4)}} \lesssim (m_j^{(2)})^{3/2}.
\]
The sixth line follows from the definition of \( \kappa_j \); the seventh line uses the inequality \( 2\sqrt{a} \leq 1 + a \); and the final line follows from the definition of \( \kappa \).

Combining equations \eqref{eqn:bound2_bp}, \eqref{eqn:bound3_bp}, and \eqref{eqn:bound4_bp}, the Lindeberg condition is bounded by
\begin{equation}
  \label{eqn:boundlinderburg_final_bp}
  LSH \lesssim \frac{\sqrt{p+p\kappa^{\mathrm{Tr}} + p\kappa^{\mathrm{Tr}}(\max_j\kappa_j^{\mathrm{Tr}})}}{p (\kappa^{\mathrm{Tr}}+1)} \lesssim\sqrt{\frac{1}{p(\kappa^{\mathrm{Tr}})^2}+\frac{1}{p\kappa^{\mathrm{Tr}}}+\frac{\max_j\kappa_j^{\mathrm{Tr}}}{p\kappa^{\mathrm{Tr}}}} = o(1),
\end{equation}
provided that both $p(\kappa^{\mathrm{Tr}})^2 \to \infty$ and $p\kappa^{\mathrm{Tr}} = \sqrt{p}\sqrt{p(\kappa^{\mathrm{Tr}})^2} \to \infty$. So far, we have verified Lindeberg's condition, which implies 

\begin{equation}
\label{eqn:thm1final1_bp}
  \frac{1}{s_p}\sum^p_{j=1}\frac{\widehat{ \gamma}^{\mathrm{Tr}}_j(U_j^{\mathrm{Ou}}+\alpha_j)}{\sigma_{\gamma j, \mathrm{Ou}}^2} \xrightarrow{d.}\mathcal{N}(0,1).
\end{equation}

Using the pointwise convergence result: $\left(\sum^p_{j=1}\frac{\widehat{ \gamma}^{\mathrm{Tr}}_j \widehat{ \gamma}^{\mathrm{Ou}}_j}{\sigma_{\gamma j, \mathrm{Ou}}^2}\right)/\left(\sum^p_{j=1}\frac{ \gamma^{\mathrm{Tr}}_j \gamma^{\mathrm{Ou}}_j }{\sigma_{\gamma j, \mathrm{Ou}}^2}\right)\xrightarrow{p.} 1 $ and let $V_1^{-1} = \frac{\left(\sum^p_{j=1}\frac{ \gamma^{\mathrm{Tr}}_j \gamma^{\mathrm{Ou}}_j }{\sigma_{\gamma j, \mathrm{Ou}}^2}\right)}{s_p}$, we arrive

$$
V_1^{-1}(\widehat{\beta}^{MR-Wald} - \beta_0) = \frac{\frac{1}{s_p}\sum^p_{j=1}\frac{\widehat{ \gamma}^{\mathrm{Tr}}_j(U_j^{\mathrm{Ou}}+\alpha_j)}{\sigma_{\gamma j, \mathrm{Ou}}^2} }{\left(\sum^p_{j=1}\frac{\widehat{ \gamma}^{\mathrm{Tr}}_j \widehat{ \gamma}^{\mathrm{Ou}}_j}{\sigma_{\gamma j, \mathrm{Ou}}^2}\right)/\left(\sum^p_{j=1}\frac{ \gamma^{\mathrm{Tr}}_j \gamma^{\mathrm{Ou}}_j }{\sigma_{\gamma j, \mathrm{Ou}}^2}\right)}\xrightarrow{d.}\mathcal{N}(0,1),
$$
which proved the final result.

\subsection{Idiosyncratic Pleiotropy}
We provide the proof for Theorem \ref{thm:idio_sycratic} here.

\subsubsection{{Proof of Theorem \ref{thm:idio_sycratic} (a)}}
We prove statement (a) in the following two steps:

\begin{itemize}
  \item[] \textbf{Claim 1}: $\widetilde{b} \xrightarrow{p.} \beta_1^*$;
  \item[] \textbf{Claim 2}: $\widetilde{b\beta } \xrightarrow{p.} \beta_1^*\beta_0$.
\end{itemize}

If Claims 1-2 holds, then
\[
\frac{\widetilde{b\beta }}{\widetilde{b}} 
= \left( \frac{\widetilde{b\beta }}{\beta_1^*\beta_0} \middle/ \frac{\widetilde{b}}{\beta_1^*} \right) \beta_0 
\xrightarrow{p} \beta_0,
\]
by Slutsky's theorem.

We first establish Claim 1. To this end, we verify conditions (i)--(iii), which are sufficient for the consistency of extremum estimators as stated in \citet[Theorem 2.7]{newey1994large}:
\begin{itemize}
  \item[(i)] The population objective function \( g_1(\beta) \) is uniquely minimized at \( \beta_1^* \);
  \item[(ii)] The empirical objective function \( \widehat{g}_1(\beta) \) is convex;
  \item[(iii)] For each $\beta$, $\widehat{g}_1(\beta) \xrightarrow{p.}{g}_1(\beta)$.
\end{itemize}

By Theorem 2.7 of \citet{newey1994large}, these conditions imply $\widetilde{b}$ exist with probability 1 and that \( \widetilde{b} \xrightarrow{p.} \beta_1^* \).

We now verify each condition. Condition (i) is implied by assumption C1, and condition (ii) follows from the convexity of the quantile loss function. We now turn to condition (iii). Define
\[
e_j = w_j^{(1)}(\widehat{\gamma}_j^{\mathrm{Ou}} - \widehat{\gamma}_j^{\mathrm{Tr}} \beta).
\]
Then
\[
\operatorname{Var}(|e_j|) \leq \mathbb{E}(e_j^2) = \operatorname{Var}(e_j) + \mathbb{E}^2(e_j).
\]

We compute
\begin{equation}
  \label{eqn:varejandeej}
  \begin{split}
  \operatorname{Var}(e_j) &= (w_j^{(1)})^2(\sigma^2_{\gamma j, \mathrm{Ou}} + {\sigma^2_{\gamma j, \mathrm{Tr}} \beta^2} ) \lesssim (\sigma^2_{\gamma j, \mathrm{Ou}} + {\sigma^2_{\gamma j, \mathrm{Tr}} \beta^2} ),\\
  \mathbb{E}^2(e_j) &= (w_j^{(1)}( {\gamma}_j^{\mathrm{Ou}}  - {\gamma}_j^{\mathrm{Tr}} \beta))^2\lesssim (\gamma_j^{\mathrm{Ou}})^2+(\gamma_j^{\mathrm{Tr}})^2(\beta)^2.
  \end{split}
\end{equation}
The inequalities hold under assumption C2(the bound among variances of variables) and the inequality \( (a + b)^2 \leq 2(a^2 + b^2)\).

From \eqref{eqn:varejandeej}, it follows that
\begin{equation}
\label{eqn:varblound/p2}
\begin{split}
\frac{1}{p^2} \sum_{j=1}^p \operatorname{Var}(|e_j|)  
&\leq \frac{1}{p^2} \sum_{j=1}^p \mathbb{E}(e_j^2) \\
&= \frac{1}{p^2} \sum_{j=1}^p (\operatorname{Var}(e_j) + \mathbb{E}^2(e_j))\\
&\lesssim\frac{1}{p^2} \sum_{j=1}^p(\sigma^2_{\gamma j, \mathrm{Ou}} + {\sigma^2_{\gamma j, \mathrm{Tr}} \beta^2}+(\gamma_j^{\mathrm{Ou}})^2+(\gamma_j^{\mathrm{Tr}})^2(\beta)^2).
\end{split}
\end{equation}
For the first term of \eqref{eqn:varblound/p2}, given assumptions C2 and C3:
$$
\frac{1}{p^2}\sum_{j=1}^p(\sigma^2_{\gamma j, \mathrm{Ou}} + {\sigma^2_{\gamma j, \mathrm{Tr}} \beta^2}) \lesssim O(\frac{1}{np})=o(1).
$$
For the second term of \eqref{eqn:varblound/p2}, given assumption C3:
$$
\frac{1}{p^2} \sum_{j=1}^p\{(\gamma_j^{\mathrm{Ou}})^2+(\gamma_j^{\mathrm{Tr}})^2(\beta)^2\}  \lesssim\frac{||\gamma^{\mathrm{Tr}}||_2^2 + ||\gamma^{\mathrm{Ou}}||_2^2 }{p^2}\lesssim\frac{p||\gamma^{\mathrm{Tr}}||_\infty^2 + p||\gamma^{\mathrm{Ou}}||_\infty^2 }{p^2} = o(1).$$

Therefore, for any \( \epsilon > 0 \),
\begin{equation*}
\mathbb{P}(|\widehat{g}_1(\beta) - g_1(\beta)| > \epsilon) 
= \mathbb{P}\left(\left|\frac{1}{p} \sum_{j=1}^p \left(|e_j| - \mathbb{E}(|e_j|)\right)\right| > \epsilon \right) 
\leq \frac{1}{\epsilon^2} \cdot \frac{1}{p^2} \sum_{j=1}^p \operatorname{Var}(|e_j|) = o(1),
\end{equation*}
which establishes the desired pointwise convergence (iii.).

We can invoke \citet[Theorem 2.7]{newey1994large} to conclude that $\widetilde{b}$ exists with probability goes to $1$ and \( \widetilde{ b} \xrightarrow{p.} \beta_1^* \). 

We now establish the Claim 2. We consider a new function $ \widehat{g}_3(\beta) =  \frac{1}{p} \sum_{j=1}^p w_j^{(2)} |\widehat{\gamma}_j^{\text{Ou}}\beta_0 - \widehat{\gamma}_j^{\text{Tr}} \beta|$ and $g_3(\beta) = \mathbb{E}(\widehat{g}_3(\beta))$. Since $w_j^{(2)}/w_j^{(1)} = c$ by assumption B2, we have $g_3(\beta) / g_2(\beta/\beta_0) = c$. Also condition C1, $g_3$ has a unique minimizer $\beta_0\beta_1^*$.

We will verify the following conditions (iv), (v) and (vi) for $g_3$ and $\widehat{g}_{2}$: 

\begin{itemize}
  \item[(iv.)] The population objective function \( g_3(\beta) \) is uniquely minimized at \( \beta_1^*\beta_0\), as discussed above;
  \item[(v.)] The empirical objective function \( \widehat{g}_2(\beta) \) is convex, as garanteed by the definition \eqref{eqn:MR-Wald-idio}.
  \item[(vi.)] \( \widehat{g}_2(\beta) \) converges in probability to \( g_3(\beta) \) for each \( \beta \), which will discuss as follows.
\end{itemize}

We first bound the difference between $g_2$ and $g_3$:
\begin{equation}
\label{eqn:g_2g_3}
\begin{split}
|g_2(\beta) - g_3(\beta)|
&\leq \frac{1}{p}\sum_{j=1}^p  w_j^{(2)} \mathbb {E}|\widehat{\Gamma}_j^{\mathrm{Ou}}-\widehat{\gamma}_j^{\text{Ou}}\beta_0|\\
&\leq \frac{1}{p}\sum_{j=1}^p w_j^{(2)} \mathbb {E}|\widehat{\Gamma}_j^{\mathrm{Ou}}-{\Gamma}_j^{\mathrm{Ou}}+(\widehat{\gamma}_j^{\text{Ou}}-{\gamma}_j^{\text{Ou}})\beta_0 + ({\Gamma}_j^{\text{Ou}}-{\gamma}_j^{\text{Ou}}\beta_0)|\\
&\lesssim \frac{1}{p} \sum_{j=1}^p  \mathbb {E}|U_j + \alpha_j|\\
&= \frac{1}{p} \sum_{j=1}^p (\mathbb {E}|U_j| + \mathbb{E}|\alpha_j - \alpha_j^*| + |\alpha_j^*|)\\
&\lesssim \frac{1}{p} \sum_{j=1}^p (\sigma_{uj} + \tau_0)  + \frac{||\alpha^*||_1}{p}\\
&\lesssim O(\frac{1}{\sqrt{n}}) + o(1) \\
&=o(1)
\end{split}
\end{equation}

where $w_j^{(2)}\lesssim 1$ by assumption C2, and the last second inequality follows the fact       $\mathbb{E}|U_j|\leq \sqrt{\mathbb{E}|U_j|^2} = \sigma_{uj}$ and $\mathbb{E}|\alpha_j-\alpha_j^*|\leq \sqrt{\mathbb{E}|\alpha_j-\alpha_j^*|^2} = \tau_{0}$, and the last inequality follows by assumption C2.

We next bound the difference between $\widehat{g}_2$ and $g_2$. Define
\[
t_j = w_j^{(2)}(\widehat{\Gamma}_j^{\mathrm{Ou}} - \widehat{\gamma}_j^{\mathrm{Tr}} \beta).
\]
We compute
\begin{equation}
  \label{eqn:varejande_tj}
  \begin{split}
  \operatorname{Var}(t_j) &= (w_j^{(2)})^2(\sigma^2_{\Gamma j, \mathrm{Ou}} + {\sigma^2_{\gamma j, \mathrm{Tr}} \beta^2} + {\tau_0^2}) \lesssim \sigma^2_{\Gamma j, \mathrm{Ou}} + {\sigma^2_{\gamma j, \mathrm{Tr}} \beta^2},\\
  \mathbb{E}^2(t_j) &= (w_j^{(2)}( {\gamma}_j^{\mathrm{Ou}} \beta_0 - {\gamma}_j^{\mathrm{Tr}} \beta + \alpha_j^*))^2\lesssim (\gamma_j^{\mathrm{Ou}})^2(\beta_0)^2+(\gamma_j^{\mathrm{Tr}})^2(\beta)^2+(\alpha_j^*)^2.
  \end{split}
\end{equation}
The inequalities hold under assumption C2, the bound \( \tau_0^2 \lesssim \max_j \sigma^2_{\Gamma j, \mathrm{Ou}} \) and the inequality \( (a + b+c)^2 \leq 3(a^2 + b^2+c^2)\).

From \eqref{eqn:varejande_tj}, it follows that
\begin{equation}
\label{eqn:varblound/p2_tj}
\begin{split}
\frac{1}{p^2} \sum_{j=1}^p \operatorname{Var}(|t_j|)  
&\leq \frac{1}{p^2} \sum_{j=1}^p \mathbb{E}(t_j^2) \\
&= \frac{1}{p^2} \sum_{j=1}^p (\operatorname{Var}(t_j) + \mathbb{E}^2(t_j))\\
&\lesssim\frac{1}{p^2} \sum_{j=1}^p(\sigma^2_{\Gamma j, \mathrm{Ou}} + {\sigma^2_{\gamma j, \mathrm{Tr}} \beta^2}+(\gamma_j^{\mathrm{Ou}})^2(\beta_0)^2+(\gamma_j^{\mathrm{Tr}})^2(\beta)^2+(\alpha_j^*)^2).
\end{split}
\end{equation}
For the first term of \eqref{eqn:varblound/p2_tj}, given assumption C2, we have:
$$
\frac{1}{p^2}\sum_{j=1}^p(\sigma^2_{\Gamma j, \mathrm{Ou}} + {\sigma^2_{\gamma j, \mathrm{Tr}} \beta^2}) \lesssim O(\frac{1}{np})=o(1).
$$
For the second term of \eqref{eqn:varblound/p2_tj}, given assumption C3, we have:
$$
\frac{1}{p^2} \sum_{j=1}^p\{(\gamma_j^{\mathrm{Ou}})^2(\beta_0)^2+(\gamma_j^{\mathrm{Tr}})^2(\beta)^2\}  =O(\frac{||\gamma^{\mathrm{Tr}}||_2^2 + ||\gamma^{\mathrm{Ou}}||_2^2 }{p^2})\lesssim\frac{p||\gamma^{\mathrm{Tr}}||_\infty^2 + p||\gamma^{\mathrm{Ou}}||_\infty^2 }{p^2}=o(1) $$
For the final term of \eqref{eqn:varblound/p2_tj}, given assumption C3, we have: 
$$
\frac{1}{p^2} \sum_{j=1}^p(\alpha_j^*)^2 = \left(\frac{||\alpha^*||_2}{p}\right)^2 = \left(\frac{||\alpha^*||_1}{p}\right)^2=o(1) 
$$

Therefore, for any \( \epsilon > 0 \),
\begin{equation}
\label{eqn:123}
\mathbb{P}(|\widehat{g}_2(\beta) - g_2(\beta)| > \epsilon/2) 
= \mathbb{P}\left(\left|\frac{1}{p} \sum_{j=1}^p \left(|t_j| - \mathbb{E}(|t_j|)\right)\right| > \epsilon/2 \right) 
\leq \frac{1}{4\epsilon^2} \cdot \frac{1}{p^2} \sum_{j=1}^p \operatorname{Var}(|e_j|) = o(1),
\end{equation}
We also have 
$$
\mathbb{P}(|\widehat{g}_2(\beta) - g_3(\beta)| > \epsilon) 
\leq \mathbb{P}(|\widehat{g}_2(\beta) - g_2(\beta)| > \epsilon/2) + \mathbb{P}(|{g}_2(\beta) - g_3(\beta)| > \epsilon).
$$
The above inequality and equations \eqref{eqn:g_2g_3} and \eqref{eqn:123} to establish the desired pointwise convergence (vi.).

We can invoke \citet[Theorem 2.7]{newey1994large} with conditions (iv)-(vi) to conclude that $\widetilde{b\beta }$ exists with probability goes to $1$ and \( \widetilde{b\beta } \xrightarrow{p} \beta_2^*\beta_0\). We have shown the claim 2. Thus the theorem \eqref{thm:idio_sycratic} (a) holds by combining the Slutsky's theorem and claims 1-2.

\subsubsection{{Proof of Theorem \ref{thm:idio_sycratic} (b)}
}
From the previous section, we know that $\widetilde{b\beta } \xrightarrow{p.} \beta_2^* \beta_0$ and $\widetilde{b} \xrightarrow{p.} \beta_2$. We now outline the steps to derive the asymptotic variance of the estimator $\widetilde{b\beta }/\widetilde{b} - \beta_0$:

\begin{itemize}
  \item[Step 1.] Show the Bahadur representation  of $\widetilde{b} - \beta_1^*$.
  \item [Step 2.] Show the Bahadur representation of $\widetilde{b\beta } - \beta_1^*\beta_0$.
  \item [Step 3.] Show the asymptotic normality of the ratio estimator \ref{eqn:MR-Wald-idio}.
\end{itemize}
The first and second step involve applying the Bahadur representation for misspecified quantile regression using independent but non-identically distributed samples, with samples $\{\widehat{\Gamma}_j^{\text{Ou}}, \widehat{\gamma}_j^{\text{Tr}}\}_{j=1}^p$ (and $\{\widehat{\gamma}_j^{\text{Ou}}, \widehat{\gamma}_j^{\text{Tr}}\}_{j=1}^p$). This step therefore requires careful investigation.

We first consider the Bahadur representation for ${\widetilde{b}} - \beta_1^*$,  where 
$$
\widetilde{b} = \argmin_{\beta\in \mathbb{R}}   \frac{1}{p} \sum_{j=1}^p w_j^{(1)} |\widehat{\gamma}_j^{\text{Ou}} - \widehat{\gamma}_j^{\text{Tr}} \beta|.
$$

Define 
\begin{align*}
&\Psi_p (\beta) = \frac{1}{p} \sum_{j=1}^p w_j^{(1)} \widehat{\gamma}_j^{\mathrm{Tr}} (\frac{1}{2} - \mathbf{1}\{\widehat{\gamma}_j^{\mathrm{Ou}} \leq \widehat{\gamma}_j^{\mathrm{Tr}} \beta \}), \;\; \Psi (\beta)  = \mathbb{E}(\Psi_p(\beta)).
\end{align*}
We are going to show that 
\begin{equation}
  \label{eqn:bahadur1}
  \begin{split}
  &\sqrt{p}(\widetilde{b} - \beta_1^*) \;\;= -\sqrt{p}J_2^{-1}\Psi_p(\beta_1^*)  +o_p(1),\text{ where }J_1 =\frac{1}{p}\sum^p_{j=1}w_j^{(1)}(f_j^{(1)}(\widehat{\gamma}_j^{\mathrm{Tr}}\beta_1^*)(\widehat{\gamma}_j^{\mathrm{Tr}})^2).
  \end{split}
\end{equation}


Define $u_j = \widehat{\gamma}_j^{\text{Ou}} - \widehat{\gamma}_j^{\text{Tr}} \beta_1^*$. Due to the Knight's equality \citep{knight1998limiting}, we have 
\begin{equation*}
\begin{split}
\widehat{g}_1(\beta_1^* + \delta) - \widehat{g}_1(\beta_1^*) &= \frac{1}{p} \sum_{j=1}^p w_j^{(1)} |\widehat{\gamma}_j^{\text{Ou}} - \widehat{\gamma}_j^{\text{Tr}} (\beta_1^* + \delta)| - \frac{1}{p} \sum_{j=1}^p w_j^{(1)} |\widehat{\gamma}_j^{\text{Ou}} - \widehat{\gamma}_j^{\text{Tr}} \beta_1^*| \\
&= \frac{1}{p} \sum_{j=1}^p\delta w_j^{(1)} \widehat{\gamma}_j^{\text{Tr}}(\frac{1}{2} - I(u_j\leq 0))  + \frac{1}{p}\sum_j^p\int^{\widehat{\gamma}_j^{\mathrm{Tr}}\delta}_0 ( I(u_j<s) - I(u_j<0)ds\\
&= Z^{(1)}_{p}(\delta) + Z^{(2)}_{p}(\delta), 
\end{split}
\end{equation*}
with 
$$
Z^{(1)}_{p}(\delta) = \frac{1}{p} \sum_{j=1}^p\delta w_j^{(1)} \widehat{\gamma}_j^{\text{Tr}}(\frac{1}{2} - I(u_j\leq 0)),  \;\; Z^{(2)}_{p}(\delta) = \frac{1}{p}\sum_j^p\int^{\widehat{\gamma}_j^{\mathrm{Tr}}\delta}_0 ( I(u_j<s) - I(u_j<0)ds.
$$

Decompose the $Z^{(2)}_{p}(\delta):= \frac{1}{p}\sum_j^{p}Z^{(2)}_{p,j}(\delta)$ where $Z^{(2)}_{p,j}(\delta) = \int^{\widehat{\gamma}_j^{\mathrm{Tr}}\delta}_0 ( I(u_j<s) - I(u_j<0)ds$. We have 
\begin{equation}
  \label{eqn:Z^{(2)}_{p,j}}
  \begin{split}
   \mathbb{E}(Z^{(2)}_{p,j}\mid \widehat{\gamma}_j^{\mathrm{Tr}})
&=\int^{\widehat{\gamma}_j^{\mathrm{Tr}}\delta}_0 F_j^{(1)}(\widehat{\gamma}_j^{\mathrm{Tr}}\beta_1^*+s) - F_j^{(1)}(\widehat{\gamma}_j^{\mathrm{Tr}}\beta_1^*)ds\\
&=\frac{1}{2}f_j^{(1)}(\widehat{\gamma}_j^{\mathrm{Tr}}\beta_1^*)(\widehat{\gamma}_j^{\mathrm{Tr}})^2\delta^2 + v_{0,j}(\delta|\widehat{\gamma}_j^{\mathrm{Tr}}),
\end{split}
\end{equation}
where $v_{0,j}(\delta|\widehat{\gamma}_j^{\mathrm{Tr}}) =\int^{\widehat{\gamma}_j^{\mathrm{Tr}}\delta}_0 \{F_j^{(1)}(\widehat{\gamma}_j^{\mathrm{Tr}}\beta_1^*+s) - F_j^{(1)}(\widehat{\gamma}_j^{\mathrm{Tr}}\beta_1^*)-f_j^{(1)}(\widehat{\gamma}_j^{\mathrm{Tr}}\beta_1^*)s\}ds$, and $F_j^{(1)}$ and $f_j^{(1)}$ are the CDF and PDF of $\widehat{\gamma}_j^{\mathrm{Ou}}$.

We define the quantity $v_{j}(\delta\mid\widehat{\gamma}_j^{\mathrm{Tr}})$ as follows:
\begin{equation}
  \label{eqn:varZ^{(2)}_{p,j}}
  \begin{split}
   v_{j}(\delta\mid\widehat{\gamma}_j^{\mathrm{Tr}}):&=\mathbb{V}ar(Z^{(2)}_{p,j}\mid \widehat{\gamma}_j^{\mathrm{Tr}})\\
&\leq \mathbb{E}(\{Z_{p,j}^{(2)}\}^2\mid\widehat{\gamma}_j^{\mathrm{Tr}})\\
&\leq \widehat{\gamma}_j^{\mathrm{Tr}}\delta\mathbb{E}(\{Z_{p,j}^{(2)}|\widehat{\gamma}_j^{\mathrm{Tr}}\}.
\end{split}
\end{equation}
The second inequality in equation \eqref{eqn:varZ^{(2)}_{p,j}} follows by the fact that $Z_{p,j}^{(2)}\leq \int^{\widehat{\gamma}_j^{\mathrm{Tr}}\delta}_0 1= \widehat{\gamma}_j^{\mathrm{Tr}}\delta$.

We verify two facts to establish the Bahadur representation. For fixed $t$, we have
\begin{equation}
\label{eqn:v0o1}
\begin{split}
  &\sum_{j=1}^{p} v_{0,j} (t/\sqrt{p}\mid\widehat{\gamma}_j^{\mathrm{Tr}}) \\=& \sum^p_{j=1}\int^{\widehat{\gamma}_j^{\mathrm{Tr}}t/\sqrt{p}}_0  \{F_j^{(1)}(\widehat{\gamma}_j^{\mathrm{Tr}}\beta_1^*+s) - F_j^{(1)}(\widehat{\gamma}_j^{\mathrm{Tr}}\beta_1^*)-f(\widehat{\gamma}_j^{\mathrm{Tr}}\beta_1^*)s\}ds
  \\
  =&\frac{1}{p}\sum^p_{j=1}\int^{\widehat{\gamma}_j^{\mathrm{Tr}}t}_0  \sqrt{p}\{F_j^{(1)}(\widehat{\gamma}_j^{\mathrm{Tr}}\beta_1^*+\frac{s}{\sqrt{p}}) - F_j^{(1)}(\widehat{\gamma}_j^{\mathrm{Tr}}\beta_1^*)-f(\widehat{\gamma}_j^{\mathrm{Tr}}\beta_1^*)\frac{s}{\sqrt{p}}\}ds\\
  =&o(1)
\end{split}
\end{equation}
where the last equality is given by Taylor extension and the property of $f_j^{(1)}$, as provided by assumption C4 (the uniform continuity of the density functions).

We also have the following bound:

\begin{equation}
\begin{split}
\label{eqn:boundvj1}
  \sum_{j=1}^{p}v_j (t/\sqrt{p}\mid\widehat{\gamma}_j^{\mathrm{Tr}})
  &\leq (\max_j\widehat{\gamma}_j)\frac{t}{\sqrt{p}}\sum_{j=1}^p \mathbb{E}(\{Z_{p,j}^{(2)}|\widehat{\gamma}_j^{\mathrm{Tr}}\}
  \\
  &\leq (\max_j\widehat{\gamma}_j)\frac{t}{\sqrt{p}}\{\sum_{j=1}^p\frac{1}{2p}f_j(\widehat{\gamma}_j^{\mathrm{Tr}}\beta_1^*)(\widehat{\gamma}_j^{\mathrm{Tr}})^2 + \sum^p_{j=1}v_{0,j}(\delta|\widehat{\gamma}_j^{\mathrm{Tr}})\}\\
  &=o_p(1),
\end{split}
\end{equation}
where the last equality follows by \eqref{eqn:v0o1} and the fact 
\begin{equation*}
\begin{split}
\frac{\max_j\widehat{\gamma}_j^{\mathrm{Tr}}}{\sqrt{p}} 
&\leq \frac{||\gamma^{\mathrm{Tr}}||_\infty}{\sqrt{p}}+\frac{||\widehat{\gamma}^{\mathrm{Tr}} - \gamma^{\mathrm{Tr}}||_\infty}{\sqrt{p}}\\
&\leq o(1)+\frac{||\widehat{\gamma}^{\mathrm{Tr}} - \gamma^{\mathrm{Tr}}||_2}{\sqrt{p}}\\
&= o(1) + O_p(\sqrt{1/n})=o_p(1),  
\end{split}  
\end{equation*}
by assumption C5, C2  and $\mathbb{P}(\frac{||\widehat{\gamma}^{\mathrm{Tr}} - \gamma^{\mathrm{Tr}}||_2}{\sqrt{p}}>\frac{M}{\sqrt{n}})\leq \frac{n}{M^2}\mathbb{E}(||\widehat{\gamma}^{\mathrm{Tr}} - \gamma^{\mathrm{Tr}}||_2^2/p) \lesssim \frac{1}{M^2}$, as we can select large M to make the probability arbitrarily small.

With equations \eqref{eqn:v0o1}, and \eqref{eqn:boundvj1}, by using Theorem 2.3 and Equation 2.9 of \cite{hjort2011asymptotics},
we have 
$$\sqrt{p}(\widetilde{b} - \beta_1^*) = -\sqrt{p}J_1^{-1}\Phi_p(\beta_1^*)  +o_p(1),\text{ where }J_1 = \frac{1}{p}\sum^p_{j=1}w_j^{(1)}(f_j^{(1)}(\widehat{\gamma}_j^{\mathrm{Tr}}\beta_1^*)(\widehat{\gamma}_j^{\mathrm{Tr}})^2).$$
So far, we have shown the Bahadur representation \eqref{eqn:bahadur1}. \bigskip

We now proceed with Step 2: the Bahadur representation for ${\widetilde{b\beta }} - \beta_1^*\beta_0$,  where 
$$
\widetilde{b\beta} = \argmin_{\beta\in \mathbb{R}}   \frac{1}{p} \sum_{j=1}^p w_j^{(2)} |\widehat{\Gamma}_j^{\text{Ou}} - \widehat{\gamma}_j^{\text{Tr}} \beta |,\quad 
$$
Define 
\begin{align*}
&\Phi_p (\beta) = \frac{1}{p} \sum_{j=1}^p w_j^{(1)} \widehat{\gamma}_j^{\mathrm{Tr}} (\frac{1}{2} - \mathbf{1}\{\widehat{\Gamma}_j^{\mathrm{Ou}} \leq \widehat{\gamma}_j^{\mathrm{Tr}} \beta \}), \;\; \Phi (\beta)  = \mathbb{E}(\Phi_p(\beta)).
\end{align*}
We are going to show that 
\begin{equation}
  \label{eqn:bahadur2}
  \begin{split}
  &\sqrt{p}(\widetilde{b\beta} - \beta_1^*\beta_0) = -\sqrt{p}J_2^{-1}\Phi_p(\beta_1^*\beta_0)  +o_p(1),\text{ where }J_2 = \frac{1}{p}\sum^p_{j=1}w_j^{(2)}(f_j^{(2)}(\widehat{\gamma}_j^{\mathrm{Tr}}\beta_1^*\beta_0)(\widehat{\gamma}_j^{\mathrm{Tr}})^2). \\
  \end{split}
\end{equation}


Define $u_j = \widehat{\Gamma}_j^{\text{Ou}} - \widehat{\gamma}_j^{\text{Tr}} \beta_1^*\beta_0$. Due to the Knight's equality \citep{knight1998limiting}, we have 
\begin{equation*}
\begin{split}
\widehat{g}_2(\beta_1^*\beta_0 + \delta) - \widehat{g}_2(\beta_1^*\beta_0) &= \frac{1}{p} \sum_{j=1}^p w_j^{(2)} |\widehat{\Gamma}_j^{\text{Ou}} - \widehat{\gamma}_j^{\text{Tr}} (\beta_1^*\beta_0 + \delta)| - \frac{1}{p} \sum_{j=1}^p w_j^{(2)} |\widehat{\Gamma}_j^{\text{Ou}} - \widehat{\gamma}_j^{\text{Tr}} \beta_1^*\beta_0 | \\
&= \frac{1}{p} \sum_{j=1}^p\delta w_j^{(1)} \widehat{\gamma}_j^{\text{Tr}}(\frac{1}{2} - I(u_j\leq 0))  + \frac{1}{p}\sum_j^p\int^{\widehat{\gamma}_j^{\mathrm{Tr}}\delta}_0 ( I(u_j<s) - I(u_j<0)ds\\
&= Z^{(1)}_{p}(\delta) + Z^{(2)}_{p}(\delta), 
\end{split}
\end{equation*}
with 
$$
Z^{(1)}_{p}(\delta) = \frac{1}{p} \sum_{j=1}^p\delta w_j^{(1)} \widehat{\gamma}_j^{\text{Tr}}(\frac{1}{2} - I(u_j\leq 0)),  \;\; Z^{(2)}_{p}(\delta) = \frac{1}{p}\sum_j^p\int^{\widehat{\gamma}_j^{\mathrm{Tr}}\delta}_0(I(u_j<s) - I(u_j<0)ds.
$$
Decompose the $Z^{(2)}_{p}(\delta):= \frac{1}{p}\sum_j^{p}Z^{(2)}_{p,j}(\delta)$ where $Z^{(2)}_{p,j}(\delta) = \int^{\widehat{\gamma}_j^{\mathrm{Tr}}\delta}_0 ( I(u_j<s) - I(u_j<0)ds$. We have 
\begin{equation}
  \label{eqn:Z^{(2)}_{p,j}_2}
  \begin{split}
   \mathbb{E}(Z^{(2)}_{p,j}\mid \widehat{\gamma}_j^{\mathrm{Tr}})
&=\int^{\widehat{\gamma}_j^{\mathrm{Tr}}\delta}_0 F_j^{(2)}(\widehat{\gamma}_j^{\mathrm{Tr}}\beta_1^*\beta_0+s) - F_j^{(2)}(\widehat{\gamma}_j^{\mathrm{Tr}}\beta_1^*\beta_0)ds\\
&=\frac{1}{2}f_j^{(2)}(\widehat{\gamma}_j^{\mathrm{Tr}}\beta_1^*\beta_0)(\widehat{\gamma}_j^{\mathrm{Tr}})^2\delta^2 + v_{0,j}(\delta\mid\widehat{\gamma}_j^{\mathrm{Tr}}),
\end{split}
\end{equation}
where $v_{0,j}(\delta\mid\widehat{\gamma}_j^{\mathrm{Tr}}) =\int^{\widehat{\gamma}_j^{\mathrm{Tr}}\delta}_0 \{F_j^{(2)}(\widehat{\gamma}_j^{\mathrm{Tr}}\beta_1^*\beta_0+s) - F_j^{(2)}(\widehat{\gamma}_j^{\mathrm{Tr}}\beta_1^*\beta_0)-f_j^{(2)}(\widehat{\gamma}_j^{\mathrm{Tr}}\beta_1^*\beta_0)s\}ds$, and $F_j^{(2)}$ and $f_j^{(2)}$ are the CDF and PDF of $\widehat{\Gamma}_j^{\mathrm{Tr}}$.

We define the quantity $v_{j}(\delta\mid\widehat{\gamma}_j^{\mathrm{Tr}})$ as follows, which will be important to construct the Bahadur representation: 
\begin{equation}
  \label{eqn:varZ^{(2)}_{p,j}_2}
  \begin{split}
   v_{j}(\delta\mid\widehat{\gamma}_j^{\mathrm{Tr}}):&=\mathbb{V}ar(Z^{(2)}_{p,j}\mid \widehat{\gamma}_j^{\mathrm{Tr}})\\
&\leq \mathbb{E}(\{Z_{p,j}^{(2)}\}^2\mid\widehat{\gamma}_j^{\mathrm{Tr}})\\
&\leq \widehat{\gamma}_j^{\mathrm{Tr}}\delta\;\mathbb{E}(\{Z_{p,j}^{(2)}\mid\widehat{\gamma}_j^{\mathrm{Tr}}\}.
\end{split}
\end{equation}
The second inequality in equation \eqref{eqn:varZ^{(2)}_{p,j}_2} follows by the fact that $Z_{p,j}^{(2)}\leq \int^{\widehat{\gamma}_j^{\mathrm{Tr}}\delta}_0 1\leq \widehat{\gamma}_j^{\mathrm{Tr}}\delta$.

We verify two facts to establish the Bahadur representation. For fixed $t$, we have
\begin{equation}
\label{eqn:v0o1_2}
\begin{split}
  &\sum_{j=1}^{p} v_{0,j} (t/\sqrt{p}\mid\widehat{\gamma}_j^{\mathrm{Tr}}) \\=& \sum^p_{j=1}\int^{\widehat{\gamma}_j^{\mathrm{Tr}}t/\sqrt{p}}_0  \{F_j^{(2)}(\widehat{\gamma}_j^{\mathrm{Tr}}\beta_1^*+s) - F_j^{(2)}(\widehat{\gamma}_j^{\mathrm{Tr}}\beta_1^*)-f^{(2)}(\widehat{\gamma}_j^{\mathrm{Tr}}\beta_1^*)s\}ds
  \\
  =&\frac{1}{p}\sum^p_{j=1}\int^{\widehat{\gamma}_j^{\mathrm{Tr}}t}_0  \sqrt{p}\{F_j^{(2)}(\widehat{\gamma}_j^{\mathrm{Tr}}\beta_1^*+\frac{s}{\sqrt{p}}) - F_j^{(2)}(\widehat{\gamma}_j^{\mathrm{Tr}}\beta_1^*)-f^{(2)}(\widehat{\gamma}_j^{\mathrm{Tr}}\beta_1^*)\frac{s}{\sqrt{p}}\}ds\\
  =&o(1)
\end{split}
\end{equation}
where the last equality is given by Taylor extension and the property of $f_j^{(2)}$, as provided by assumption C4 (the uniform continuity of the density function).
\begin{equation}
\begin{split}
\label{eqn:v0o1v2}
  \sum_{j=1}^{p}v_j (t/\sqrt{p}\mid\widehat{\gamma}_j^{\mathrm{Tr}})
  &\leq (\max_j\widehat{\gamma}_j)\frac{t}{\sqrt{p}}\sum_{j=1}^p \mathbb{E}(\{Z_{p,j}^{(2)}\mid\widehat{\gamma}_j^{\mathrm{Tr}}\}
  \\
  &\leq (\max_j\widehat{\gamma}_j)\frac{t}{\sqrt{p}}\{\sum_{j=1}^p\frac{1}{2}f_j(\widehat{\gamma}_j^{\mathrm{Tr}}\beta_1^*)(\widehat{\gamma}_j^{\mathrm{Tr}})^2\frac{1}{p} + \sum^p_{j=1}v_{0,j}(\delta|\widehat{\gamma}_j^{\mathrm{Tr}})\}\\
  &=o_p(1).
\end{split}
\end{equation}
where the last equality follows by \eqref{eqn:v0o1_2} and the fact 
$
\frac{\max_j\widehat{\gamma}_j^{\mathrm{Tr}}}{\sqrt{p}} =o_p(1), $ as we discussed in step 1.

With equations \eqref{eqn:v0o1_2} and \eqref{eqn:v0o1v2}
By using Theorem 2.3 and Equation 2.9 of \cite{hjort2011asymptotics},
we have 
$$\sqrt{p}(\widetilde{\beta} - \beta_1^*) = -\sqrt{p}J_1^{-1}\Phi_p(\beta_1^*)  +o_p(1),\text{ where }J_1 = \frac{1}{p}\sum^p_{j=1}w_j^{(1)}(f_j^{(1)}(\widehat{\gamma}_j^{\mathrm{Tr}}\beta_1^*)(\widehat{\gamma}_j^{\mathrm{Tr}})^2).$$
So far, we have shown the Bahadur representation \eqref{eqn:bahadur2}.

\bigskip

We now proceed with Step 3. From Steps 1 and 2, we have the following decomposition
\begin{equation}
  \label{eqn:numerator}
  \begin{split}
  \sqrt{p}(\widetilde{\beta}^{MR-Wald}-\beta_0) =& \sqrt{p}(\widetilde{b\beta} - \widetilde{b}\beta_0)/\widetilde{b} \\
  =&\sqrt{p}(J_1^{-1}\beta_0\Phi_p(\beta_1^*)  -J_2^{-1}\Psi_p(\beta_1^*\beta_0))/\widetilde{b} +o_p(1)\\
  = -\frac{1}{\sqrt{p}}\sum_{j=1}^p &\left\{J_1^{-1}\beta_0w_j^{(1)}\widehat{\gamma}_j^{\mathrm{Tr}}(\frac{1}{2}-\mathbb{I}(\widehat{\gamma}_j^{\mathrm{Ou}}\leq\widehat{\gamma}_j^{\mathrm{Tr}}\beta_1^*))\right.\\   -&\left.J_2^{-1}w_j^{(2)}\widehat{\gamma}_j^{\mathrm{Tr}}(\frac{1}{2}-\mathbb{I}(\widehat{\Gamma}_j^{\mathrm{Ou}}\leq\widehat{\gamma}_j^{\mathrm{Tr}}\beta_1^*\beta_0))\right\}/\widetilde{b} + o_p(1)
  \end{split}
\end{equation}
Due to Lindeberg-Feller CLT and Slutsky's theorem and assumption C5, we have 
$\sqrt{p}(\widetilde{\beta}^{MR-Wald}-\beta_0)\xrightarrow{d.}N(0,V_3)$, where 
$$
V_3 = (\beta_0^2\bar{J}_1^{-2} \sum^p_{j=1}\zeta^{(1)}_j+\bar{J}_2^{-2} \sum^p_{j=1}\zeta^{(2)}_j-2\beta_0\bar{J}_1^{-1}\bar{J}_2^{-1}\sum^p_{j=1}\zeta^{(12)}_j)/(\beta_1^*)^2,
$$
with
$\zeta^{(1)}_j = \mathbb{E}[\{w_j^{(1)}\widehat{\gamma}_j^{\mathrm{Tr}}(\frac{1}{2} - \mathbb{I}(\widehat{\gamma}_j^{\mathrm{Ou}}\leq \widehat{\gamma}_j^{\mathrm{Tr}}\beta_1^*))\}^2]$,
$\zeta^{(2)}_j = \mathbb{E}[\{w_j^{(2)}\widehat{\gamma}_j^{\mathrm{Tr}}(\frac{1}{2} - \mathbb{I}(\widehat{\Gamma}_j^{\mathrm{Ou}}\leq \widehat{\gamma}_j^{\mathrm{Tr}}\beta_1^*\beta_0))\}^2]$, and   $\zeta^{(12)}_j = \mathbb{E}[\{w_j^{(1)}\widehat{\gamma}_j^{\mathrm{Tr}}(\frac{1}{2} - \mathbb{I}(\widehat{\gamma}_j^{\mathrm{Ou}}\leq \widehat{\gamma}_j^{\mathrm{Tr}}\beta_1^*))\}\{w_j^{(2)}\widehat{\gamma}_j^{\mathrm{Tr}}(\frac{1}{2} - \mathbb{I}(\widehat{\Gamma}_j^{\mathrm{Ou}}\leq \widehat{\gamma}_j^{\mathrm{Tr}}\beta_1^*\beta_2))\}]$.

\subsection{Directional Pleiotropy}
\label{sec:app_directional_pleiotropy}
We now extend the MR-Wald estimator in equation~\eqref{eqn:mr_walds} to accommodate directional pleiotropy under Assumption~P3, where the pleiotropic effects $\alpha_j$ are independently and identically distributed as a normal distribution. In the following, we introduce the extended estimator, establish its theoretical properties, and examine its numerical performance in turn.
\subsubsection{The estimator}
In the construction of the MR-Wald estimator in equation~\eqref{eqn:mr_walds}, we use the IVW estimator to estimate the product \( b\beta  \) based on the regression of \( \widehat{\Gamma}_j^{\text{Ou}} \) on \( \widehat{\gamma}_j^{\text{Tr}} \), denoted as \( \widehat{b\beta } \). To account for directional pleiotropy, we propose using the MR-Egger estimator to estimate \( b\beta  \), which adjusts for the directional parameter \( \mu \). The quantity \( \widetilde{b} \) is also estimated using the MR-Egger method by regressing \( \widehat{\gamma}_j^{\text{Ou}} \) on \( \widehat{\gamma}_j^{\text{Tr}} \). Specifically, we define:
\begin{equation}
\begin{split}
\label{eqn:mr_wald_d}
  &\widehat{\beta}^{\text{MR-Wald-D}} = {\widetilde{b\beta }}/{\widetilde{b}},\\
  &(\widetilde{b\beta }, \widetilde{\eta}) = \arg\min_{\beta, \eta} \sum_{j=1}^p \frac{(\widehat{\Gamma}_j^{\text{Ou}} - \beta \widehat{\gamma}_j^{\text{Tr}} - \eta)^2}{\sigma^2_{y j, \text{Ou}}},\\
  &(\widetilde{b},\widetilde{\eta}) = \arg\min_{b, \eta}\sum_{j=1}^p \frac{(\widehat{\gamma}_j^{\text{Ou}} - b\widehat{\gamma}_j^{\text{Tr}}-\eta)^2}{\sigma^2_{x j, \text{Ou}}}.
\end{split}
\end{equation}
Here, the superscript \textit{-D} denotes an extension of the MR-Wald estimator that accounts for directional pleiotropy. By following the same procedure as in the proof of the main theorem, one can show that $\widehat{\beta}^{\text{MR-Wald-D}}$ is a consistent estimator of $\beta_0$ and is asymptotically normal. The technical details are omitted for brevity.

\section{Additional numerical results}
\label{sec:supp_numerical}
We next assess the performance of the proposed estimator \eqref{eqn:mr_wald_d} in the presence of directional pleiotropy. The data-generating mechanism is analogous to that described in the manuscript and is outlined below.

We consider a simulation setting with \( p = 200 \) SNPs and a sample size of \( n = 100{,}000 \), The simulation involves two independent samples: one for the treatment GWAS and the other for the outcome GWAS.

\paragraph{Treatment GWAS.} 
For each individual \( i \in \{1, \dots, n\} \) and SNP \( j \in \{1, \dots, p\} \), 
the genotype is generated as \( Z_{i,j}^{(2)} \sim \text{Binomial}(2, 0.3) \). 
The treatment variable is given by
\(
D^{(2)}_i = \sum_{j=1}^{p} \gamma_j^{\mathrm{Tr}} Z_{i,j}^{(2)} + U_i + \epsilon_{d,i},
\)
where \( \gamma_j^{\mathrm{Tr}} \sim \text{Unif}(0.05, 0.10) \), and \( U_i, \epsilon_{d,i} \overset{\text{iid}}{\sim} \mathcal{N}(0, 1) \). 
The marginal SNP-treatment associations \( \widehat{\gamma}_j^{\mathrm{Tr}} \) are estimated by regressing \( D^{(2)} \) on \( Z_j^{(2)} \).

\paragraph{Outcome GWAS.} 
Genotypes are independently generated as \( Z_{i,j}^{(1)} \sim \text{Binomial}(2, 0.3) \). 
The treatment variable is defined by
\(
D^{(1)}_i = \sum_{j=1}^{p} \gamma_j^{\mathrm{Ou}} Z_{i,j}^{(1)} + U_i + \epsilon_{d,i},
\)
where \( \gamma_j^{\mathrm{Ou}} = g(\gamma_j^{\mathrm{Tr}}) \) for a given function \( g(\cdot) \), and \( U_i, \epsilon_{d,i} \overset{\mathrm{iid}}{\sim} \mathcal{N}(0,1) \). 
The outcome variable is generated as
\(
Y^{(1)}_i = 0.5 D^{(1)}_i + U_i + \epsilon_{y,i} + \sum_{j=1}^{p} \alpha_j Z_{i,j}^{(1)},
\)
where \( \epsilon_{y,i} \overset{\mathrm{iid}}{\sim}   \mathcal{N}(0,1) \), and the pleiotropic effects satisfy \( \alpha_j \sim \mathcal{N}(\alpha_j^*, \tau^2_0) \). 
The marginal SNP-outcome associations \( \widehat{\Gamma}_j^{\mathrm{Ou}} \) and SNP-treatment associations \( \widehat{\gamma}_j^{\mathrm{Ou}} \) 
are estimated by regressing \( Y^{(1)} \) and \( D^{(1)} \), respectively, on \( Z_j^{(1)} \). Under this data-generating mechanism, the true parameter satisfies
\[
\Gamma_j^{\mathrm{Ou}} = \beta_0 \gamma_j^{\mathrm{Ou}} + \alpha_j, 
\quad \alpha_j \sim \mathcal{N}(\mu, \tau_0^2).
\]
We examine the performance of our estimator under directional pleiotropy, where $\mu =0.05$ and $\tau_0=0.02$ We compare the following six methods: 

\begin{itemize} \item[E1:] \textbf{MR-Wald (proposed):} \(\widehat{\beta} = \widehat{\beta b} / \widehat{b}\), where \(\widehat{\beta b}\) is obtained from the weighted least squares regression of \(\widehat{\Gamma}_j^{\mathrm{Ou}}\) on \(\widehat{\gamma}_j^{\mathrm{Tr}}\), and \(\widehat{b}\) from the weighted least squares regression of \(\widehat{\gamma}_j^{\mathrm{Ou}}\) on \(\widehat{\gamma}_j^{\mathrm{Tr}}\). \item[E2:] \textbf{MR-Wald-D (proposed):} \(\widehat{\beta} = \widetilde{\beta b} / \widetilde{b}\), where \(\widetilde{\beta b}\) is defined in \eqref{eqn:mr_wald_d}
. \item[E3:] \textbf{Weighted Median \citep{bowden2016consistent}:} defined as the weighted median of the ratio estimates \( \widehat{\beta}_j = {\widehat{\Gamma}^{\mathrm{Ou}}_j}/{\widehat{\gamma}^{\mathrm{Tr}}_j}, \quad j = 1, \dots, p, \) with weights \(w_j = 1 / \mathrm{Var}(\widehat{\beta}_j)\). \item[E4:] \textbf{MR-Egger \citep{burgess2017interpreting}:} estimating \((\beta, \alpha)\) by minimizing \( \sum_{j=1}^p w_j \bigl(\widehat{\Gamma}_j - \widehat{\gamma}_j^{\mathrm{Tr}} \beta - \alpha \bigr)^2, \) which corresponds to the IVW estimator with an intercept term. \item[E5:] \textbf{RAPS \citep{zhao2018statistical}:} a robust adjusted profile likelihood estimator constructed under the assumption that \((\widehat{\Gamma}_j, \widehat{\gamma}_j^{\mathrm{Tr}})\) follows a bivariate normal distribution. \item[E6:] \textbf{DIVW \citep{ye2021debiased}:} a de-biased IVW estimator that corrects for weak instrument bias in the standard IVW method. \end{itemize}

To allow heterogeneity between $\gamma_j^{\mathrm{Tr}}$ and $\gamma_j^{\mathrm{Ou}}$, we set $\gamma_j^{\mathrm{Ou}} = g(\gamma_j^{\mathrm{Tr}})$ 
and consider four specifications for $g$: the identity $g(\gamma) = \gamma$, 
a scaled shift $g(\gamma) = (\gamma+0.1)/2$, a nonlinear transformation $g(\gamma) = \sin(5\pi\gamma)/5$, 
and a nonparametric estimator $\widehat{g}_{\mathrm{Afr}}$ derived from BMI-SNP associations in European and African samples 
(see Section~\ref{sec:realdata_new}). 
Figure~\ref{fig:gfun_graphical} in the manuscript illustrates these functions.

Table~\ref{tab:simresulttc} reports the simulation results under several forms of heterogeneous pleiotropic effects. Across all scenarios, the proposed estimator $\widehat{\beta}^{\text{MR-Wald-D}}$ exhibits uniformly negligible bias and achieves the smallest or near smallest RMSE among all competing methods. Although the associated confidence intervals are noticeably wider than those of the alternative estimators, $\widehat{\beta}^{\text{MR-Wald-D}}$ consistently attains close to nominal coverage, with empirical coverage probabilities around 95

By contrast, the conventional MR Wald, weighted median, RAPS, and dIVW estimators display substantial bias under directional pleiotropy, with bias magnitudes comparable to their corresponding RMSEs. Despite producing relatively short confidence intervals, these methods fail to achieve adequate coverage, with empirical coverage rates essentially equal to zero in all scenarios. This behavior highlights that ignoring directional pleiotropy can lead to severely biased causal effect estimates and unreliable inference.

The MR Egger estimator improves upon these methods by explicitly accounting for directional pleiotropy, leading to reduced bias and improved RMSE relative to MR Wald, weighted median, RAPS, and dIVW. However, MR Egger does not address weak instrument bias, and consequently its coverage probabilities remain below the nominal level in several scenarios, particularly under stronger heterogeneity.

Overall, these results demonstrate that the proposed $\widehat{\beta}^{\text{MR-Wald-D}}$ estimator provides a favorable performance in the presence of heterogeneous directional pleiotropy, delivering reliable inference even when existing methods either suffer from severe bias or fail to achieve nominal coverage. 

\begin{table}
\centering
\caption{Simulation results for various estimators under Scenarios (v). Reported summary statistics include the bias, root-mean-square error (RMSE), confidence interval (CI) length, and CI coverage rate (nominal level 95 \%). We report $(\text{Bias}  /\beta_0)\times100\%$, $(\text{RMSE}/\beta_0)\times100\%$, $(\text{CI length}/\beta_0)\times100\%$, and CI coverage rate. Cell colors indicate performance, with yellow denoting poor and green denoting favorable results.}
\vspace{-1.5em}

\begin{flushleft}
\begin{tikzpicture}[baseline=0ex]
  \begin{axis}[
    hide axis,
    colorbar horizontal,
    point meta min=0,
    point meta max=1,
    colormap={mycolormap}{
      rgb255(0cm)=(255,255,0);
      rgb255(0.33cm)=(204,204,0);
      rgb255(0.66cm)=(0,153,0);
      rgb255(1cm)=(0,255,0)
    },
    colorbar style={
      height=0.12cm,
      width=5cm,
      xtick={0,1},
      xticklabels={Poor,Favorable},
      ticklabel style={font=\scriptsize}
    }]
  \addplot [draw=none] coordinates {(0,0)};
  \end{axis}
\end{tikzpicture}
\end{flushleft}
\label{tab:simresulttc}
\vspace{-4em}
\scalebox{1}{
\begin{tabular}{ccABFD}
\cmidrule(r){1-6}
$g(\gamma)$&Method&\multicolumn{4}{c}{Scenario (i)}  \\ 
\cmidrule(r){1-6}
\cmidrule(r){1-6}
&&\multicolumn{1}{c}{Bias} & \multicolumn{1}{l}{RMSE} & \multicolumn{1}{l}{\hspace{-2em}CI length} & \multicolumn{1}{l}{CI} \\
&MR-Wald&129.8 & 129.8 & 16.9 & 0.0\\
&MR-Wald-D&0.1 & 25.1 & 93.3 & 93.2\\
$\gamma$&W.Median&107.0 & 107.2 & 17.0 & 0.0\\
&MR-Egger&-21.5 & 29.9 & 77 & 79.9\\
&RAPS&129.7 & 129.7 & 19.6 & 0.0\\
&DIVW&129.7 & 129.8 & 9.8 & 0.0\\
\cmidrule(r){1-6} 
&MR-Wald&112.9 & 113.0 & 13.6 & 0.0\\
&MR-Wald-D&0.5 & 47.9 & 189.3 & 95.5\\  
$\frac{\gamma+0.1}{2}$&W.Median&117.7 & 117.9 & 17.7 & 0.0\\
&MR-Egger&-60.3 & 63.4 & 76.4 & 13.4\\
&RAPS&144.7 & 144.7 & 21.3 & 0.0\\
&DIVW&144.8 & 144.9 & 10.2 & 0.0\\
\cmidrule(r){1-6} 
&MR-Wald&54.8 & 54.9 & 6.6 & 0.0
\\&MR-Wald-D&0.0 & 20.1 & 79.3 & 95.7\\
$\frac{\sin(5\pi \gamma)}{5}$&W.Median&232.3 & 232.4 & 22 & 0.0\\
&MR-Egger&-8.0 & 21.3 & 79.2 & 93.1\\
&RAPS&267.0 & 267.1 & 24.7 & 0.0\\
&DIVW&266.7 & 266.8 & 13.3 & 0.0\\
\cmidrule(r){1-6}
&MR-Wald&190.3 & 190.3 & 28.3 & 0.0\\
&MR-Wald-D&-0.4 & 21.8 & 83.6 & 95.2\\
$\widehat{g}_{Afr}(\gamma)$&W.Median&78.7 & 78.9 & 16.0 & 0.0\\
&MR-Egger&-14 & 24.3 & 76.7 & 88.4\\
&RAPS&97.9 & 98.0 & 18.4 & 0.0\\
&DIVW&98.0 & 98.1 & 9.1 & 0.0\\
\cmidrule(r){1-6}
\end{tabular}
}
\end{table}

\newpage

\setlength{\bibsep}{0pt}
\putbib

\end{bibunit}
\end{document}